%% file: SMP-14-023_temp.tex
\pdfoutput=1

\documentclass[11pt,twoside,a4paper,cmspaper,final,collab]{cms-tdr}
\def\svnVersion{394679}\def\cmsCernNoTag{CERN-EP/2016-231}\def\cmsCernDate{\today}\def\cmsMessage{Published in Physical Review D as \href{http://dx.doi.org/10.1103/PhysRevD.95.052002}{\doi{10.1103/PhysRevD.95.052002.}}}

\begin{document}\cmsNoteHeader{SMP-14-023}

\hyphenation{had-ron-i-za-tion}
\hyphenation{cal-or-i-me-ter}
\hyphenation{de-vices}
\RCS$Revision: 394439 $
\RCS$HeadURL: svn+ssh://svn.cern.ch/reps/tdr2/papers/SMP-14-023/trunk/SMP-14-023.tex $
\RCS$Id: SMP-14-023.tex 394439 2017-03-16 15:46:10Z ahortian $
\newlength\cmsFigWidth
\ifthenelse{\boolean{cms@external}}{\setlength\cmsFigWidth{0.85\columnwidth}}{\setlength\cmsFigWidth{0.4\textwidth}}
\ifthenelse{\boolean{cms@external}}{\providecommand{\cmsLeft}{top\xspace}}{\providecommand{\cmsLeft}{left\xspace}}
\ifthenelse{\boolean{cms@external}}{\providecommand{\cmsRight}{bottom\xspace}}{\providecommand{\cmsRight}{right\xspace}}
\providecommand{\BLACKHAT} {\textsc{BlackHat}\xspace}
\providecommand{\FEWZ}{\textsc{fewz}\xspace}
\providecommand{\MT}{\ensuremath{M_\mathrm{T}}\xspace}
\providecommand{\MGvATNLO}{\MADGRAPH{5}\_a\textsc{mc@nlo}\xspace}
\cmsNoteHeader{SMP-14-023}
\title{Measurements of differential cross sections for associated production of a \texorpdfstring{$\PW$ boson and jets in proton-proton collisions at $\sqrt{s}=8\TeV$}{W boson and jets in proton-proton collisions at sqrt(s) = 8 TeV}}

\date{\today}

\abstract{
  Differential cross sections for a $\PW$ boson produced in association with jets are measured in a data sample of proton-proton collisions at a center-of-mass energy of 8\TeV recorded with the CMS detector and corresponding to an integrated luminosity of 19.6\fbinv. The $\PW$ bosons are identified through their decay mode $\PW\to \mu\nu$. The cross sections are reported as functions of jet multiplicity, transverse momenta, and the scalar sum of jet transverse momenta ($\HT$) for different jet multiplicities. Distributions of the angular correlations between the jets and the muon are examined, as well as the average number of jets as a function of $\HT$ and as a function of angular variables.
The measured differential cross sections are compared with tree-level and higher-order recent event generators, as well as
next-to-leading-order and next-to-next-to-leading-order
theoretical predictions.
The agreement of the generators with the measurements builds confidence in their use for the simulation of $\PW$+jets background processes in searches for new physics at the LHC.
}

\hypersetup{%
pdfauthor={CMS Collaboration},%
pdftitle={Measurements of differential cross sections for associated production of a W boson and jets in proton-proton collisions at sqrt(s)=8 TeV},%
pdfsubject={CMS},%
pdfkeywords={CMS, physics, W+Jets, vector boson, jets, electroweak, standard model}}

\maketitle
\section{Introduction}
\label{introduction}

This paper presents measurements of differential cross sections for associated production of a $\PW$ boson and jets in proton-proton collisions at a center-of-mass energy of 8\TeV at the CERN LHC. Measurements of the production of vector bosons in association with jets provide stringent tests of perturbative quantum chromodynamics (QCD). In addition, the production of a $\PW$ boson in association with jets ($\PW$+jets) is the main background source for rarer standard model (SM) processes, such as top quark production and Higgs boson production in association with a $\PW$ boson, and it is also a prominent background to several searches for physics beyond the SM.

The studies described here focus on the production of $\PW$+jets with the subsequent decay of the $\PW$ boson into a muon and a neutrino. The final-state topology is characterized by one isolated muon with high transverse momentum \pt, significant missing transverse energy $\MET$, and up to seven jets. Because of higher trigger thresholds and additional systematic uncertainties, the decay channel of the $\PW$ boson into an electron and a neutrino is not considered.
Differential cross sections are extracted as functions of jet multiplicity, the \pt of the jets, the scalar sum of the jet transverse momenta, $\HT$, and the pseudorapidity $\eta$ of the jets.
Measurements of differential cross sections as functions of angular correlation variables are also performed. The average number of jets per event, $\langle N_\text{jets}\rangle$, is further studied as a function of $\HT$ and angular variables.

These measurements are based on a sample of proton-proton collisions at $\sqrt{s}=8\TeV$ recorded with the CMS detector, corresponding to an integrated luminosity of 19.6\fbinv. In order to perform differential measurements of the $\PW$+jets cross section, a high-purity sample of muonic $\PW$ boson decays is selected and the distributions are corrected back to the stable-particle level by means of a regularized unfolding procedure.
The measured fiducial differential cross sections are compared to the predictions of Monte Carlo (MC) event generators for the $\PW + n$~jets hard-scattering process, with final states of different parton multiplicities matched to parton showers. The generators used are {\MADGRAPH} 5~\cite{MG5} interfaced with \PYTHIA ~\cite{PYTHIA}, which uses a leading-order (LO) matrix element calculation, and \MGvATNLO~\cite{madgraphamcnlo} and {\SHERPA 2}~\cite{Hoeche:2012yf}, which use next-to-leading-order (NLO) matrix element calculations.
The differential cross sections are also compared with the NLO parton-level predictions of \BLACKHAT{}+\SHERPA~\cite{Bern:2013zja} and with a next-to-next-to-leading-order calculation ($N_\text{jetti}$ NNLO) for the production of $\PW + 1$ jet~\cite{Boughezal:2015dva,Boughezal:2016dtm}.

Previous measurements of $\PW+$jets production were performed by the CDF~\cite{CDFwplusjets} and D0~\cite{Abazov2011200,PhysRevD.88.092001} Collaborations in  proton-antiproton collisions at $\sqrt{s}=1.96\TeV$ at the Fermilab Tevatron. The ATLAS~\cite{AtlasWJets4p6pb} and CMS~\cite{Wjets7Tev} Collaborations have measured $\PW+$jets production cross sections in proton-proton collisions at $\sqrt{s} = 7\TeV$ at the LHC, corresponding to integrated luminosities of 4.6 and 5.0\fbinv, respectively.

The differential $\PW$+jets cross sections at 8\TeV presented in this paper extend the kinematic reach of the 7\TeV CMS results (in \pt, $\HT$, and jet multiplicity) and expand the set of kinematic observables studied. The increased center-of-mass energy widens the observed ranges of kinematic quantities, such as the \pt of the jets in the event and $\HT$ for higher jet multiplicities, that are sensitive to higher-order processes. The larger data sample motivates the increase in number and complexity of the angular correlation variables examined to more accurately understand how particle emissions are modeled by the MC generators used in the analysis of the LHC data and by the most current NLO calculations. The quantity $\langle N_\text{jets}\rangle$ is studied as a function of $\HT$ and angular correlation variables to further explore the modeling of higher-order processes and correlations among emitted particles. In addition, the measurements are expanded to include the cross section dependence on variables, such as the dijet invariant mass in different multiplicity ranges, that are sensitive to the presence of physics beyond the SM.

This paper is organized in the following manner: Section~\ref{cms} describes the CMS detector. Section~\ref{samples} describes the MC simulated samples and the data sample used for the analysis. The identification criteria for the final-state objects and the selection criteria used to select $\PW (\to\mu\nu)$+jets events are listed in Section~\ref{eventselection}. Section~\ref{background} describes the modeling of the backgrounds. The variables used for the differential cross section measurements are detailed in Section~\ref{observables}. The procedure used for unfolding is detailed in Section~\ref{unfolding}, and Section~\ref{systematics} describes the systematic uncertainties. Finally, Section~\ref{results} gives the results and Section~\ref{summary} summarizes them.

\section{The CMS detector}
\label{cms}
The CMS detector consists of an inner tracking system and electromagnetic (ECAL) and hadron (HCAL) calorimeters surrounded by a 3.8\unit{T} superconducting solenoid. The inner tracking system consists of a silicon pixel and strip tracker, providing the required granularity and precision for the reconstruction of vertices of charged particles in the range $0 < \phi < 2\pi$ in azimuth and $\abs{\eta}<2.5$ in pseudorapidity. The crystal ECAL and the brass and scintillator HCAL are used to measure with high resolution the energies of photons, electrons, and hadrons for $\abs{\eta}<3.0$.
The three muon systems surrounding the solenoid cover the region $\abs{\eta}<2.4$ and are composed of drift tubes in the barrel region $(\abs{\eta}<1.2)$, cathode strip chambers in the endcaps $(0.9<\abs{\eta}<2.4)$, and resistive plate chambers in both the barrel region and the endcaps $(\abs{\eta}<1.6)$.
Events are recorded based on a trigger decision using information from the CMS detector subsystems. The first level of the trigger system, composed of custom hardware processors, uses information from the calorimeters and muon detectors to select events in a fixed time interval of less than 4\mus.
The final trigger decision is based on the information from all subsystems, processed by the high-level trigger (HLT), which consists of a farm of computers running a version of the reconstruction software optimized for fast processing. The HLT processor farm decreases the event rate from
100\unit{kHz} to less than 1\unit{kHz}, before data storage.
A more detailed description of the CMS detector, together with a definition of the coordinate system used and the relevant kinematic variables, can be found in Ref.~\cite{Chatrchyan:2008zzk}.

\section{Data and simulated samples}
\label{samples}

Events are retained if they pass a trigger selection requiring one isolated muon with $\pt>24\GeV$ and $\abs{\eta}<2.1$.

Signal and background processes are generated with various state-of-the-art generators and passed through detector simulation based on \GEANTfour~\cite{GEANT4} description of CMS. 
Each simulated sample is normalized to the integrated luminosity of the data sample. The simulated events are required to pass an emulation of the trigger requirements applied to the data. Trigger efficiencies in the simulation are corrected for differences with respect to the data. Simulations also include additional collisions in the same or adjacent bunch crossings (pileup, PU). To model PU, minimum-bias events generated with \PYTHIA{}6 using the Z2* tune~\cite{tunez2} are superimposed on the simulated events, matching the multiplicity of PU collisions observed in data, which has an average value of approximately 21.

The $\PW$+jets signal process is simulated with the matrix element (ME) generator \MADGRAPH 5.1.1~\cite{MG5} interfaced with \PYTHIA 6.426 using the Z2* tune for parton showering and hadronization. This sample of events, denoted \MADGRAPH{}5+\PYTHIA{}6 (denoted as MG5+PY6 in the figure legends), is produced with the CTEQ6L1 parton distribution function (PDF) set~\cite{CTEQ} and is normalized to the inclusive NNLO cross section calculated with {\FEWZ} 3.1~\cite{FEWZ}. The \MADGRAPH{}5+\PYTHIA{}6 calculation includes the production of up to four partons at LO. The jets from matrix elements are matched to parton showers following the \kt-jet MLM prescription~\cite{Matching}, where partons are clustered using the \kt algorithm~\cite{fastjetmanual} with a distance parameter of~1.
The merging of parton showers and matrix elements with the MLM scheme uses a matching scale of 20\GeV.
The factorization and renormalization scales for the 2$\to$2 hard process in the event are chosen to be the transverse mass of the $\PW$ boson produced in the central process. The \kt computed for each QCD emission vertex is used as renormalization scale for the calculation of the strong coupling constant $\alpha_{S}$ of that vertex.

Background processes include $\ttbar$, single top quark, $\cPZ/\gamma^*$+jets, diboson ($\cPZ\cPZ$/$\PW\cPZ$/$\PW\PW$) + jets, and QCD multijet production. Their contributions, with the exception of QCD multijet production, are estimated from simulation. The
simulated samples of $\ttbar$ and $\cPZ/\gamma^*$+jets events are generated with {\MADGRAPH} version 5.1.1; the single top quark samples ($s$-, $t$-, and t$\PW$-channel production) are generated with \POWHEG version 1~\cite{Nason:2004rx,Frixione:2007vw,Alioli:2010xd,Alioli:2009je}; and the diboson samples ($\PW\PW$, $\PW\cPZ$, or $\cPZ\cPZ$) are generated with \PYTHIA 6.424 using the Z2$^*$ tune.
The simulations with \MADGRAPH and \PYTHIA use the CTEQ6L1 PDFs, and the simulations with \POWHEG use the CTEQ6M PDFs. The $\cPZ/\gamma^*$+jets sample is normalized to the NNLO inclusive cross section calculated with {\FEWZ} 3.1~\cite{FEWZ}. Single top quark and diboson samples are normalized to NLO inclusive cross sections calculated with {\MCFM}~\cite{ mcfm_t, mcfm_Wt, mcfm_tch,mcfm_diboson}.
The $\ttbar$ contribution is normalized to the predicted cross section at NNLO with next-to-next-to-leading-logarithm accuracy~\cite{Czakon:2013goa}.

When comparing the measurements with the theoretical prediction, other event generators are used for the $\PW $+jets process. Those generators, which are not used for the measurement itself, are described in Section~\ref{results}.

\section{Object identification and event selection}
\label{eventselection}

The final-state particles in $\PW$+jets events are identified and reconstructed with the particle-flow (PF) algorithm~\cite{pf_algos0,pf_algos}, which optimally combines the information from the various elements of the CMS detector.

Muon PF candidates are reconstructed as tracks in the muon system that are matched to tracks reconstructed in the inner tracking system~\cite{CMS-PAPERS-MUO-10-004}. Muons are required to have $\pt > 25\GeV$ and to be reconstructed in the HLT fiducial volume $\abs{\eta}< 2.1$. The track associated with a muon candidate is required to have hits in at least six strip tracker layers, at least one pixel hit, segments from at least two muon stations, and a good quality global fit with $\chi ^2$ per degree of freedom $<10$.
In order to reduce the contamination due to muons that do not originate from the decay of a $\PW$ boson, an isolation requirement is imposed:
\ifthenelse{\boolean{cms@external}}{
\begin{multline}
I_\text{iso} = \frac{1}{\pt^{\mu}}\Biggl[\sum^\text{charged}\pt + \max\Bigl(0, \sum^\text{neutral}\pt \\+ \sum^{\gamma}\pt - 0.5\sum^\text{PU}\pt\Bigr)\Biggr] \leq 0.12,
\end{multline}
}{
\begin{equation}
I_\text{iso} = \frac{1}{\pt^{\mu}}\left[\sum^\text{charged}\pt + \max\left(0, \sum^\text{neutral}\pt + \sum^{\gamma}\pt - 0.5\sum^\text{PU}\pt\right)\right] \leq 0.12,
\end{equation}}
where the sums run over charged hadrons originating from the primary vertex of the event, neutral hadrons, photons ($\gamma$), and charged hadrons not originating from the primary vertex but from PU; only PF candidates with direction within a cone defined by $\Delta R = \sqrt{\smash[b]{(\Delta \phi )^2 + (\Delta\eta) ^2}} < 0.4$ around the direction of the muon candidate track are considered. The transverse momentum of the muon candidate is denoted by $\pt^{\mu}$. Because neutral PU particles deposit on average half as much energy as charged PU particles, the contamination in the isolation cone from neutral particles coming from PU interactions is estimated as $0.5\sum^{\text{PU}}\pt$ and it is subtracted in the definition of $I_\text{iso}$.

To reject muons from cosmic rays, the transverse impact parameter of the muon candidate with respect to the primary vertex is required to be less than 2\unit{mm}, and the longitudinal distance of the tracker track from the primary vertex is required to be less than 5\unit{mm}.
Trigger efficiency corrections, as well as muon identification and isolation efficiency corrections, are applied to the simulation as a function of \pt and $\eta$ on an event-by-event basis and are generally less than $4\%$ and $2.5\%$, respectively.

Jets and transverse missing energy \MET are also reconstructed using the PF algorithm. The missing momentum vector $\ptvecmiss$ of an event is defined as the negative of the vectorial \pt sum of the particles reconstructed with the PF algorithm; \MET is defined as the magnitude of the $\ptvecmiss$ vector~\cite{CMSMET}. Jets are reconstructed using the anti-\kt~\cite{antikt,fastjetmanual} algorithm with a distance parameter of 0.5. Reconstructed jet energies are corrected with \pt- and $\eta$-dependent correction factors to account for the following
effects: nonuniformity and nonlinearity of the ECAL and HCAL energy response to neutral
hadrons, the presence of extra particles from PU interactions, the thresholds used in jet constituent
selection, reconstruction inefficiencies, and possible biases introduced by the clustering
algorithm. Jet energy corrections are derived from simulation and adjusted using measurements of the \pt balance in dijet and $\gamma$+jet events~\cite{JetCorr}.
The jet energy resolution is approximately 15\% at 10\GeV, 8\% at 100\GeV, and 4\% at 1\TeV~\cite{JetCorr}.
Jets are required to have $\pt > 30 \GeV$,  $\abs{\eta} <2.4$, and a spatial separation from muon
candidates of $\Delta R > 0.5$. In order to reduce the contamination from PU, jets are required to be matched to the same primary vertex as the muon candidate.

The primary background process for the measurement of $\PW$+jets at high jet multiplicities (4 or more) is $\ttbar$ production. To reduce the $\ttbar$ contamination, a veto is applied that removes events containing one or more \cPqb-tagged jets.
The tagging criteria used for this veto are based on the combined secondary vertex algorithm (CSV)~\cite{btag}, which exploits the long lifetime of \cPqb~hadrons by combining information about impact parameter significance, secondary vertices, and jet kinematic properties. Differences in the \cPqb~tagging efficiency in data and simulation, as well as differences in mistagging rates, are corrected using scale factors~\cite{btag} determined as a function of \pt in multijet and $\ttbar$ events. Specifically, the tagging efficiency in simulation is decreased by randomly untagging \cPqb-tagged jets such that the data and simulated efficiencies are matched.  Additionally, a small adjustment to the mistagging rates is performed by randomly tagging untagged jets in simulated events such that the data and simulated mistagging rates agree within uncertainties.

In order to select a $\PW(\to \mu \nu)+$jets sample, events are required to contain exactly one
muon satisfying the muon selection criteria described above and one or more jets with $\pt> 30 \GeV$.
Events containing additional muons with $\pt> 15\GeV$ are vetoed.
Events are required to have $\MT >50 \GeV$,
where \MT, the transverse mass of the muon and missing transverse energy, is defined as
$\MT = \sqrt{\smash[b]{2 \pt^{\mu}   \MET \left( 1 - \cos{\Delta \phi} \right) }}$ and $\Delta \phi$ is the difference in the azimuthal angle between the direction of the muon momentum and $\ptvecmiss$.

\section{Estimation of the backgrounds}
\label{background}

Leptonic $\PW$ boson decays are characterized by a prompt, energetic, isolated lepton and a neutrino
giving rise to significant $\MET$.
Background processes with final-state signatures similar to that of $\PW$+jets are $\ttbar$, single top quark, $\cPZ$+jets, diboson ($\cPZ\cPZ$/$\PW\cPZ$/$\PW\PW$)+jets, and QCD multijet production.
All background processes except for QCD multijet production are simulated by MC event generators and are normalized as described in Section~\ref{samples}.

The multijet background is estimated using a control data sample with an inverted muon isolation requirement. In the control data sample, the muon misidentification rate is estimated in a multijet-enriched sideband region with $\MT<50$ GeV, and the shape of the multijet distribution is determined in the region with $\MT > 50$ GeV. The template for the multijet shape is rescaled according to the muon misidentification rate. This method of estimation was used in the measurement of the $\PW$+jets fiducial cross sections at 7\TeV and is described in detail in Ref.~\cite{Wjets7Tev}.

The dominant source of background comes from the $\ttbar$ process, which is reduced by the application of the \cPqb~jet veto described in Section~\ref{eventselection}. For jet multiplicities of 1 to 7, the \cPqb~jet veto rejects 62--88\% of the predicted $\ttbar$ background, while eliminating 4--22\% of the predicted $\PW$+jets signal.

\section{Measured observables}
\label{observables}

Fiducial cross sections are measured as a function of jet multiplicity, inclusively and exclusively, as a function of jet \pt and $\abs{\eta}$, and as a function of $\HT$. In terms of angular correlations between jets, cross sections are measured as a function of the difference in rapidity $\Delta y(j_i,j_k)$, and of the difference in azimuthal angle $\Delta \phi(j_i,j_k)$, between the $i$th and $k$th jets from the \pt-ordered list of jets in the event. Cross sections are also measured as a function of the differences in rapidity and in azimuthal angle between rapidity-ordered jets, most notably $\Delta y(j_F,j_B)$ and $\Delta \phi(j_F,j_B)$, the differences between the most forward and the most backward jet in the event.
Cross sections are measured as a function of $\Delta R(j_1,j_2)=\sqrt{\Delta \phi(j_1,j_2)^2 + \Delta y(j_1,j_2)^2}$ between \pt-ordered jets.
The dependence of the cross section on the invariant mass of the two leading jets for different jet multiplicities is also examined.
The difference in azimuthal angle between the muon and the leading jet is measured for different jet multiplicities.
The dependence of $\langle N_\text{jets}\rangle$ on $\HT$ and on both $\Delta y(j_1,j_2)$ and $\Delta y(j_F,j_B)$ is studied for different jet multiplicities.

\begin{figure}[htp]
\centering
\includegraphics[width=0.48\textwidth]{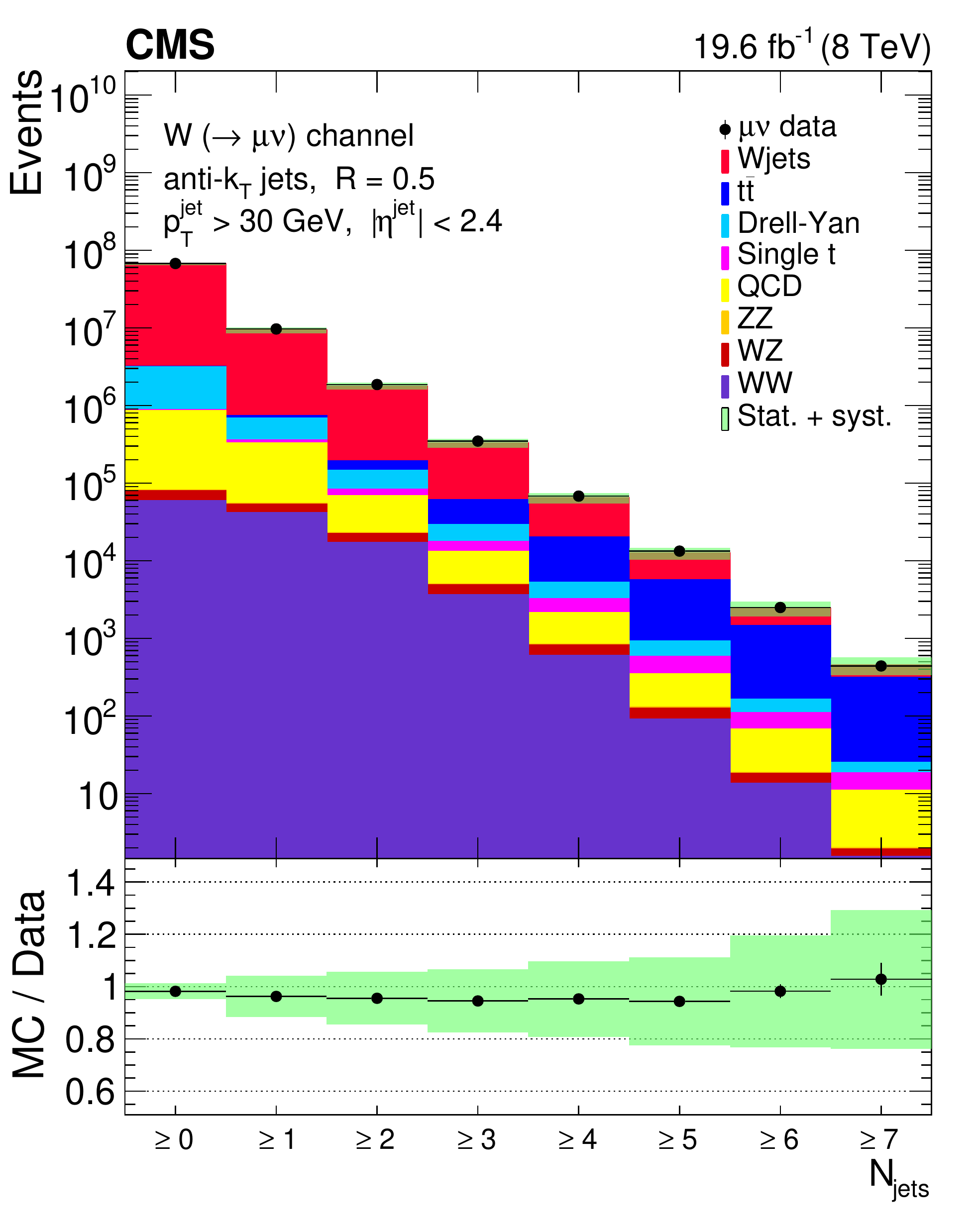}

\caption{Distribution of inclusive jet multiplicity, for reconstructed data (points) and simulated signal and backgrounds (histograms). The ratio of simulated and measured data events is shown below the distribution. The data points are shown with statistical error bars. The error band represents the total statistical and systematic uncertainty.}
\label{reco}
\end{figure}

Before correcting for detector effects and determining the cross section values, we compare the kinematic distributions reconstructed in data with the predictions for the simulated $\PW$+jets signal and the simulated background processes. The comparison of reconstructed data and simulated signal and background processes is shown in Fig.~\ref{reco} for the inclusive jet multiplicity. The uncertainty band represents the total statistical and systematic uncertainty including uncertainties in the jet energy scale and resolution, the muon momentum scale and resolution, the integrated luminosity, the pileup modeling, the normalizations of the background processes, the modeling of the \PW\cPqb\ contribution in the signal simulation, and the reconstruction, identification, and trigger efficiencies.

The number of events in each bin of exclusive reconstructed jet multiplicity for both data and simulated signal and backgrounds is listed in Table~\ref{events}. The predicted total yields agree well with the data yields for all the values of jet multiplicity.

\begin{table*}
\centering
\topcaption{Number of events in data and simulation as a function of exclusive reconstructed jet multiplicity. The purity is the number of simulated signal events (\PW +jets) divided by the total number of simulated signal and background events (Total). The ratio is the total number of simulated signal and background events divided by the number of data events.}
\begin{scotch}{l|rrrrrrrrrrr|}
 $N_\text{jets}$ & \multicolumn{1}{c}{0}& \multicolumn{1}{c}{1}& \multicolumn{1}{c}{2}& \multicolumn{1}{c}{3}& \multicolumn{1}{c}{4}& \multicolumn{1}{c}{5}& $6$ & \multicolumn{1}{c}{$\geq$7}\\ \hline
   \PW\PW +jets        & 18\,093 & 24\,420 & 13\,472 & 3057 & 515 & 77 & 12 & 1 \\
   \PW\cPZ +jets        & 8125 & 6799 & 4153 & 1042 & 183 & 30 & 4 & 0 \\
   \cPZ\cPZ +jets        & 932 & 669 & 384 & 96 & 18 & 3 & 0 & 0 \\
   QCD multijet        & 570\,722 & 228\,188 & 37\,154 & 6734 & 1076 & 171 & 40 & 9 \\
   Single top quark       & 6438 & 14\,386 & 9838 & 3444 & 877 & 196 & 34 & 7 \\
   $\cPZ/\gamma$+jets        & 1\,935\,191 & 265\,387 & 51\,613 & 9570 & 1697 & 281 & 48 & 6 \\
   \ttbar    & 1504 & 7576 & 16\,052 & 17\,377 & 10\,090 & 3487 & 1000 & 288 \\
   \PW +jets        & 54\,617\,816 & 6\,999\,393 & 1\,320\,381 & 222\,457 & 37\,822 & 5857 & 860 & 139 \\
\hline
 Total & 57\,158\,821 & 7\,546\,818 & 1\,453\,047 & 263\,777 & 52\,278 & 10\,102 & 1998 & 450 \\
\hline
 Purity        & 0.96 & 0.93 & 0.91 & 0.84 & 0.72 & 0.58 & 0.43 & 0.31 \\
\hline
 Data          & 57\,946\,098 & 7\,828\,967 & 1\,517\,517 & 279\,678 & 54\,735 & 10\,810 & 2058 & 441 \\
 Ratio         & 0.99 & 0.96 & 0.96 & 0.94 & 0.96 & 0.93 & 0.97 & 1.02 \\
\end{scotch}
\label{events}
\end{table*}

\section{Unfolding procedure}
\label{unfolding}

The fiducial cross sections are obtained by subtracting the simulated backgrounds and the estimated multijet background from the data distributions and correcting the background-subtracted data distributions back to the particle level using an unfolding procedure. This procedure takes detector effects such as detection efficiency and resolution into account. The unfolding procedure is performed using the iterative d'Agostini method~\cite{bayes2} implemented in the {RooUnfold} toolkit~\cite{RooUnfold}.
Regularization is achieved by choosing the optimal value of number of iterations, based on a $\chi^2$ comparison of the unfolded distributions, corrected with the response matrix defined below, with the reconstructed background-subtracted data distributions.
To assess the dependence of the results on the unfolding method, we also used the singular-value decomposition method~\cite{SVDUnfold}. The results from the two methods agree within the uncertainties.

A response matrix, which defines the probability of event migration between the particle-level and reconstructed phase space as well as the
overall reconstruction efficiency, is constructed using a $\PW$+jets sample simulated with
\MADGRAPH{}5+\PYTHIA{}6.
The particle-level selection defines the fiducial phase space of the measurements and is identical to the selection applied to the reconstructed objects, including the requirement of exactly one muon with $\pt>25\GeV$ and $\abs{\eta}<2.1$, jet $\pt >30\GeV$ and $\abs{\eta}<2.4$, and $\MT >50\GeV$.
The particle-level \MET is determined using the neutrino from the decay of the $\PW$ boson. The momenta of all photons in a cone of $\Delta R
< 0.1$ around the muon are added to that of the muon in order to take into account final-state radiation. The particle-level jets are clustered using the  anti-\kt~\cite{antikt} algorithm with a distance parameter of 0.5. The jet clustering algorithm uses all particles after decay and fragmentation, excluding neutrinos. The \cPqb~jet veto explained in Section~\ref{eventselection} is treated as an overall event selection requirement, and the cross section is corrected by the unfolding procedure to correspond to $\PW$ boson production in association with jets of any flavor. The contribution from $\PW\to \tau\nu$ decays resulting in a muon in the final state is estimated to be small (${\sim}$1\% of selected signal sample), and it is therefore not considered as part of the signal definition at the particle level.

\section{Systematic uncertainties}
\label{systematics}

The systematic uncertainties are evaluated by repeating all of the analysis steps (including the subtraction of backgrounds and unfolding) with systematic variations corresponding to the different sources of uncertainty. The difference in each bin between the results obtained with and without the variation is taken as the systematic uncertainty.

The dominant systematic uncertainties are those associated with the jet energy scale (JES) and jet energy resolution (JER). The JES uncertainty is propagated to the cross section measurements by varying the jet \pt scale in data by the magnitude of the uncertainty, which is parametrized as a function of \pt and $\eta$~\cite{JetCorr}. Shifting the value of \pt for each individual jet affects \MET, therefore \MET is recalculated. This variation also affects the value of \MT, which is used in the event selection. The uncertainties related to JER are assessed by varying within their uncertainties the calibration factors applied to the simulation to reproduce the resolution observed in data~\cite{JetCorr}. As in the case of JES, the changes in jet \pt due to JER are propagated to the calculation of \MET and \MT.

An uncertainty of 0.2\% in the muon momentum scale and an uncertainty of 0.6\% in muon momentum resolution are assigned~\cite{CMS-PAPERS-MUO-10-004}. The effects of these uncertainties on the measured cross sections are evaluated by varying the momentum scale and by fluctuating the muon momentum in the simulation.

The systematic uncertainty associated with the generator used to build the unfolding response matrix is assessed by weighting the simulation to agree with the data in each distribution and constructing an alternative response matrix to unfold the data. The difference between the unfolded results obtained using the weighted response matrix and the nominal results is taken as the systematic uncertainty associated with the unfolding response matrix. For the leading jet \pt cross section, the resulting uncertainty is in the range 0.02--10.8\%. The higher values in this range, as well as in other uncertainty ranges, are caused by statistical fluctuations in the data and simulation samples in certain kinematic regions.

Other sources of systematic uncertainty include the normalization of the background processes; \cPqb\ tagging efficiency; modeling of the \PW\cPqb\ contribution in the signal simulation; integrated luminosity; PU modeling; muon trigger, isolation, and identification criteria; and the finite number of simulated events used to build the response matrix.

Background normalization uncertainties are determined by varying the cross sections of the backgrounds within their theoretical uncertainties~\cite{mcfm_Zjj, mcfm_diboson, mcfm_t, mcfm_Wt, mcfm_tch}. The theoretical cross section uncertainties are 6\% for $\cPZ\cPZ$ and $\PW\cPZ$, 8\% for $\PW\PW$, and 4\% for $\cPZ$+jets for the region $M_{\mu \mu} > 50 \GeV$. For single top quark production, the uncertainties are 6\% for the $s$ and $t$ channels and 9\% for the \cPqt\PW\ channel.
The uncertainty in the $\ttbar$ modeling is assessed by comparing data and simulation in a data control region with two or more b-tagged jets. Simulated events are rescaled to match data in the control region, and the difference in the unfolded results with or without rescaling applied is taken as the systematic uncertainty related to $\ttbar$ modeling. The scale factors are about 1.26 for jet multiplicity of 2, and between 1.0 and 1.1 for jet multiplicities larger than 2, leading to uncertainties in the measured cross sections that range from 0.4\% to 27\% for jet multiplicities of 2 to 7. The estimate of the multijet background has an uncertainty based on the number of events in the inverted isolation sample and in the control regions where the normalization of the multijet background is calculated. In addition, the systematic variations applied to the backgrounds in the multijet control regions introduce variations in the multijet normalization and shape.

Uncertainties in the ratio of the \cPqb~tagging efficiencies in data and simulation are estimated~\cite{CMS-PAS-BTV-13-001}, leading to uncertainties in the measured cross sections in the range 0.4\% to 25\% for jet multiplicities of 1 to 7.

The uncertainty related to the normalization of the \PW\cPqb\ content in the signal is estimated by examining the agreement between data and simulation as a function of jet multiplicity in a control region defined by requiring exactly one \cPqb-tagged jet. The normalization of \PW\cPqb\ production is found to be underestimated in the simulation by a factor of 1.3. Enhancing the \PW\cPqb~process in simulation by this factor leads to an estimated uncertainty in the measurement of up to 0.8\% for a jet multiplicity of 7.

The uncertainty in the integrated luminosity is 2.6\%~\cite{CMS-PAS-LUM-13-001}.
The uncertainty in the modeling of PU in simulation is assigned by varying the inelastic cross section by $\pm$5\%~\cite{CMS-PAPERS-FWD-11-001}.

Uncertainties in the differences between efficiencies in data and simulation for the trigger, muon isolation, and muon identification criteria are generally less than $3\%$.

An uncertainty due to the finite number of simulated events used to construct the response matrix is estimated by randomly
varying the content of each bin of the response matrix according to a Poisson uncertainty. The standard deviation of the unfolded results is taken as an estimate of the uncertainty. It ranges from 0.1\% to 6.9\% for jet multiplicities of 1~to~7.

\begin{figure}[htb]
\centering
    \includegraphics[width=0.48\textwidth]{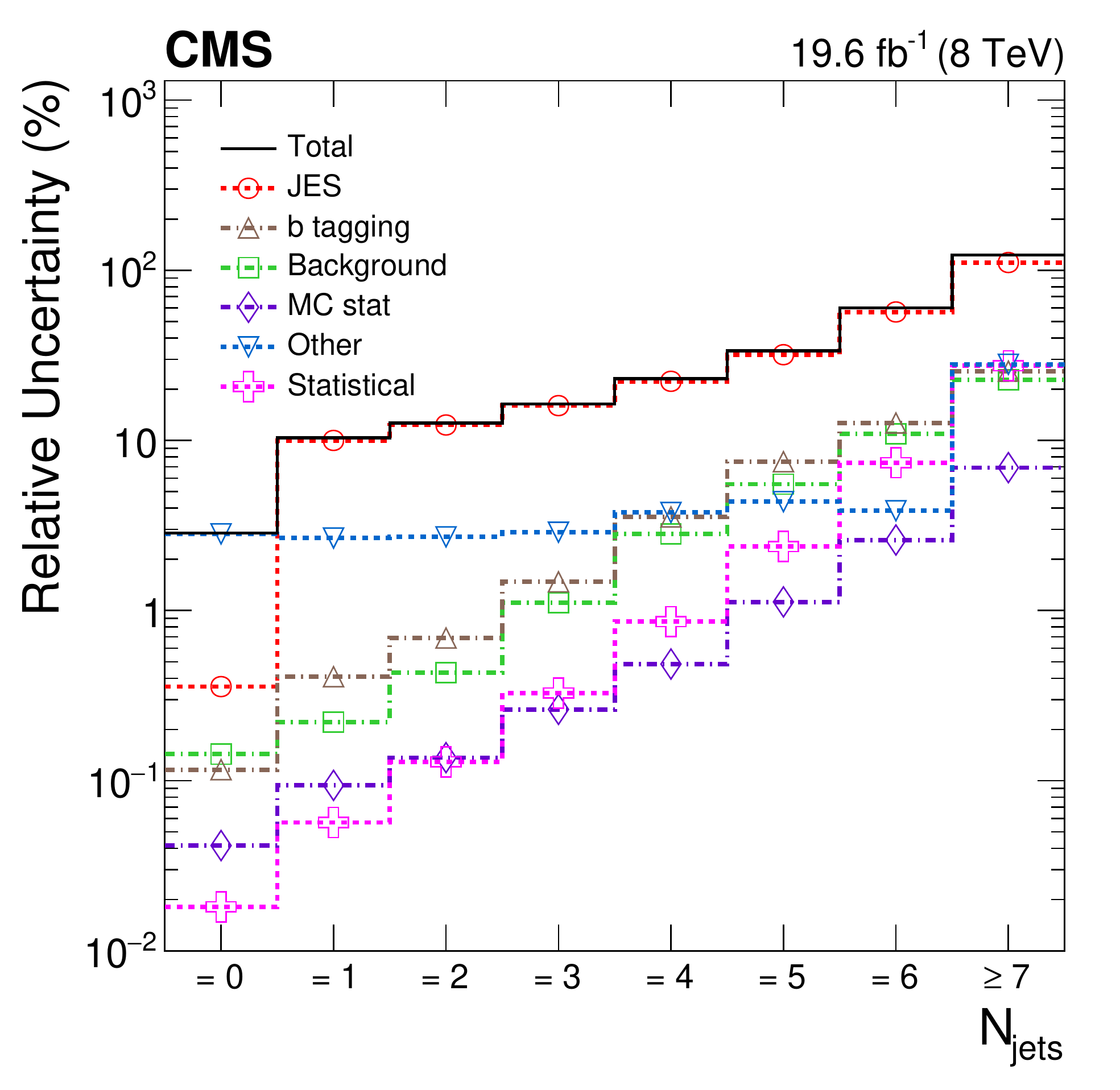}
    \caption{Systematic uncertainties in the measured cross section as a function of jet multiplicity, including uncertainties due to jet energy scale (JES), background normalization, \cPqb~tagging efficiency, finite number of simulated events used to construct the response matrix (MC stat), and other systematic uncertainties mentioned in Section~\ref{systematics}. The largest contribution to the other uncertainties is the uncertainty on the integrated luminosity, which is 2.6\%. Statistical uncertainty of the unfolded data and total uncertainty in the measured cross section are also shown.}
    \label{fig:SumSyst_ZNGoodJets_Zexc}

\end{figure}

The effect of the systematic uncertainties in the measured cross section as a function of jet multiplicity is illustrated in Fig.~\ref{fig:SumSyst_ZNGoodJets_Zexc}, and in Table~\ref{syst_Nj} for jet multiplicities of 1, 2, and 3. The total uncertainty is the sum in quadrature of all contributions.

\begin{table*}
\centering
\topcaption{Uncertainties in the measured cross section as a function of jet multiplicity, including uncertainties due to the statistical uncertainty of unfolded data (Stat), the jet energy scale (JES), pileup modeling (PU),  background normalization (BG), the jet energy resolution (JER), trigger efficiency and muon identification (LepSF), \PQb tagging efficiency,  muon momentum scale (MES) and resolution (MER), the normalization of the $\PW\PQb$ content in the signal simulation ($\PW\PQb$), the \ttbar modeling, a finite number of simulated events used to construct the response matrix (MC stat), and integrated luminosity (Int Lumi).}
  \begin{scotch}{l|ccc}
  & $N_{\text{jets}} =$ 1 &  $N_{\text{jets}} = 2$ & $N_{\text{jets}} = 3$   \\
    \hline
    Total(\%) & 10 & 13 &16 \\ \hline
    Stat(\%) & 0.057 & 0.13 &0.33 \\
    JES(\%) & 10 &12 &16\\
    PU(\%)  & 0.025 & 0.26 & 0.35 \\
    BG(\%) & 0.22 & 0.43 & 1.1 \\
    JER(\%) &  0.43 & 0.23 & 0.29 \\
    LepSF(\%) & 0.35   & 0.50& 0.72 \\
    \PQb tagging(\%) & 0.41  & 0.69 & 1.5 \\
    MES(\%) & 0.20 & 0.18 & 0.17\\
    MER(\%) & 0.015 & 0.0016 & 0.017 \\
    $\PW\PQb$(\%)  & 0.062 & 0.22& 0.38 \\
    \ttbar(\%) & 0.014  & 0.38 & 0.83 \\
    MC stat(\%) & 0.094 & 0.14 & 0.26 \\
    Int Lumi(\%) & 2.6 & 2.6 & 2.6 \\
  \end{scotch}
  \label{syst_Nj}
\end{table*}

\section{Results}
\label{results}
\begin{figure*}[btp]
\centering
    \includegraphics[width=0.48\textwidth]{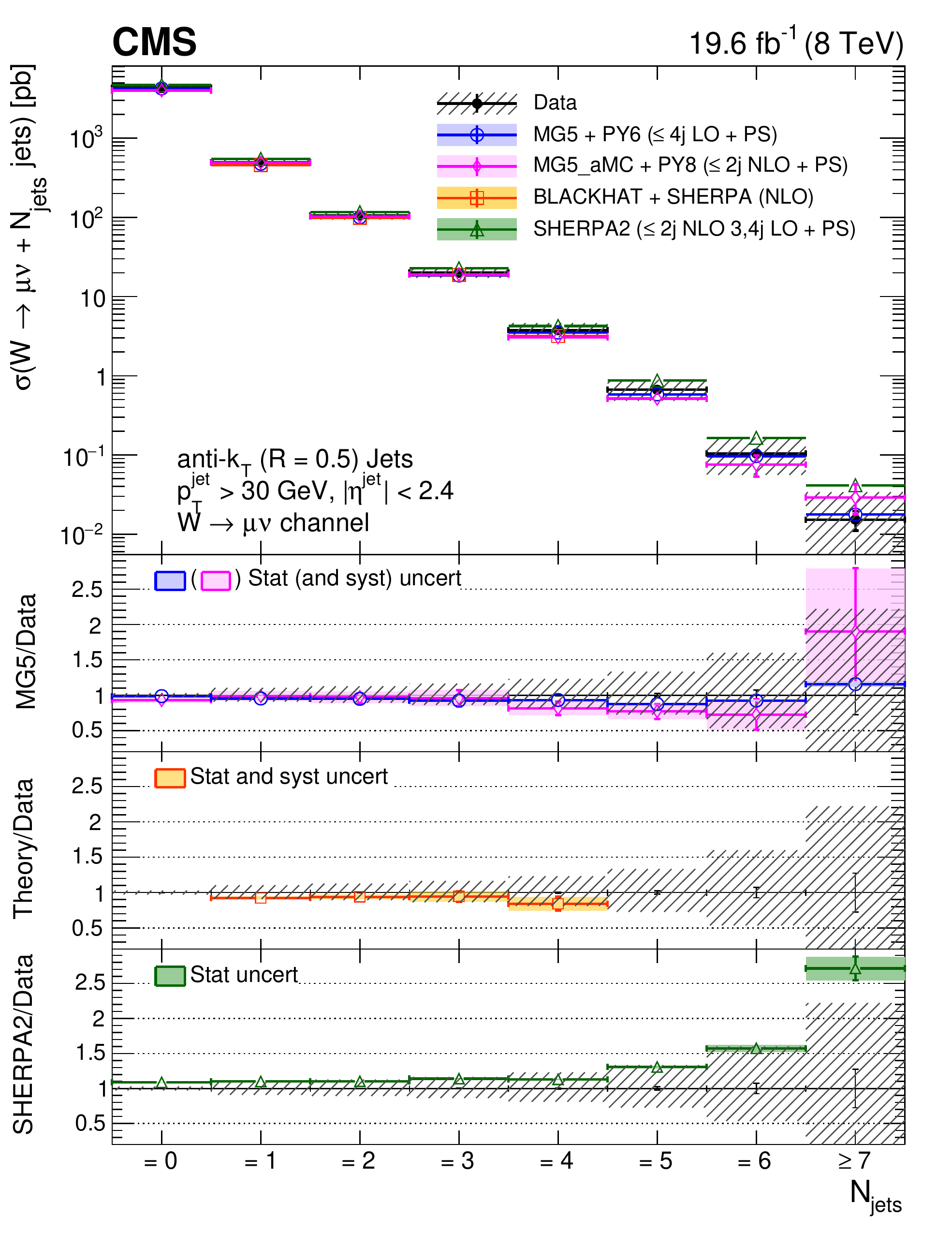}
    \includegraphics[width=0.48\textwidth]{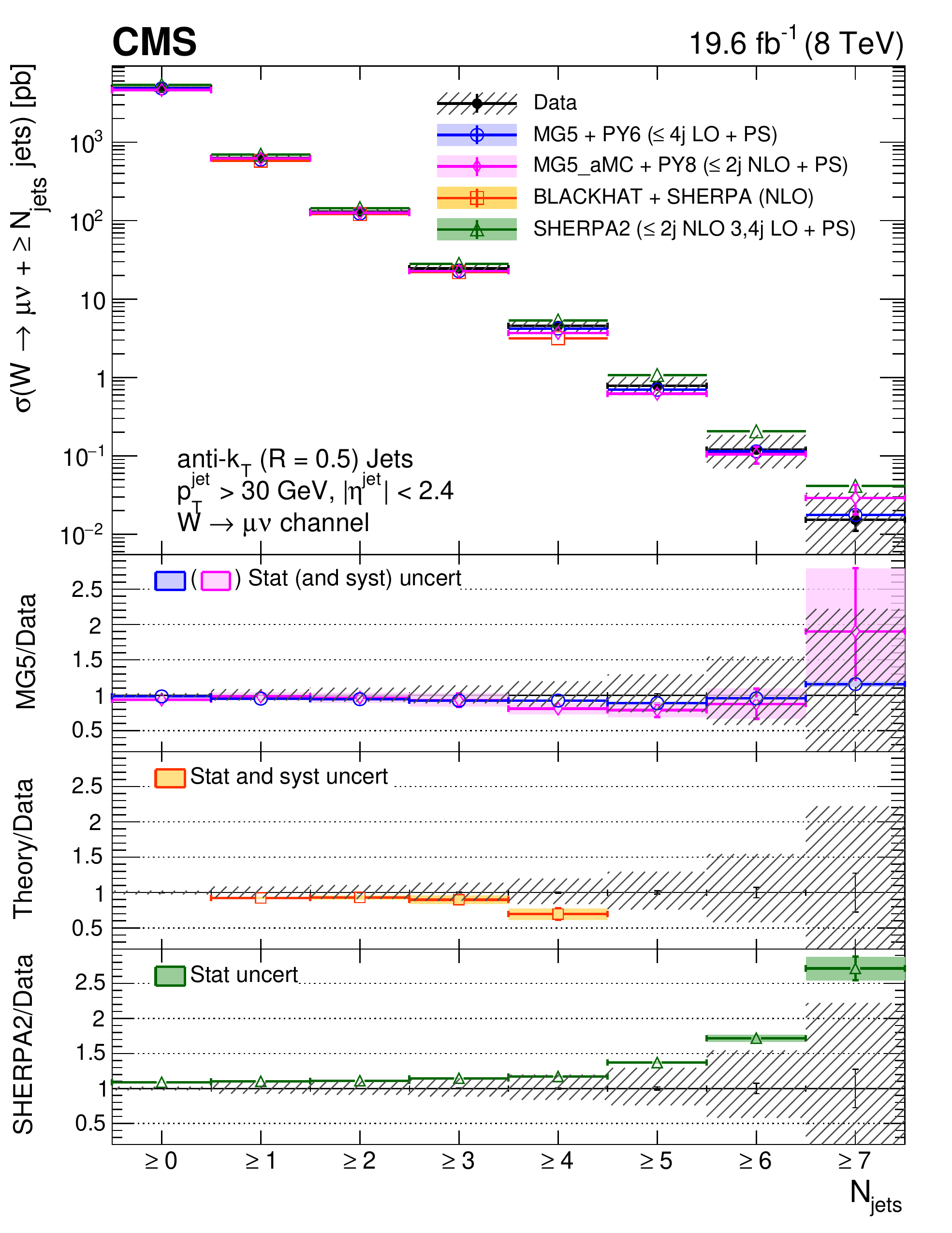}
    \caption{Measured cross section versus exclusive (left) and inclusive (right) jet multiplicity, compared to the predictions of \MADGRAPH, \MGvATNLO, {\SHERPA 2}, and \BLACKHAT{}+\SHERPA (corrected for hadronization and multiple-parton interactions), for which we currently have predictions only up to $\PW +4$ jets. Black circular
markers with the gray hatched band represent the unfolded data measurement and its
total uncertainty. Overlaid are the predictions together with their uncertainties. The lower plots show the ratio of each prediction to the
unfolded data.}
    \label{xsec_Njets}
\end{figure*}

\begin{figure*}[!htbp]
\centering
        \includegraphics[width=0.48\textwidth]{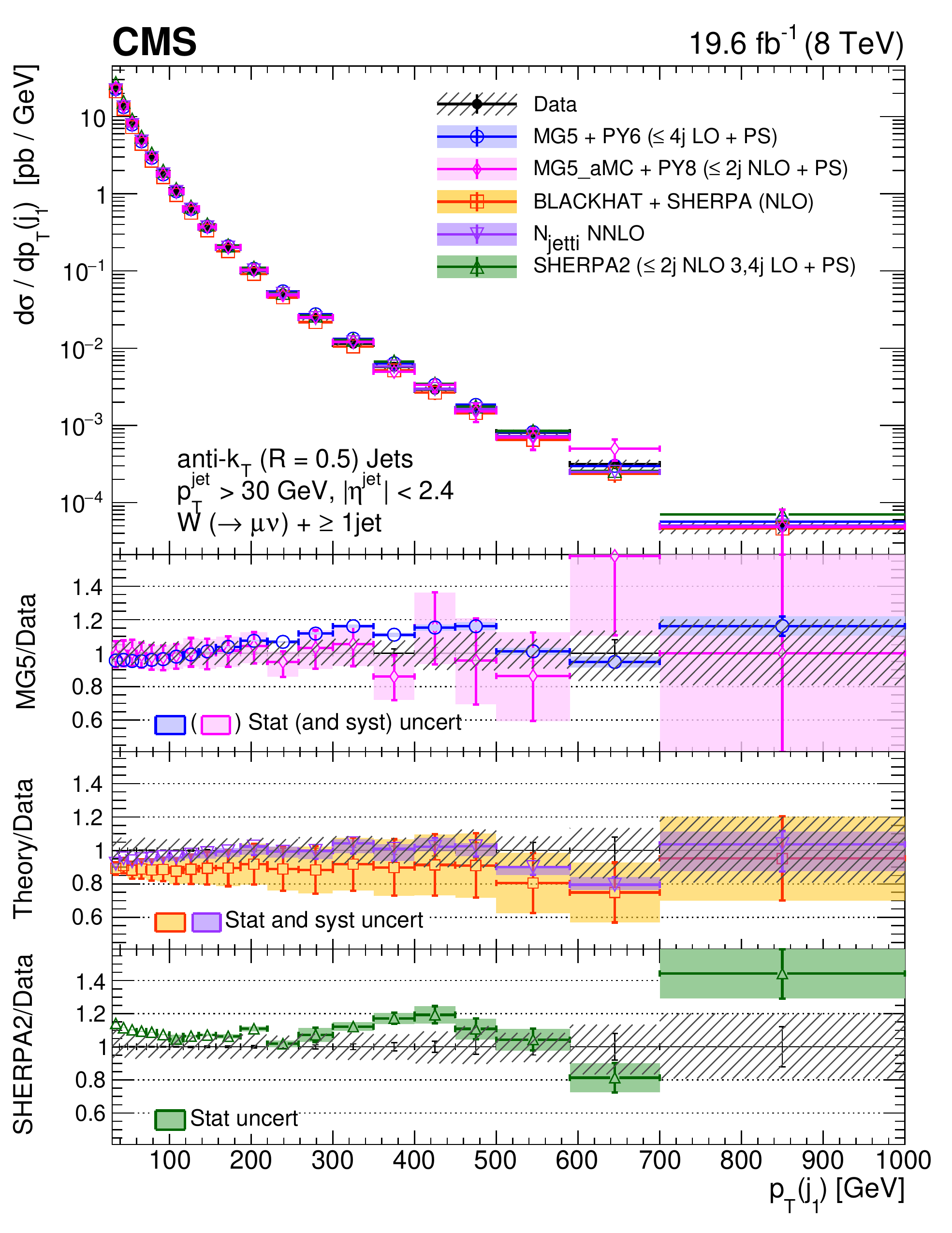}
	    \includegraphics[width=0.48\textwidth]{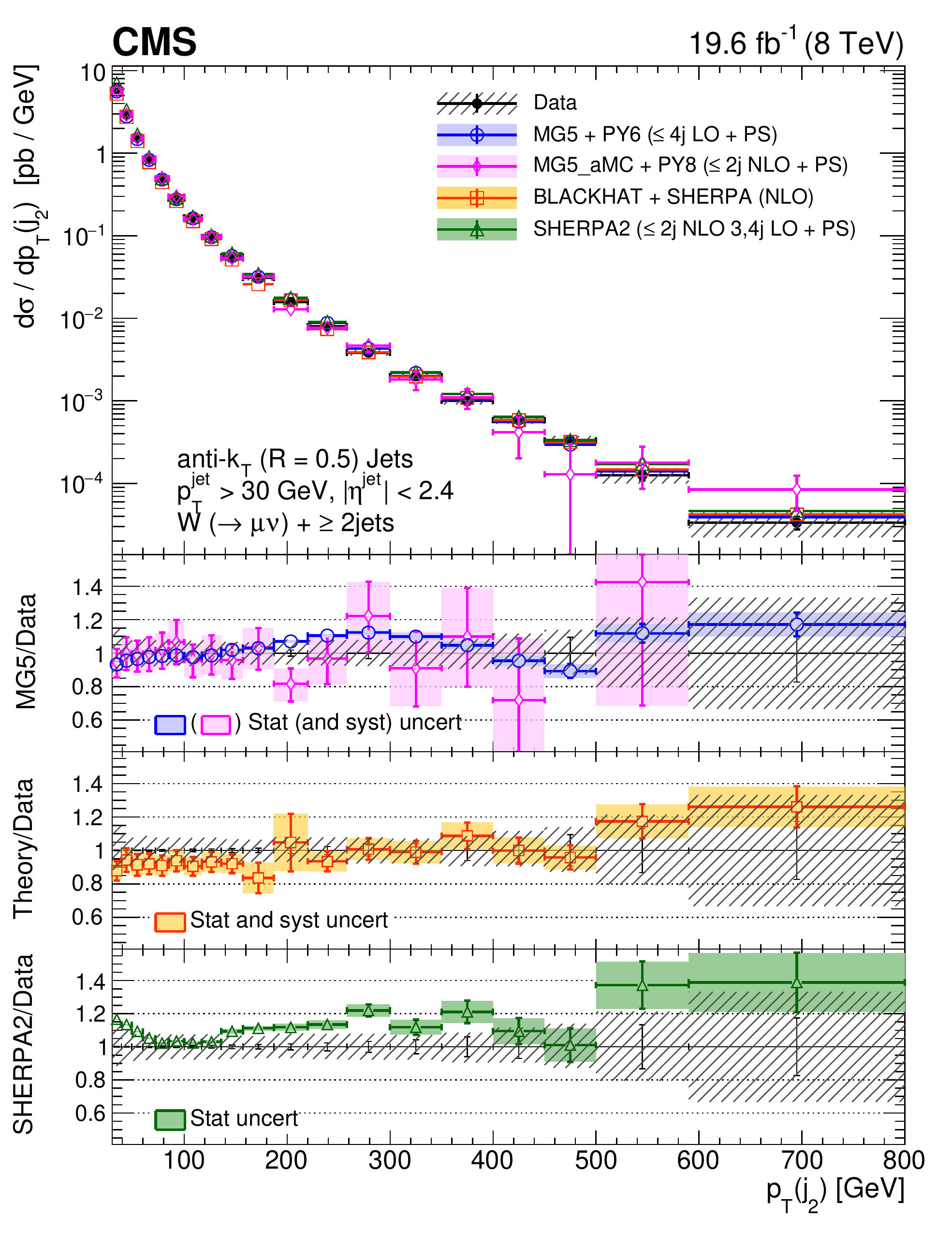}
	    \includegraphics[width=0.48\textwidth]{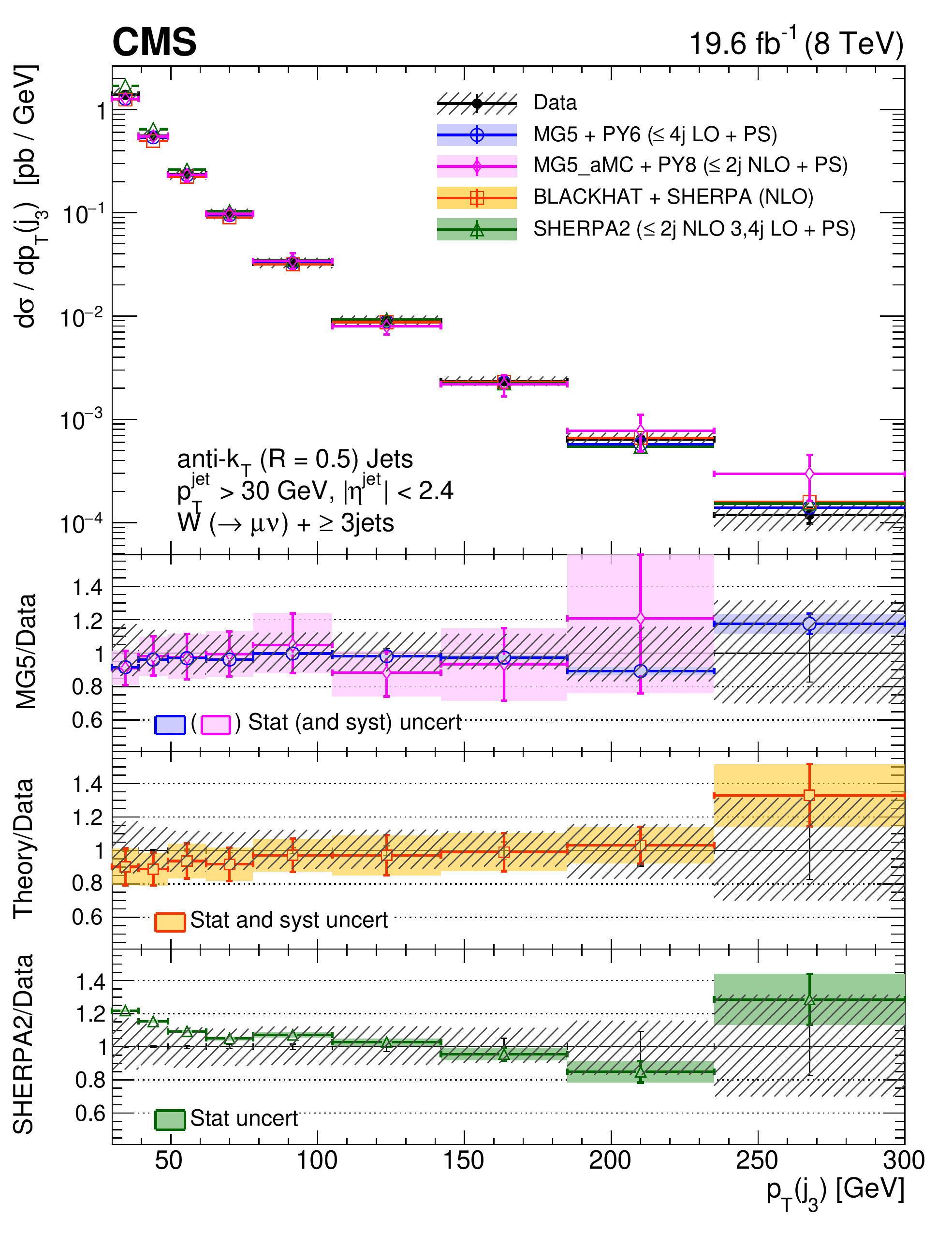}
	    \includegraphics[width=0.48\textwidth]{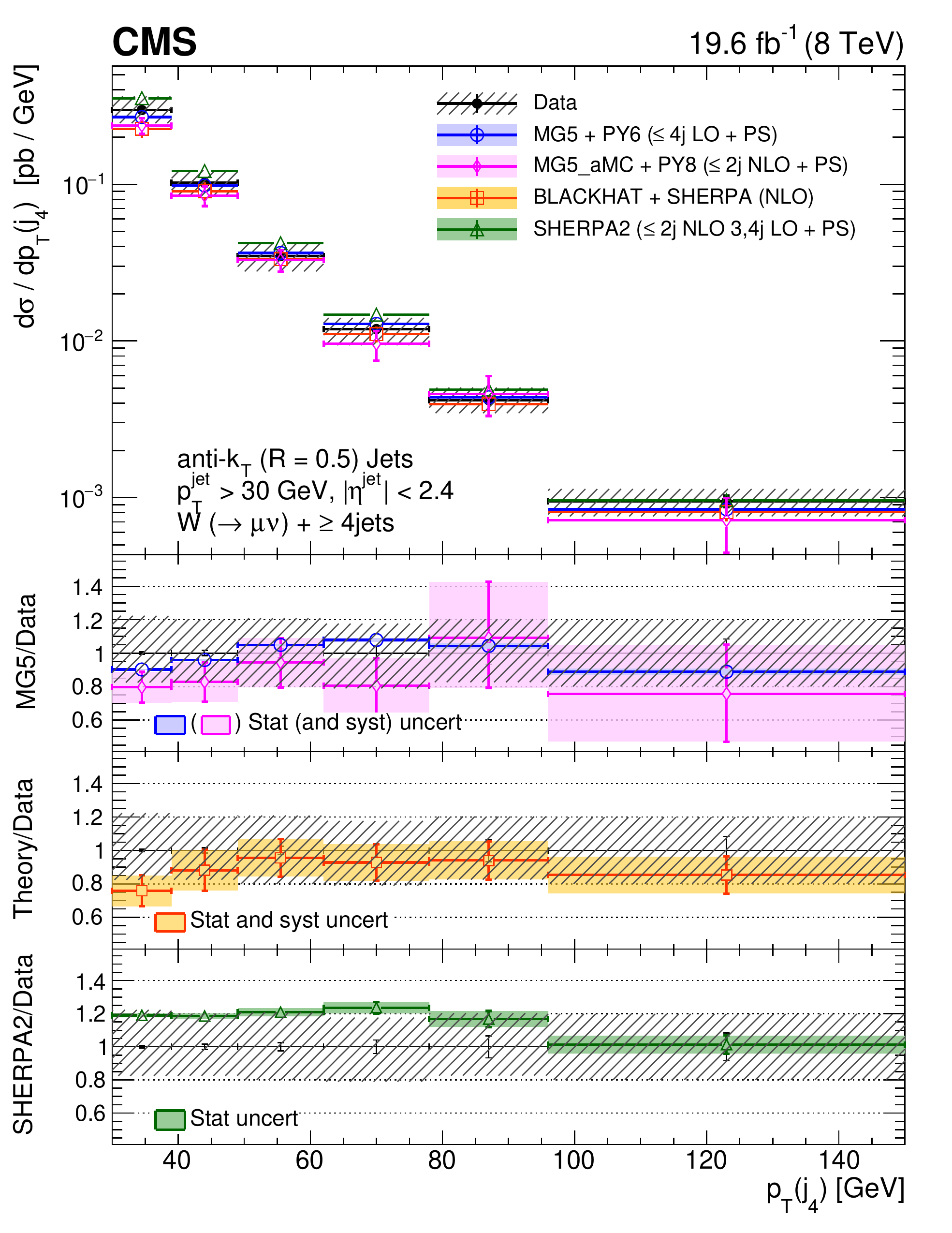}	
	    \caption{Cross sections differential in the transverse momenta of the four leading jets, compared to the predictions of \MADGRAPH, \MGvATNLO, {\SHERPA 2}, \BLACKHAT{}+\SHERPA, and NNLO inclusive one-jet production (indicated as $\rm N_{jetti}$ NNLO). The \BLACKHAT{}+\SHERPA and NNLO predictions are corrected for hadronization and multiple-parton interaction effects. Black circular markers with the gray hatched band represent the unfolded data measurements and their total uncertainties. Overlaid are the predictions together with their uncertainties. The lower plots show the ratio of each prediction to the unfolded data.}
    \label{xsec_JetPt1to4}
\end{figure*}

\begin{figure*}[!htbp]
\centering
    	 \includegraphics[width=0.48\textwidth]{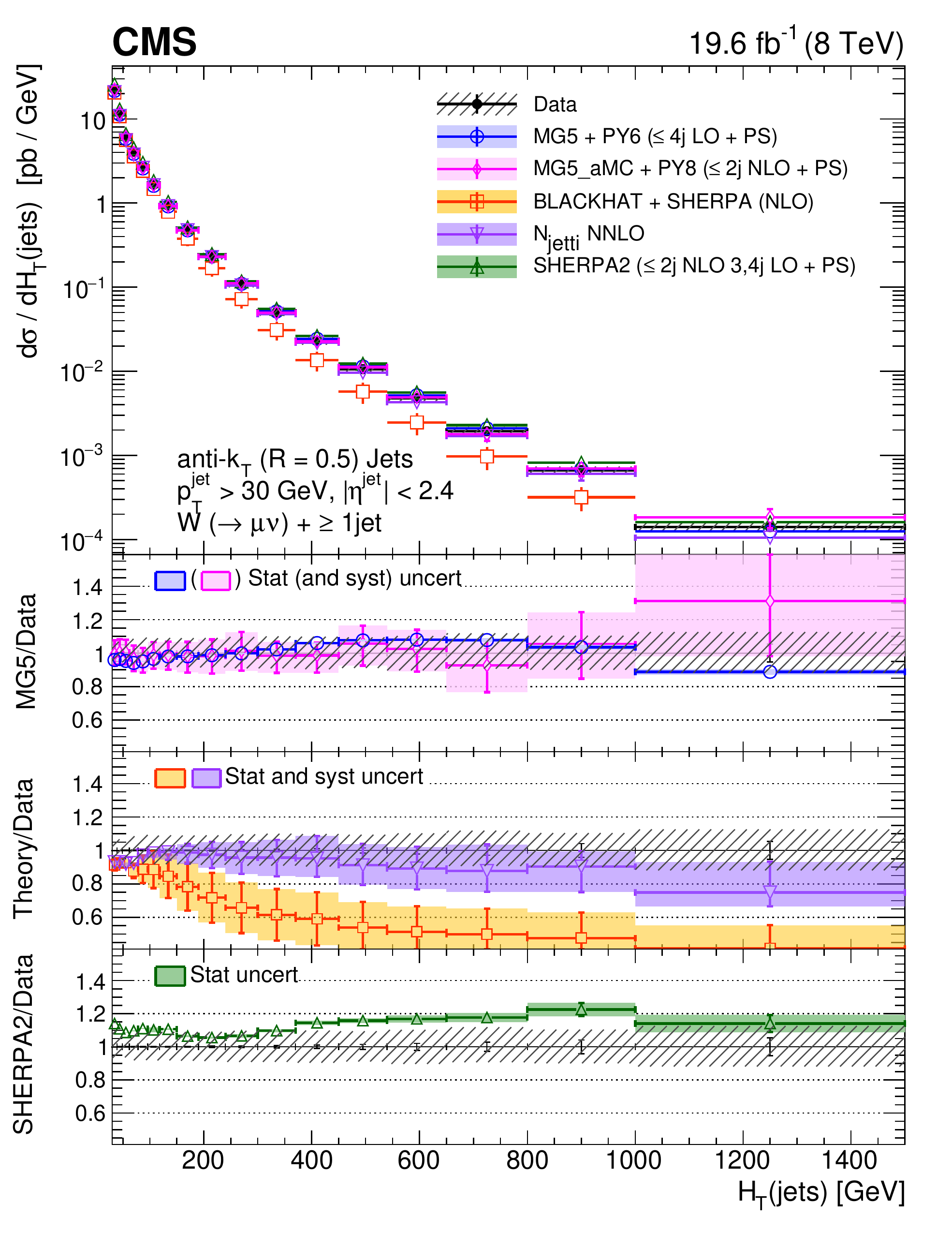}
    	 \includegraphics[width=0.48\textwidth]{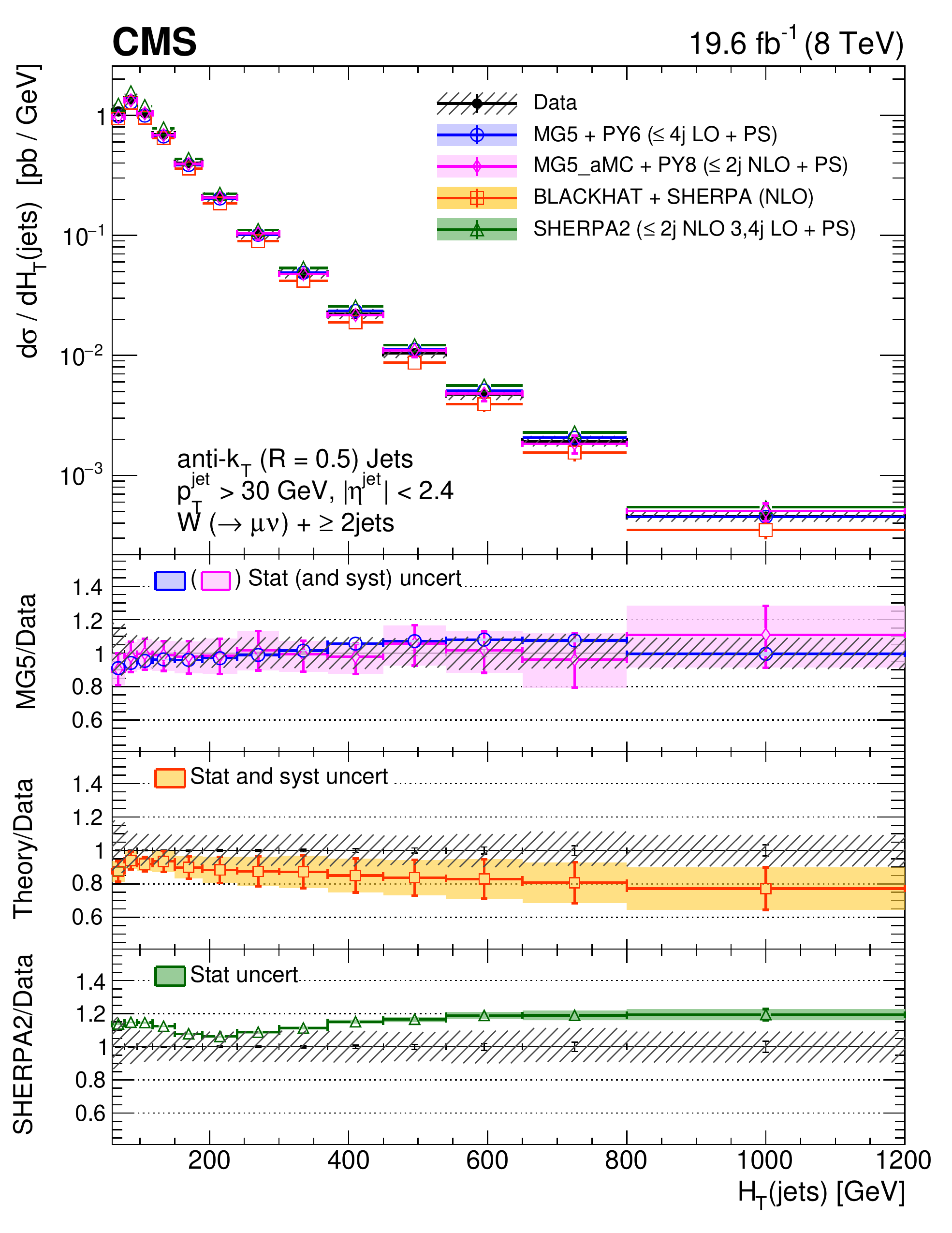}
    	 \includegraphics[width=0.48\textwidth]{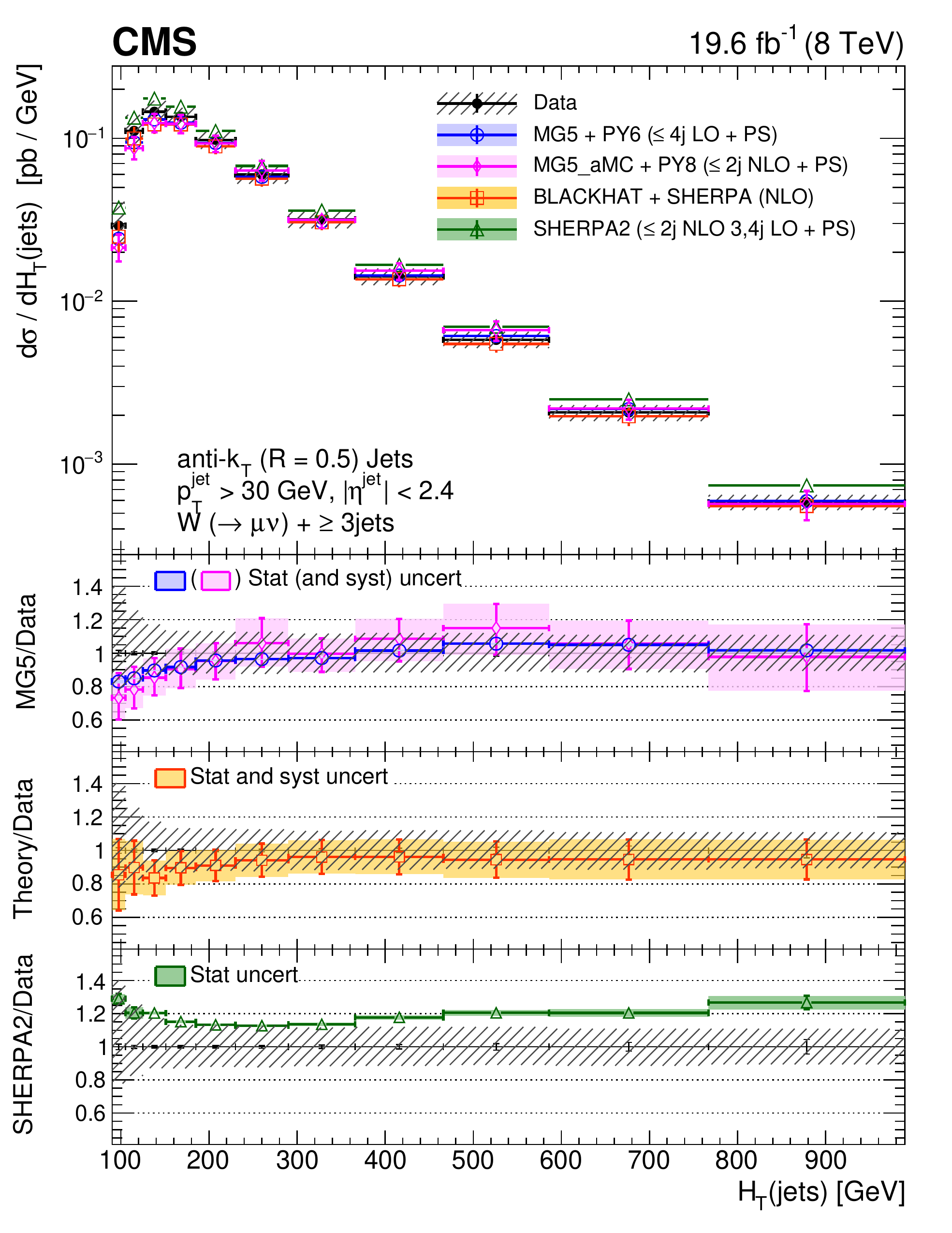}
    	 \includegraphics[width=0.48\textwidth]{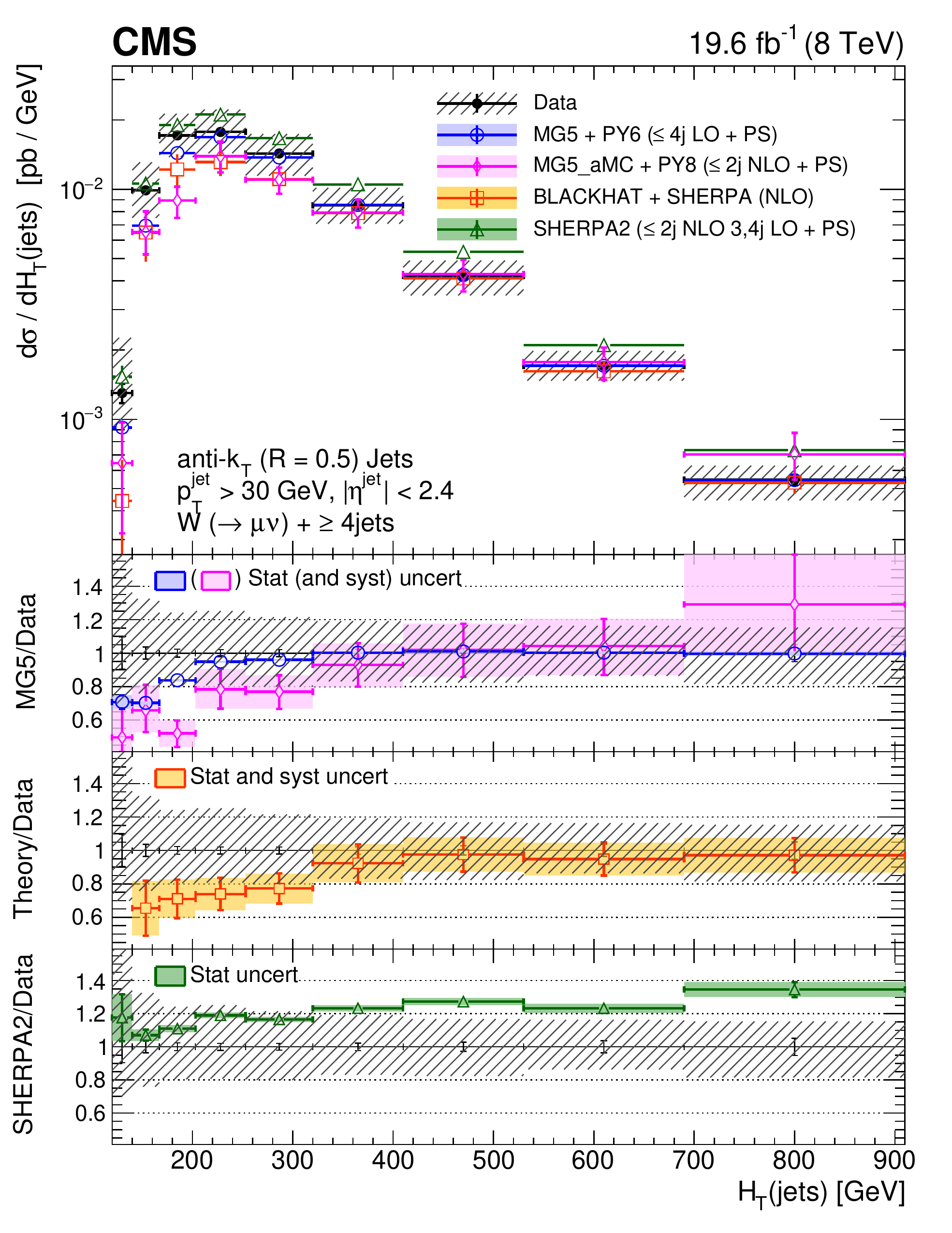}
    	\caption{Cross sections differential in $\HT$ for inclusive jet multiplicities 1--4, compared to the predictions of \MADGRAPH, \MGvATNLO, {\SHERPA 2}, \BLACKHAT{}+\SHERPA, and NNLO inclusive one-jet production (indicated as $\rm N_{jetti}$ NNLO). The \BLACKHAT{}+\SHERPA and NNLO predictions are corrected for hadronization and multiple-parton interaction effects. Black circular markers with the gray hatched band represent the unfolded data measurements and their total uncertainties. Overlaid are the predictions together with their uncertainties. The lower plots show the ratio of each prediction to the unfolded data.}
    \label{xsec_JetHT1to4}
\end{figure*}

\begin{figure*}[!htbp]
\centering
    	 \includegraphics[width=0.48\textwidth]{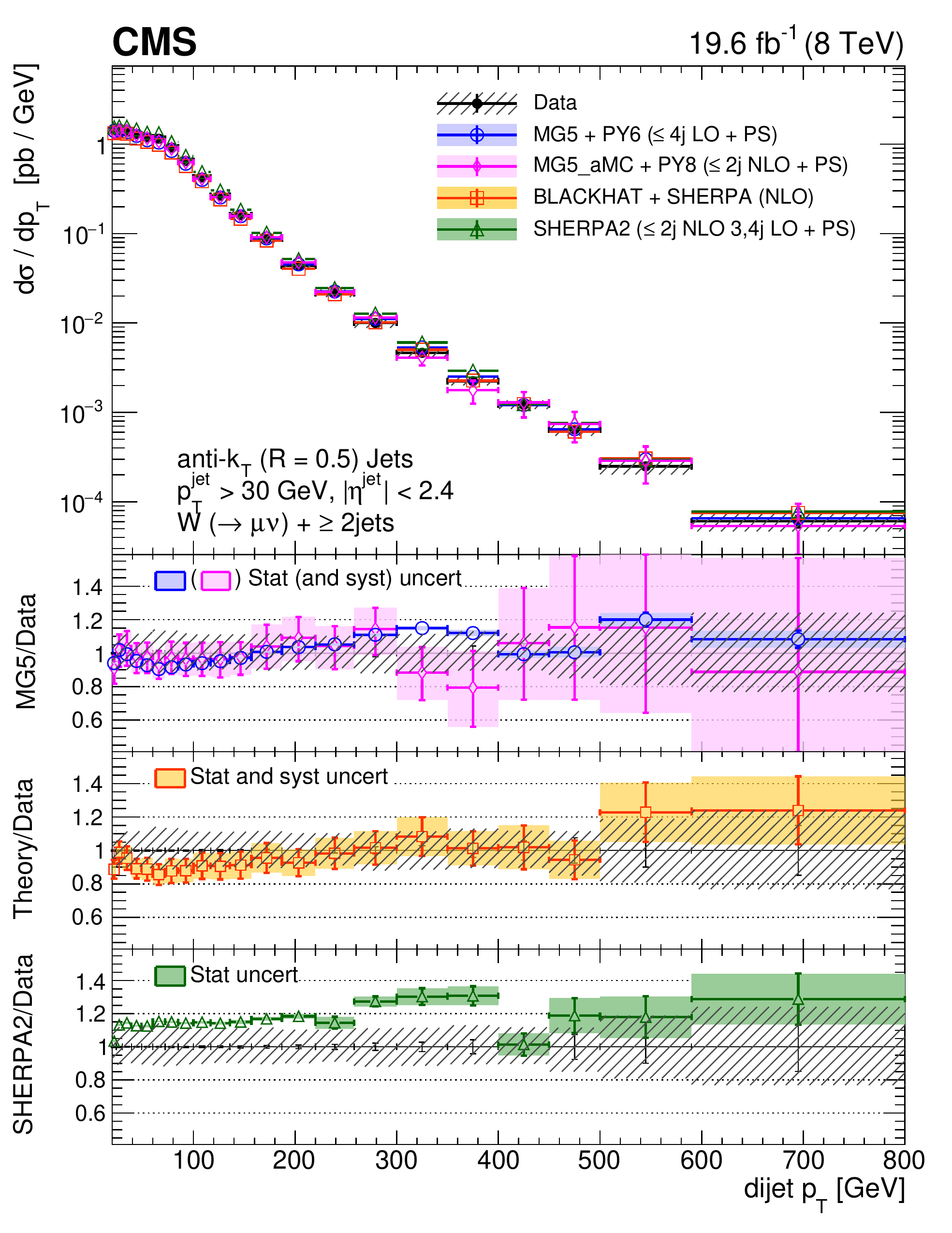}
    	 \includegraphics[width=0.48\textwidth]{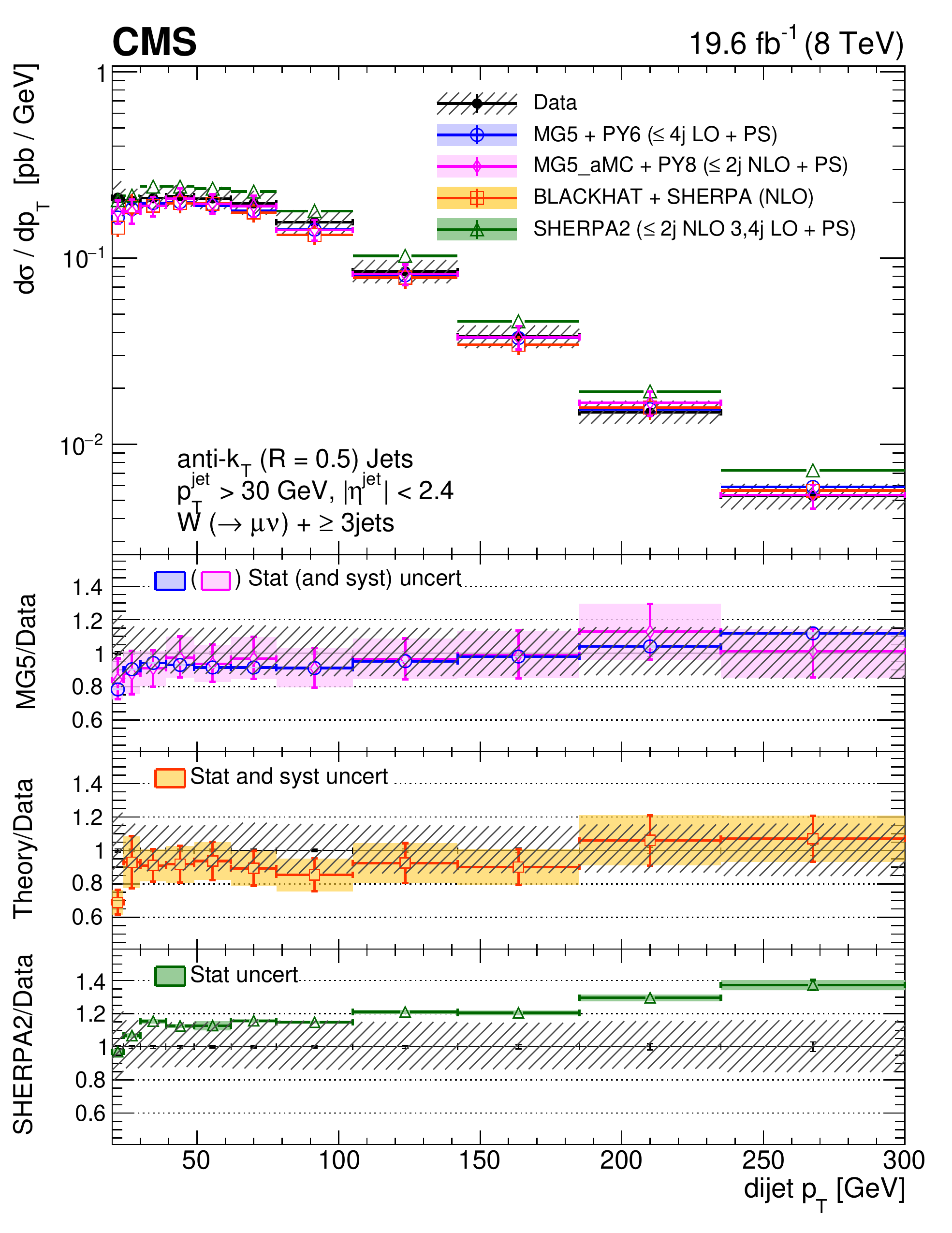}
    	 \includegraphics[width=0.48\textwidth]{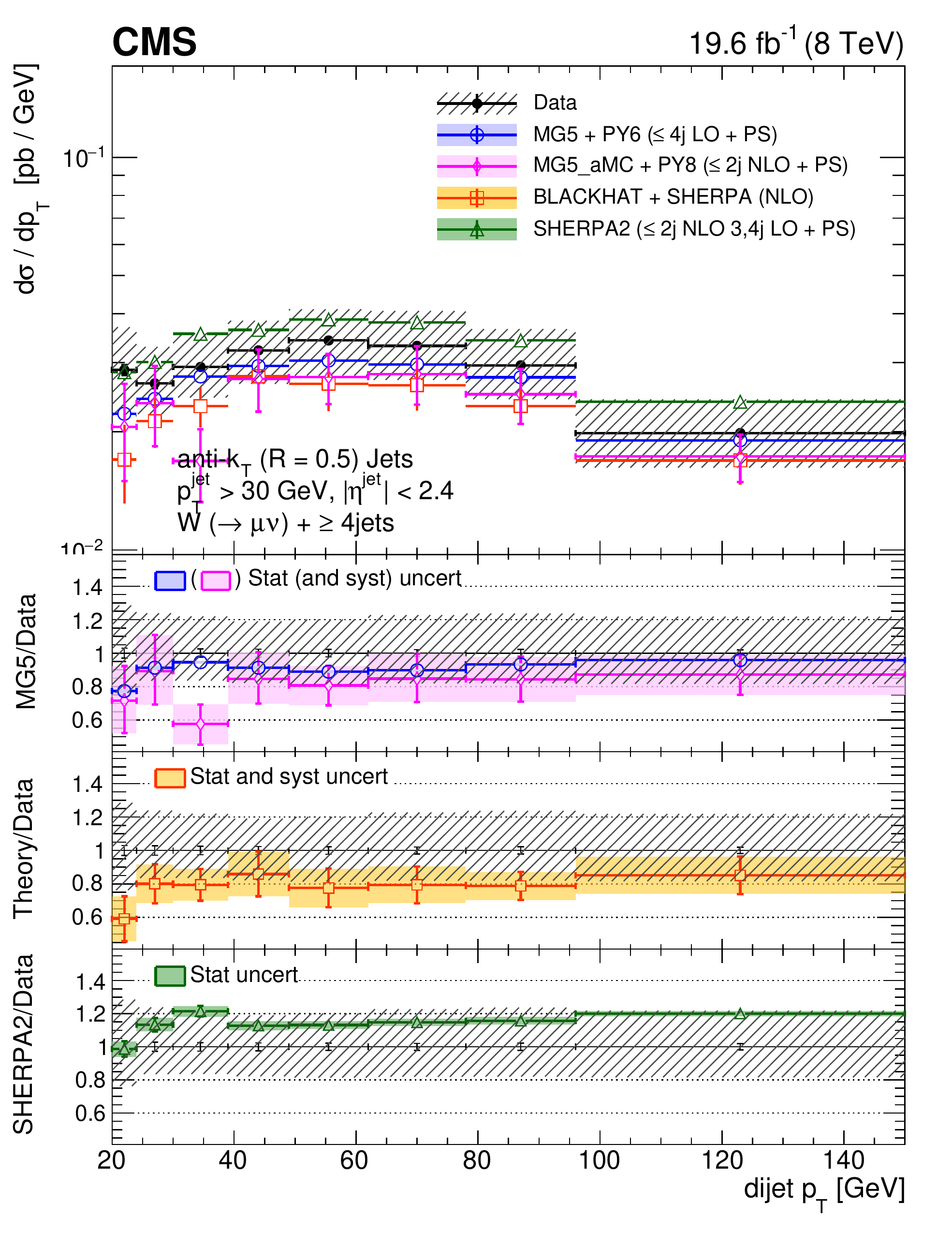}
    	\caption{Cross sections differential in dijet \pt (calculated from the two leading jets) for inclusive jet multiplicities 2--4, compared to the predictions of \MADGRAPH, \MGvATNLO, {\SHERPA 2}, and \BLACKHAT{}+\SHERPA (corrected for hadronization and multiple-parton interactions). Black circular
markers with the gray hatched band represent the unfolded data measurements and their total
uncertainties. Overlaid are the predictions together with their uncertainties. The lower plots show the ratio of each prediction to the unfolded data.}
    \label{xsec_diJetPt}
\end{figure*}

\begin{figure*}[hp]
\centering
    	 \includegraphics[width=0.48\textwidth]{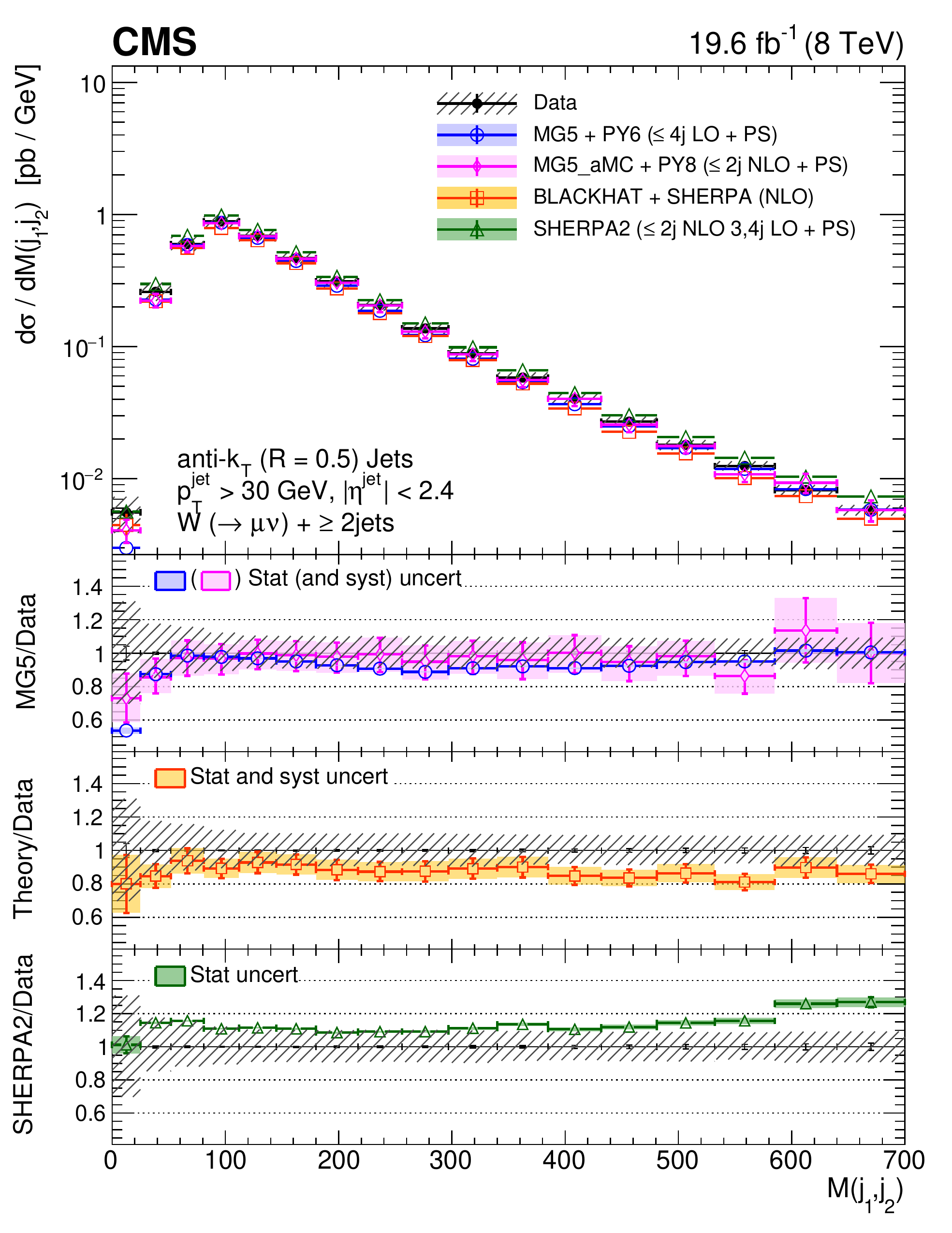}
    	 \includegraphics[width=0.48\textwidth]{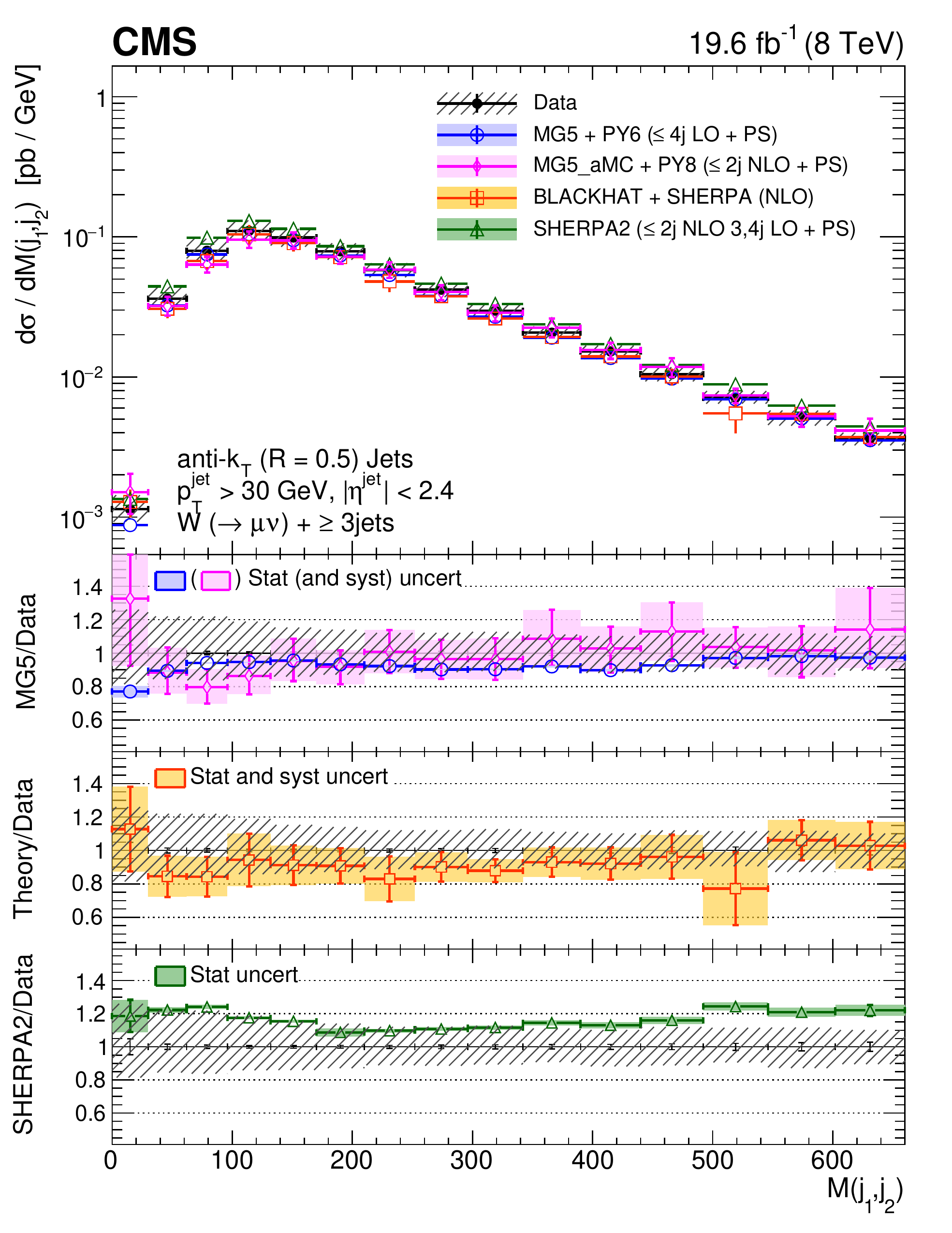}
    	 \includegraphics[width=0.48\textwidth]{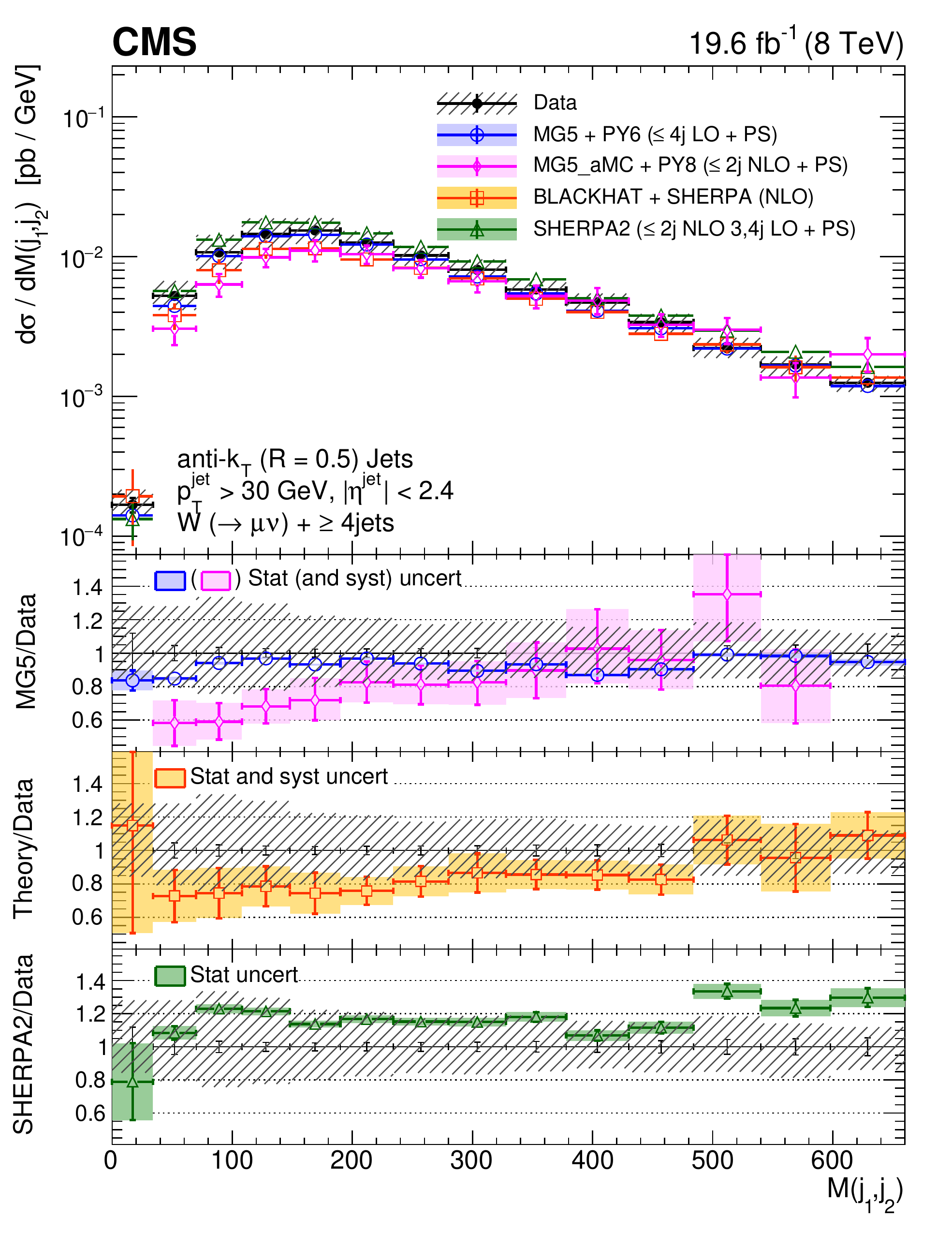}
    	\caption{Cross sections differential in dijet invariant mass (calculated from the two leading jets) for inclusive jet multiplicities 2--4, compared to the predictions of \MADGRAPH, {\MGvATNLO}, {\SHERPA 2}, and \BLACKHAT{}+\SHERPA (corrected for hadronization and multiple-parton interactions). Black circular
markers with the gray hatched band represent the unfolded data measurements and their total
uncertainties. Overlaid are the predictions together with their uncertainties. The lower plots show the ratio of each prediction to the unfolded data.}
    \label{xsec_diJetMass}
\end{figure*}

\begin{figure*}[!htbp]
\centering
    	 \includegraphics[width=0.48\textwidth]{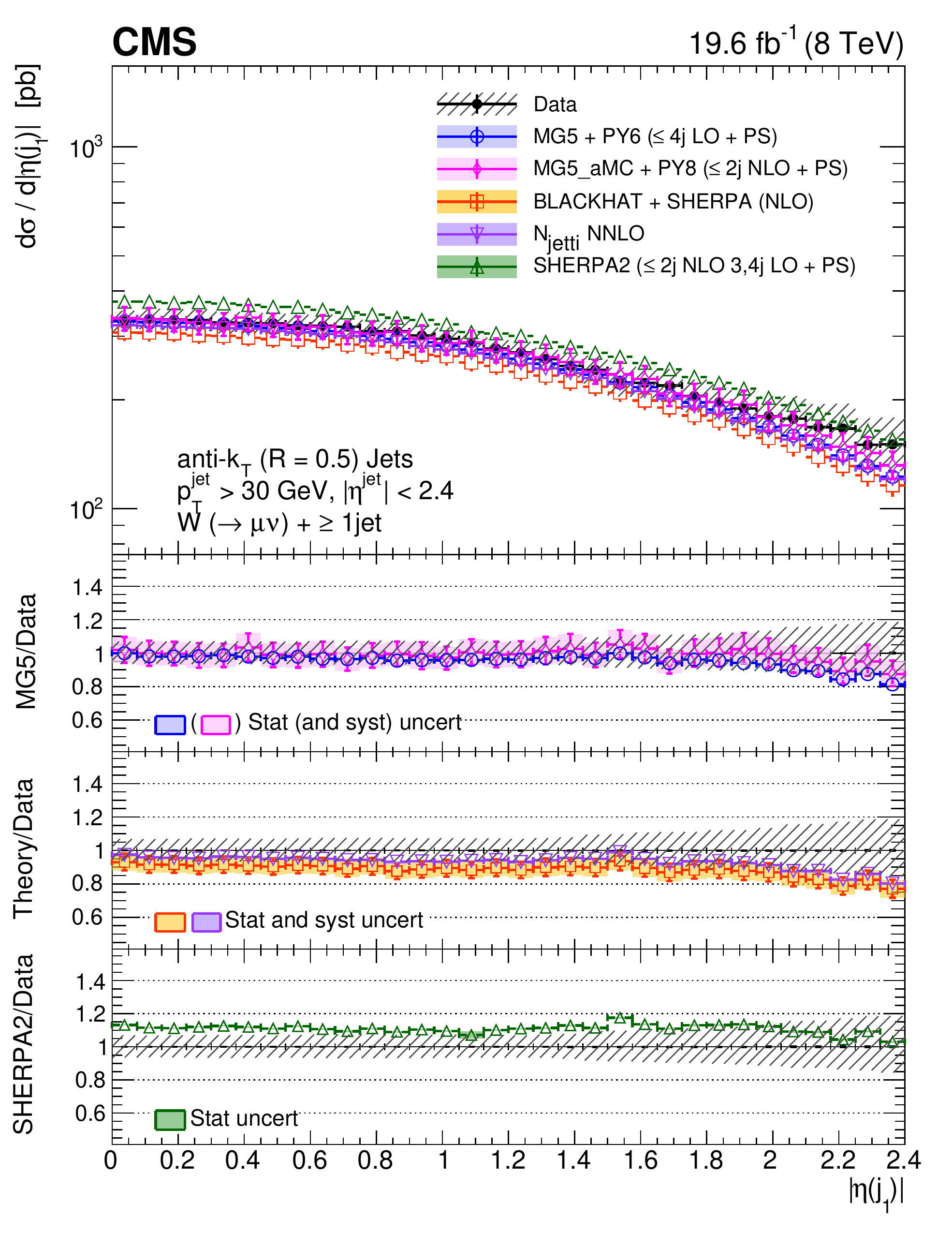}
    	 \includegraphics[width=0.48\textwidth]{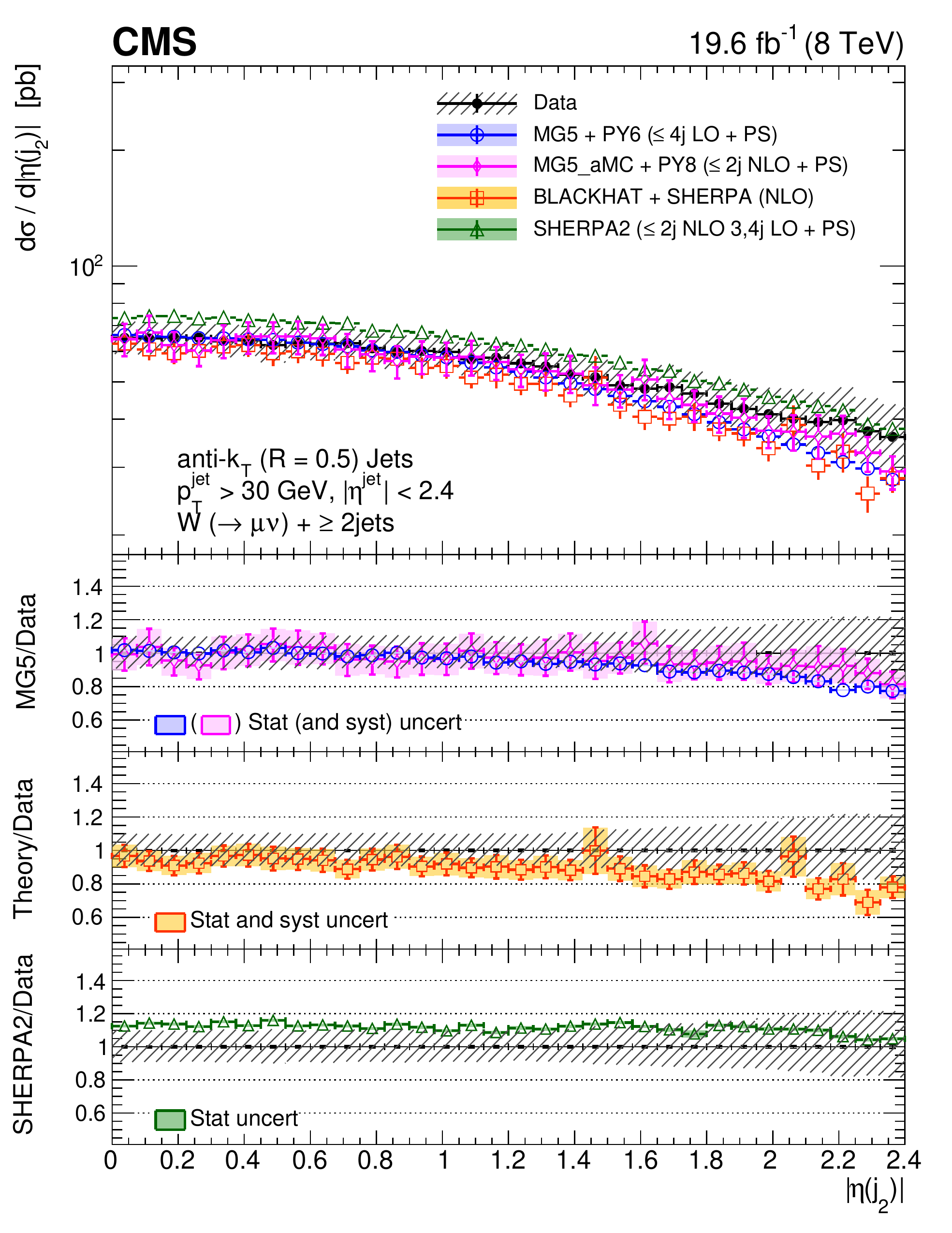}
    	 \includegraphics[width=0.48\textwidth]{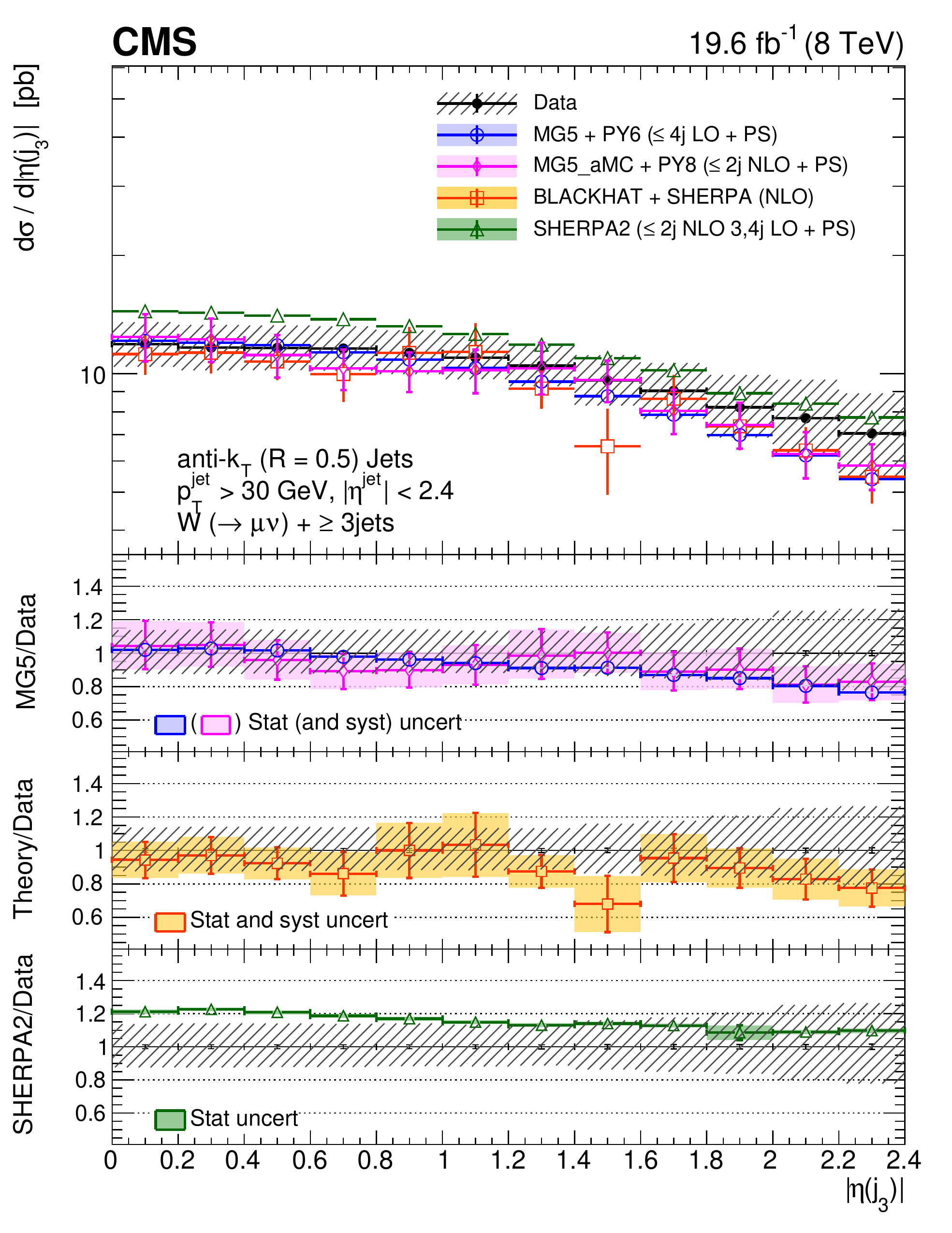}
    	 \includegraphics[width=0.48\textwidth]{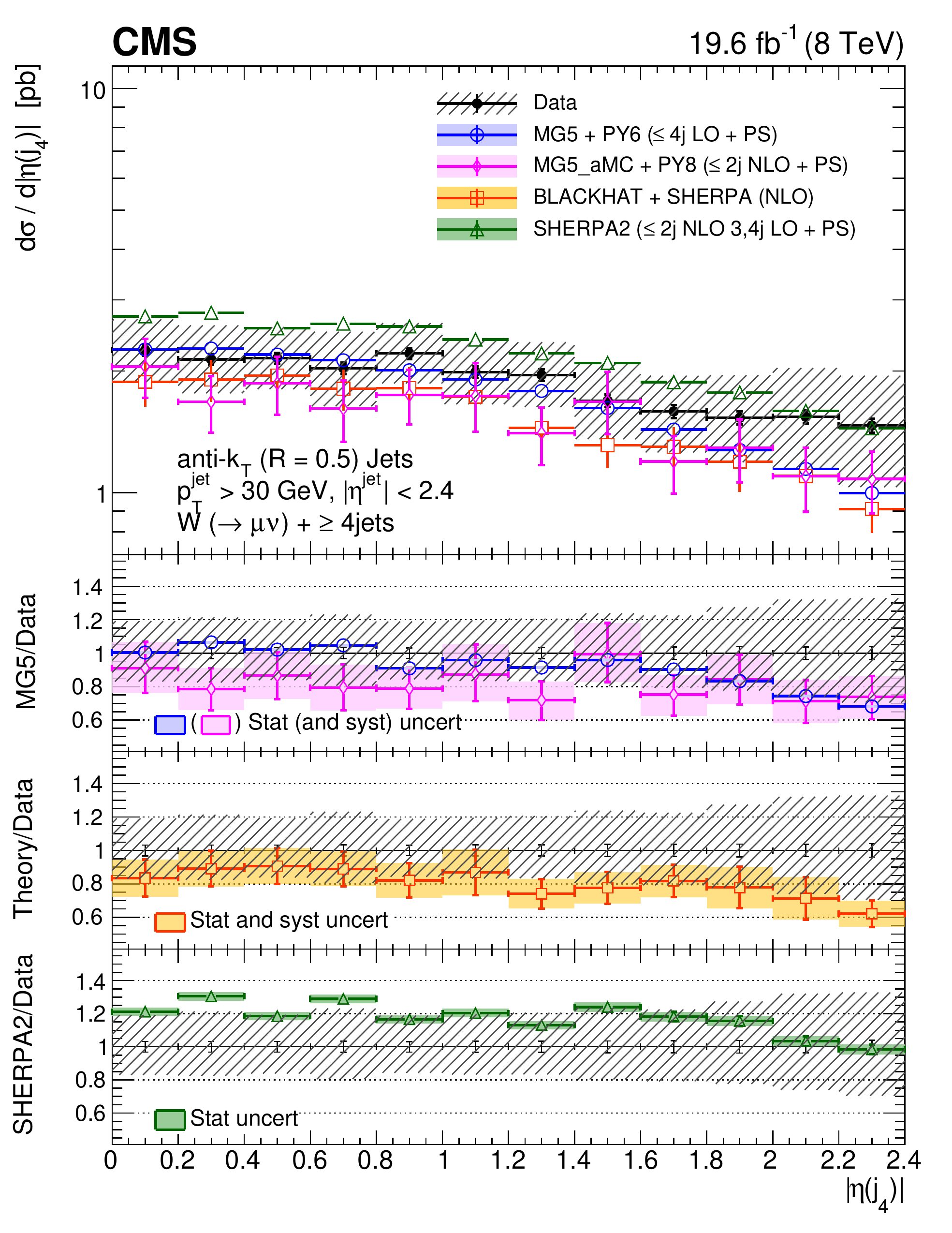}
    	\caption{Cross sections differential in the pseudorapidities of the four leading jets, compared to the predictions of \MADGRAPH, \MGvATNLO, {\SHERPA 2}, \BLACKHAT{}+\SHERPA, and NNLO inclusive one-jet production (indicated as $\rm N_{jetti}$ NNLO). The \BLACKHAT{}+\SHERPA and NNLO predictions are corrected for hadronization and multiple-parton interaction effects. Black circular markers with the gray hatched band represent the unfolded data measurements and their total uncertainties. Overlaid are the predictions together with their uncertainties. The lower plots show the ratio of each prediction to the unfolded data.}
    \label{xsec_Jeteta1to4}
\end{figure*}

\begin{figure*}[!htbp]
\centering
    	 \includegraphics[width=0.48\textwidth]{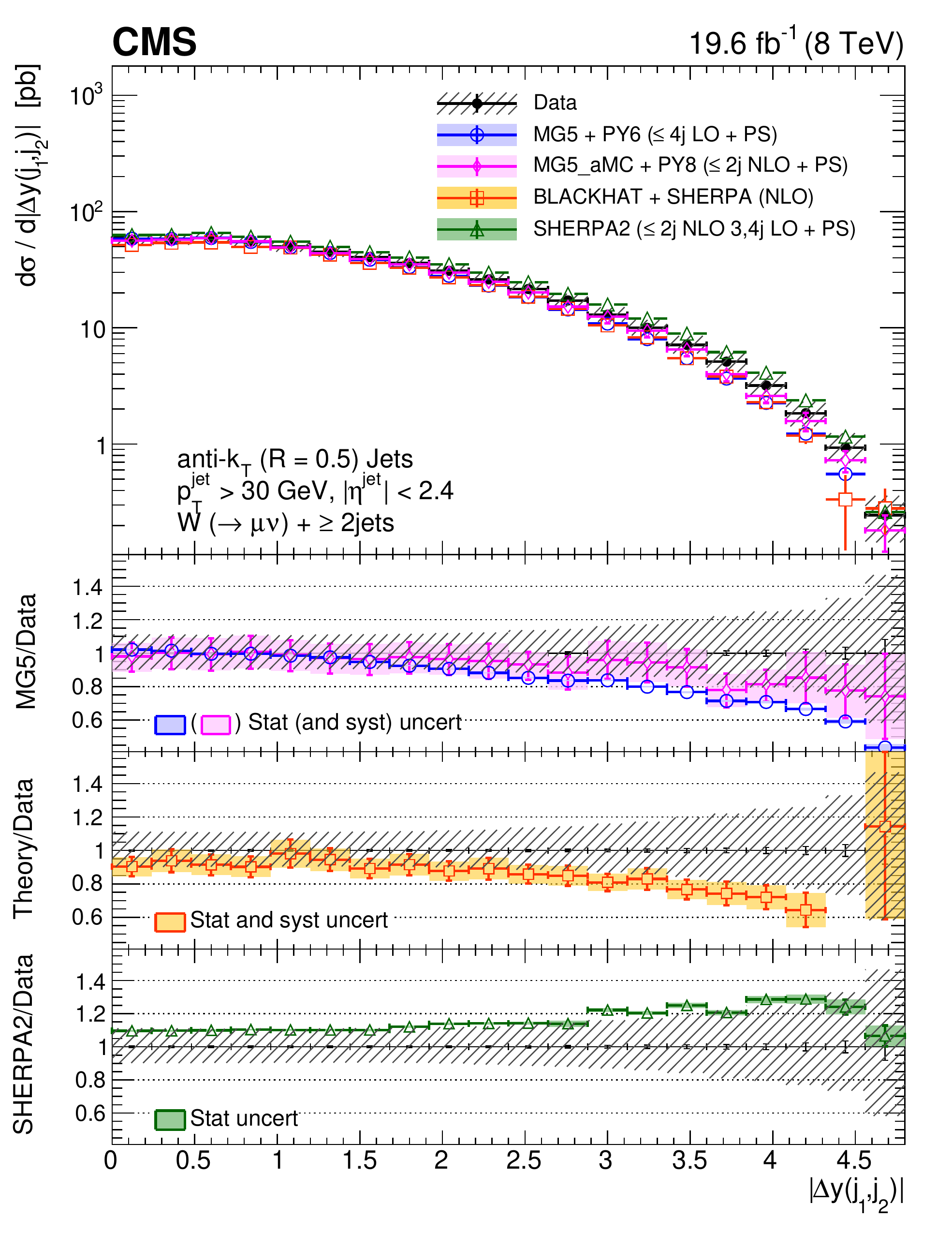}
    	 \includegraphics[width=0.48\textwidth]{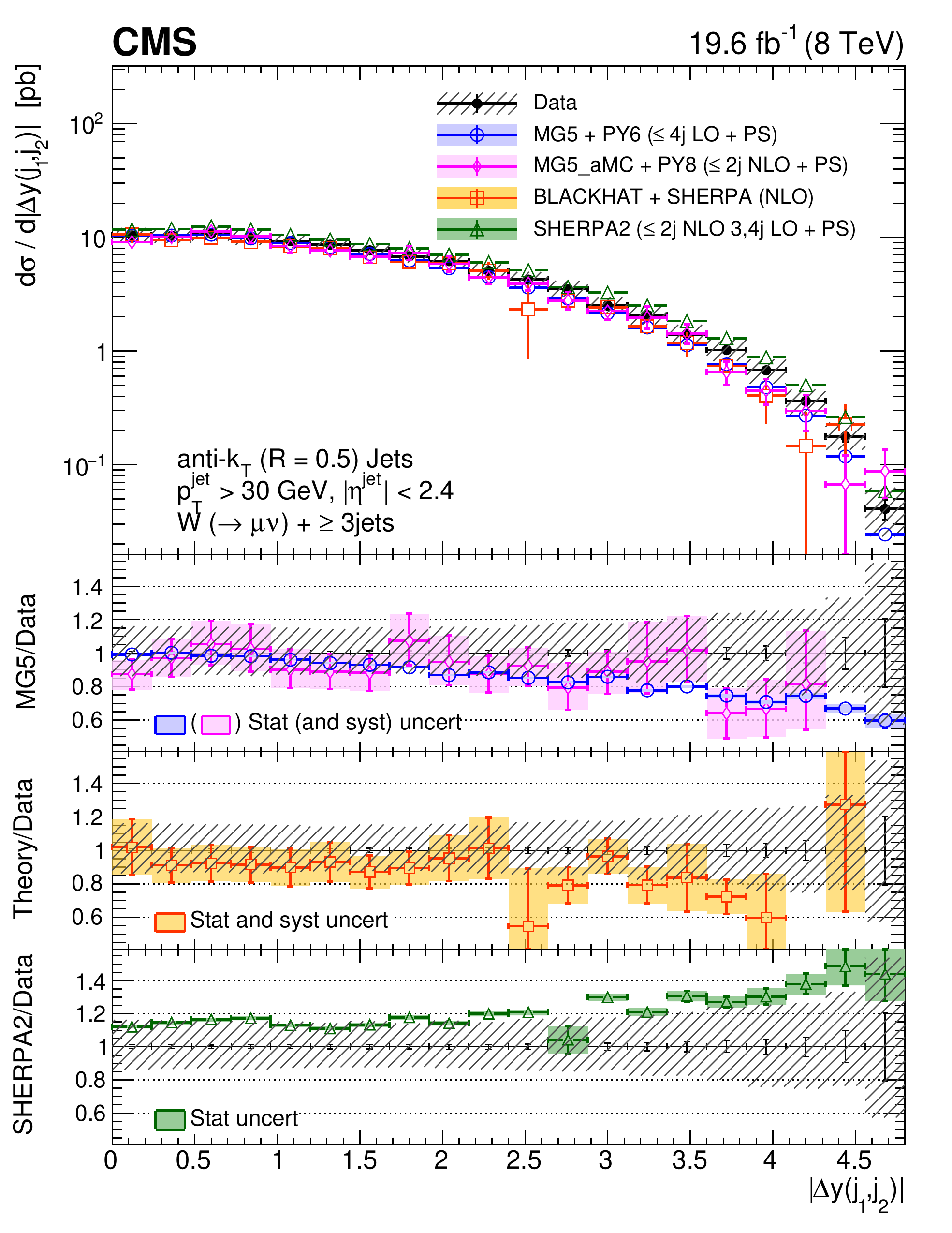}
    	 \includegraphics[width=0.48\textwidth]{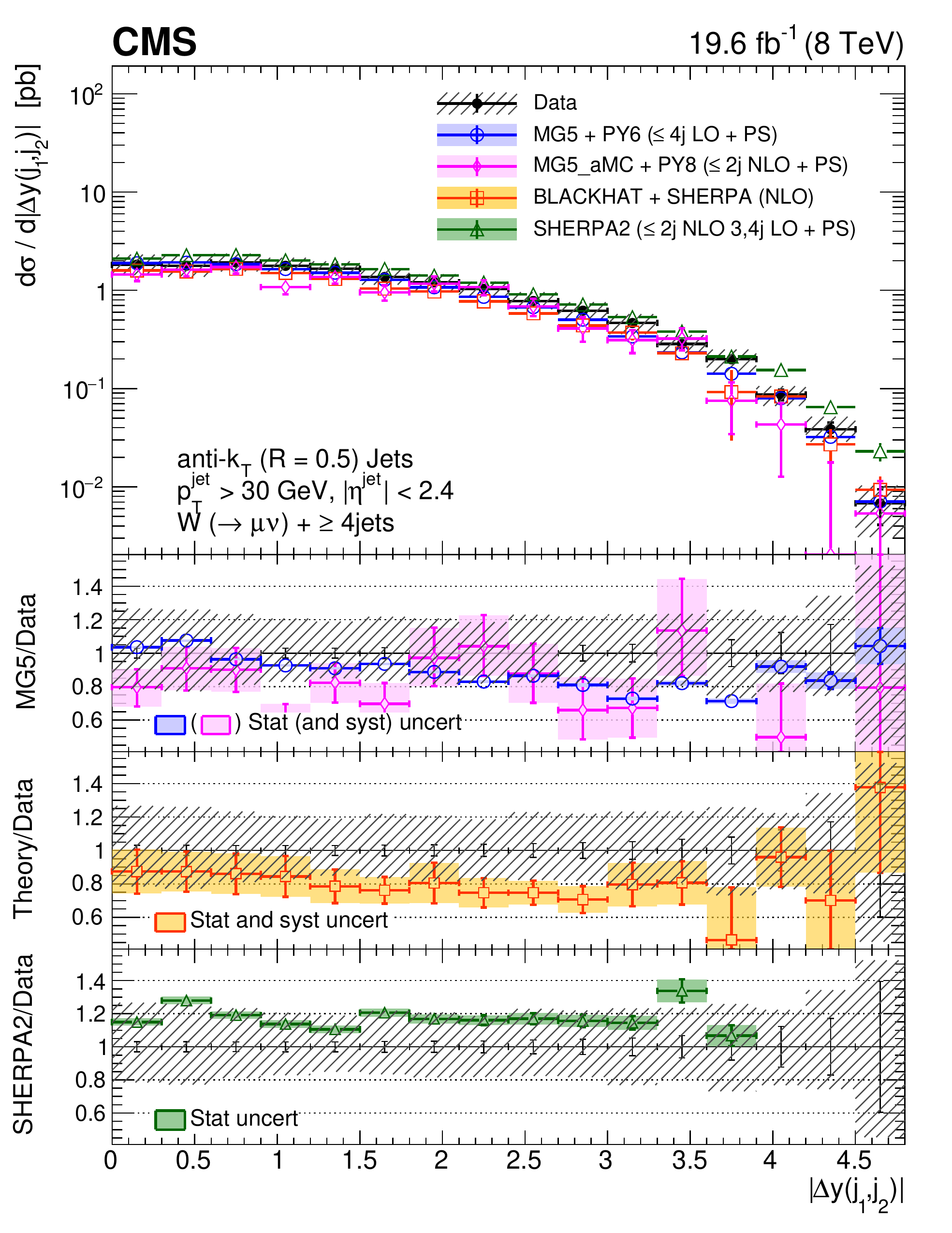}
    	\caption{Cross sections differential in $\Delta y(j_1,j_2)$ for inclusive jet multiplicities 2--4, compared to the predictions of \MADGRAPH, \MGvATNLO, {\SHERPA 2}, and \BLACKHAT{}+\SHERPA (corrected for hadronization and multiple-parton interactions). Black circular markers with the gray hatched band represent the unfolded data measurements and their total uncertainties. Overlaid are the predictions together with their uncertainties. The lower plots show the ratio of each prediction to the unfolded data.}
    \label{xsec_dRapidityJets}
\end{figure*}

\begin{figure*}[!htbp]
\centering
    	 \includegraphics[width=0.48\textwidth]{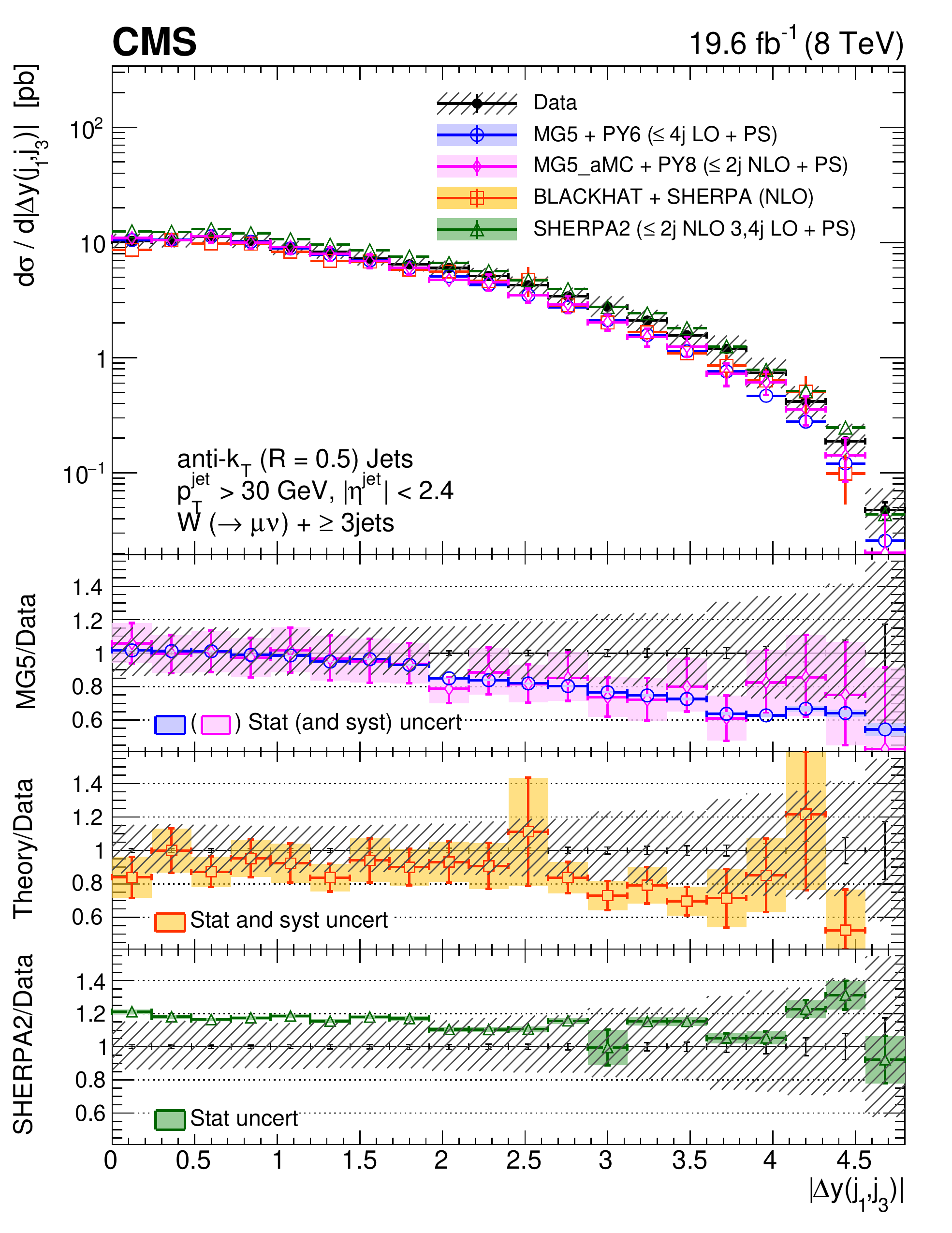}
    	 \includegraphics[width=0.48\textwidth]{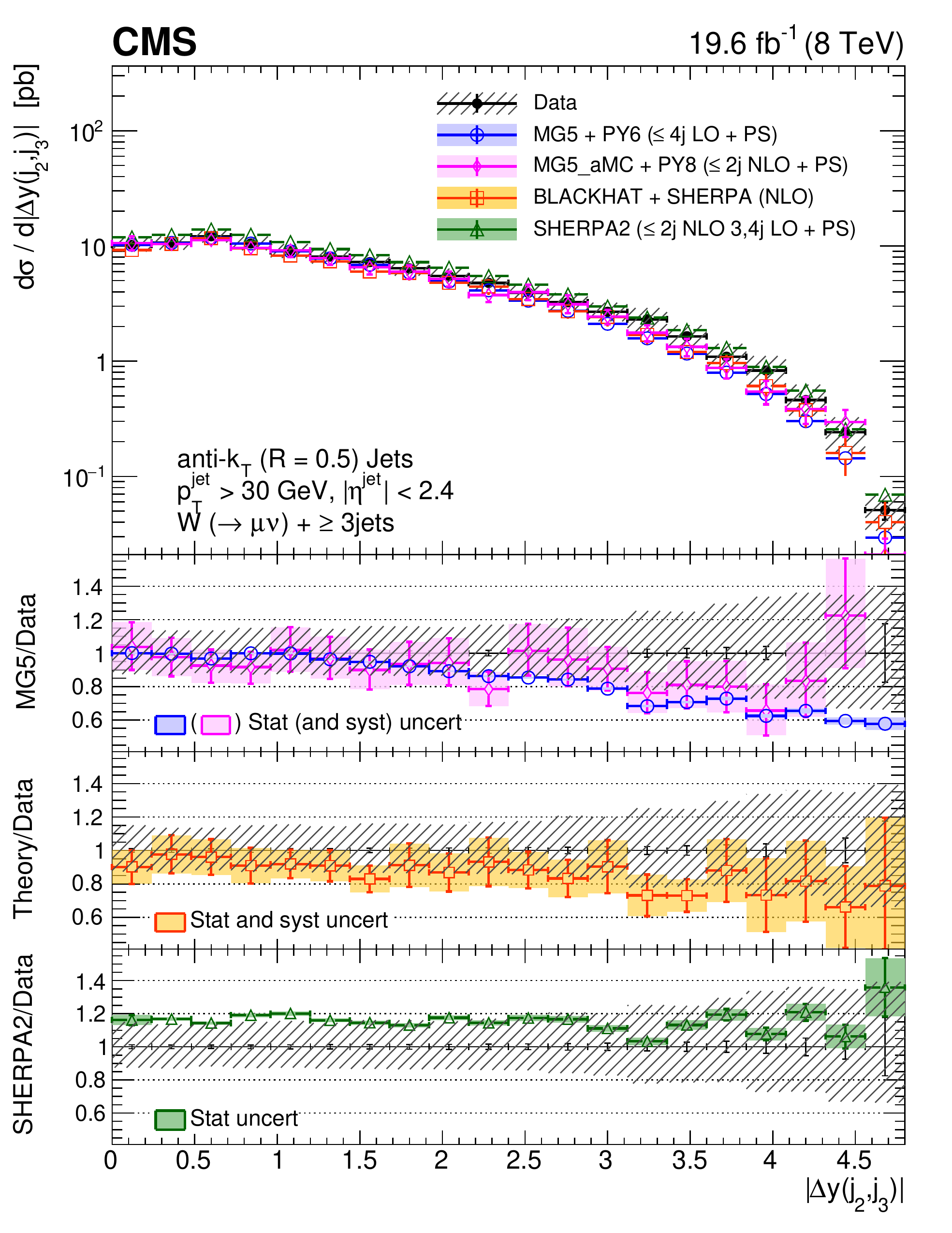}
    	\caption{Cross sections differential in $\Delta y(j_1,j_3)$ (left) and $\Delta y(j_2,j_3)$ (right) for an inclusive jet multiplicity of 3, compared to the predictions of \MADGRAPH, \MGvATNLO, {\SHERPA 2}, and \BLACKHAT{}+\SHERPA (corrected for hadronization and multiple-parton interactions). Black circular markers with the gray hatched band represent the unfolded data measurements and their total uncertainties. Overlaid are the predictions together with their uncertainties. The lower plots show the ratio of each prediction to the unfolded data.
    }
    \label{xsec_dRapidityJets13and23}
\end{figure*}

\begin{figure*}[!htbp]
\centering
    	 \includegraphics[width=0.48\textwidth]{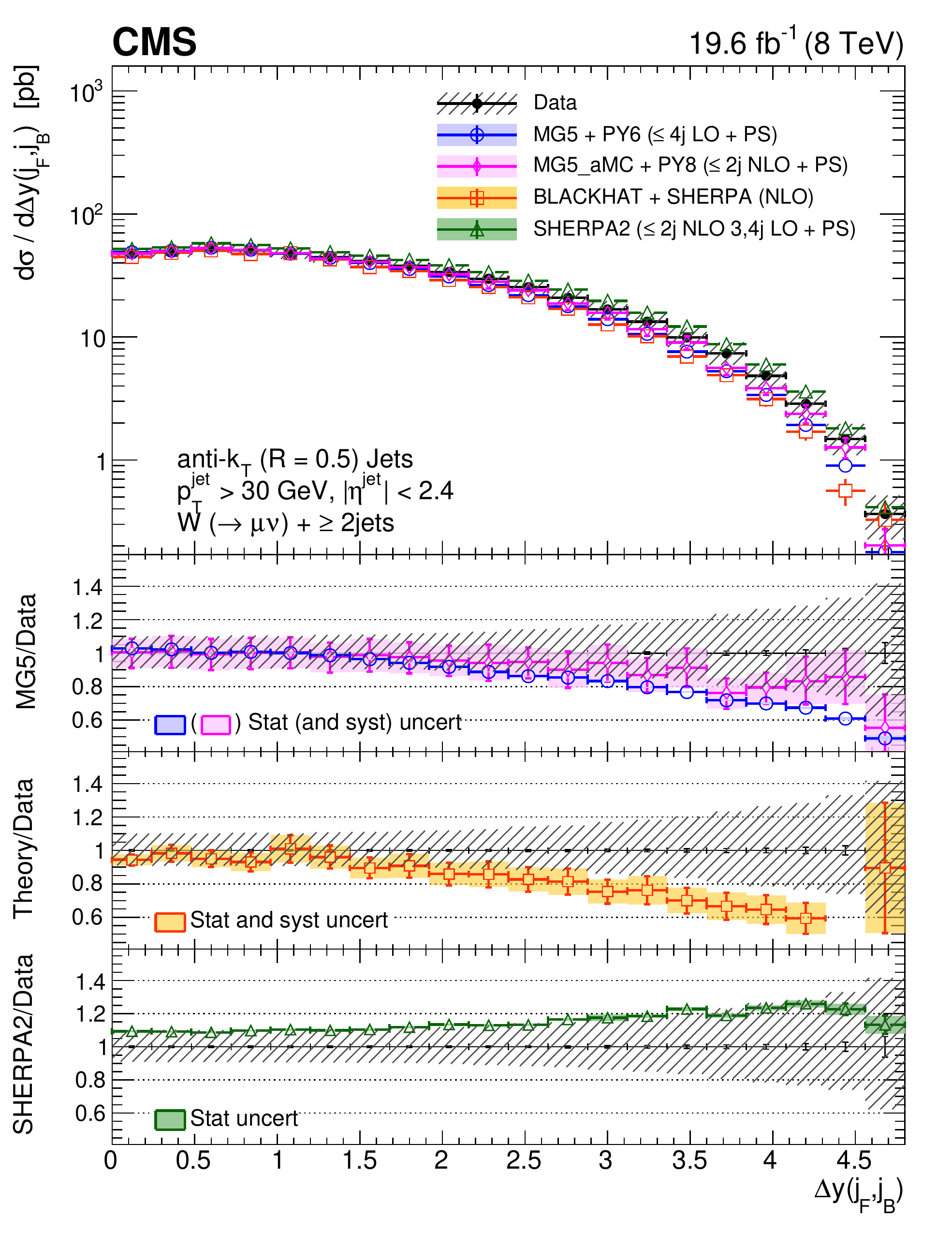}
    	 \includegraphics[width=0.48\textwidth]{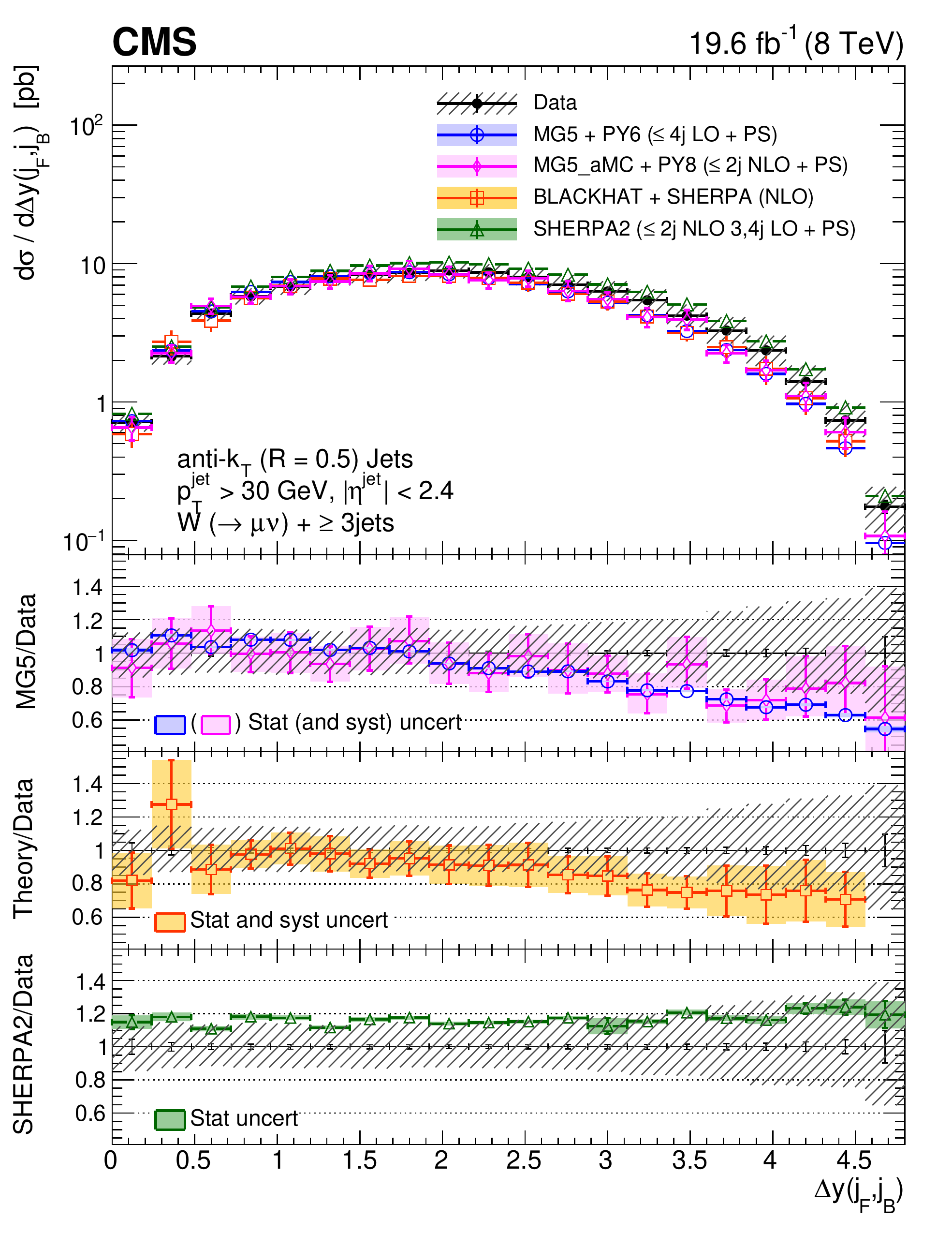}
    	 \includegraphics[width=0.48\textwidth]{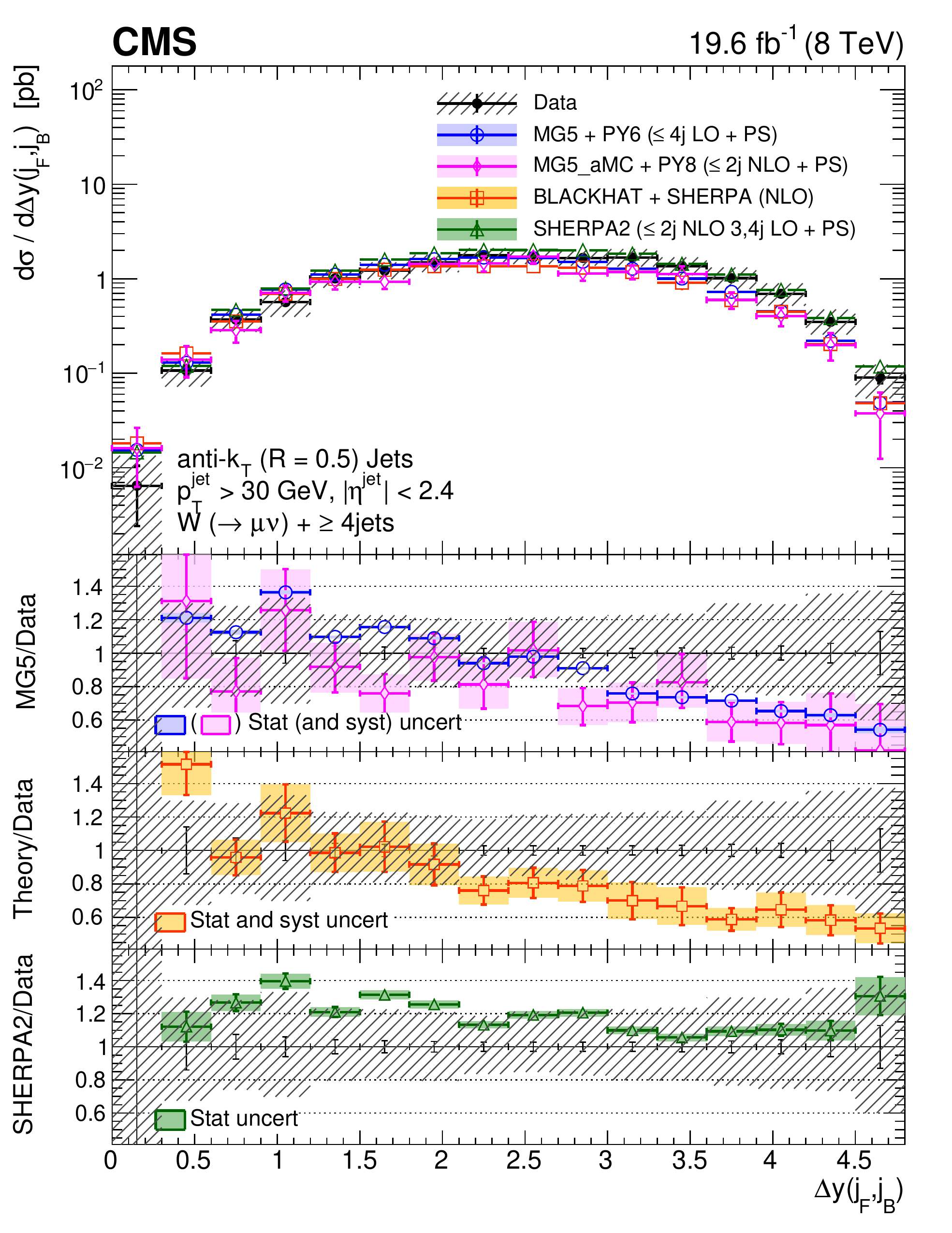}
        \caption{Cross sections differential in $\Delta y(j_F,j_B)$ for inclusive jet multiplicities 2--4, compared to the predictions of \MADGRAPH, \MGvATNLO, {\SHERPA 2}, and \BLACKHAT{}+\SHERPA (corrected for hadronization and multiple-parton interactions). Black circular markers with the gray hatched band represent the unfolded data measurements and their total uncertainties. Overlaid are the predictions together with their uncertainties. The lower plots show the ratio of each prediction to the unfolded data.}    	
    \label{xsec_dRapidityJetsFB}
\end{figure*}

\begin{figure*}[!htbp]
\centering
    	 \includegraphics[width=0.48\textwidth]{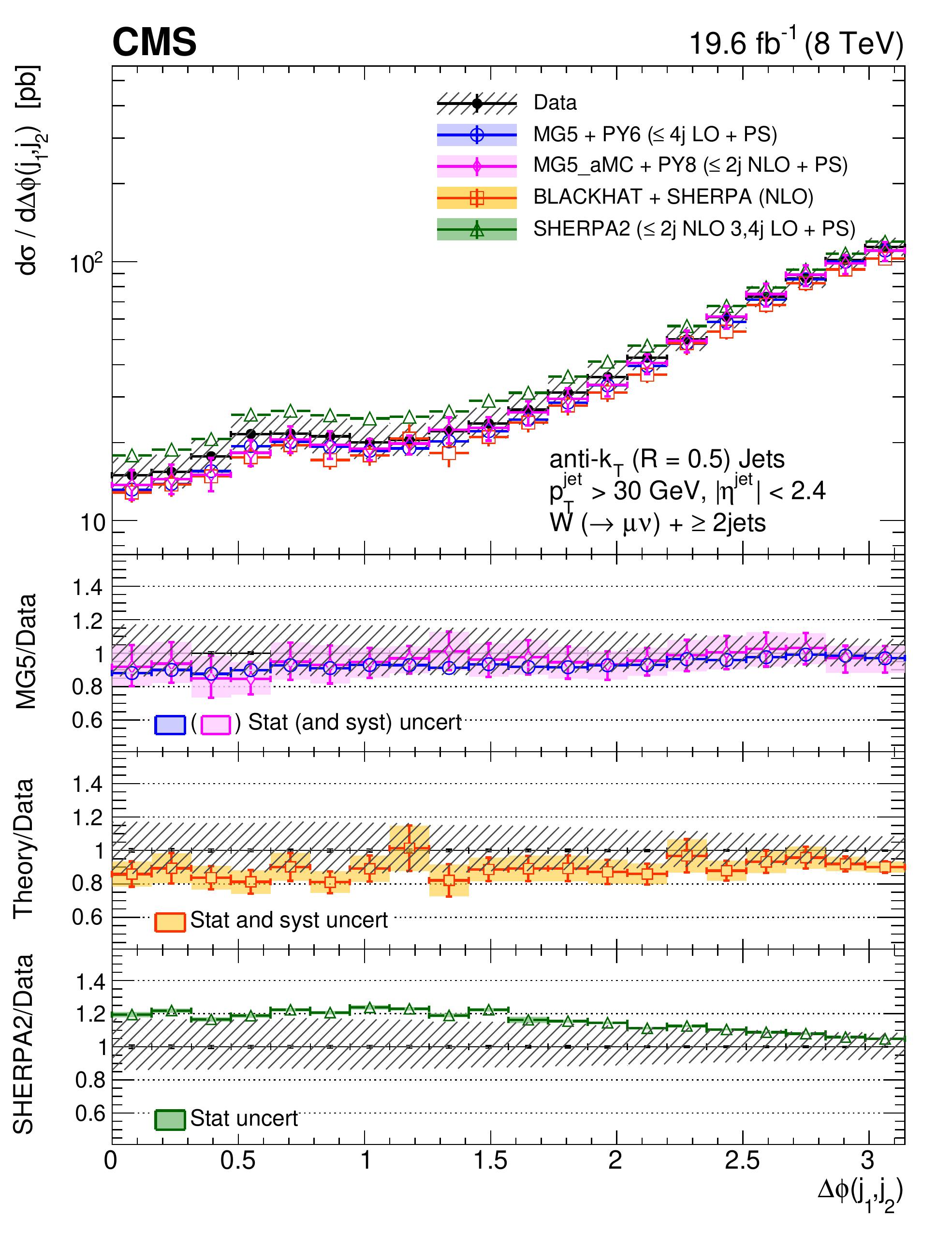}
    	 \includegraphics[width=0.48\textwidth]{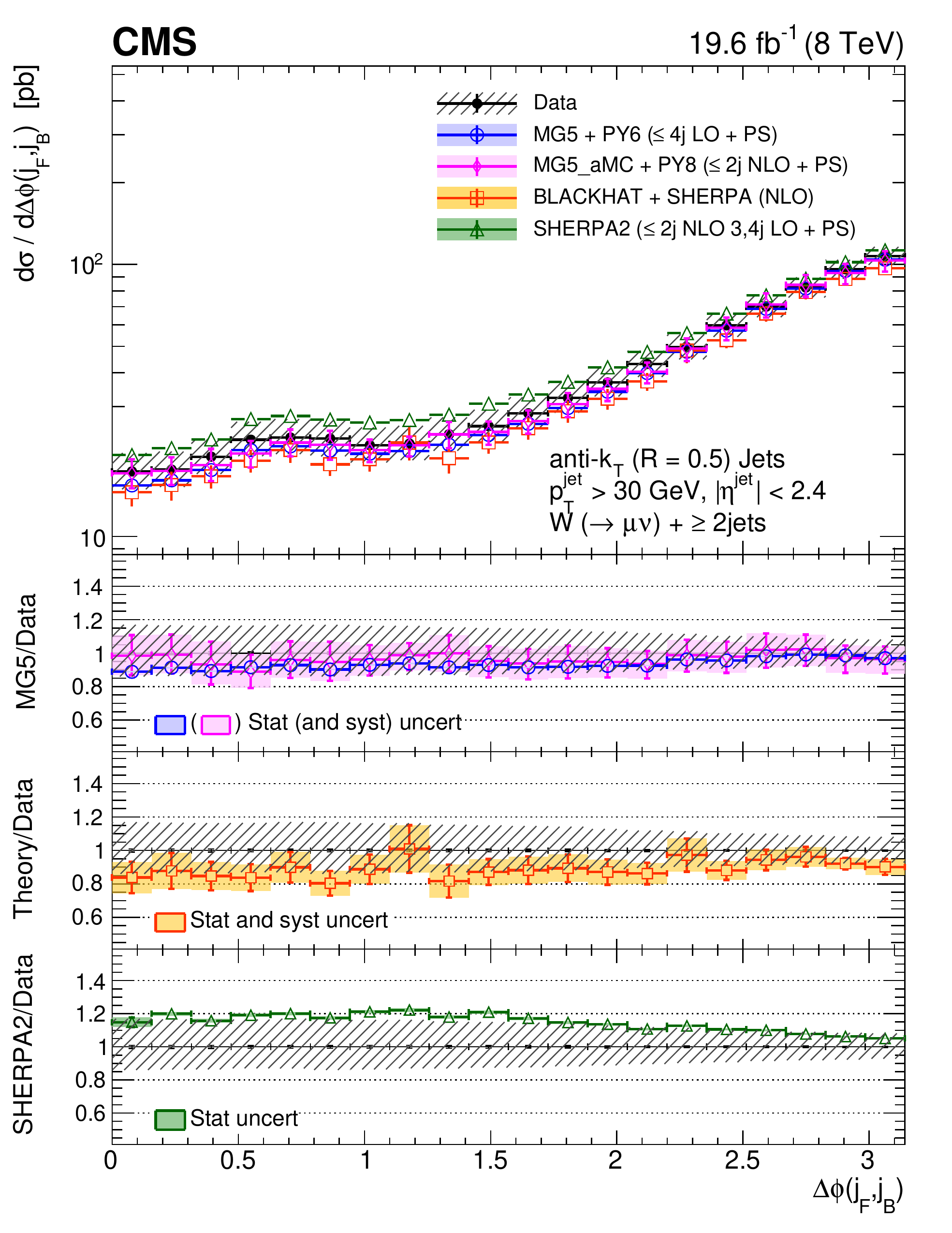}
    	\caption{Cross sections differential in $\Delta\phi(j_1,j_2)$ (left) and $\Delta\phi(j_F,j_B)$ (right) for an inclusive jet multiplicity of 2, compared to the predictions of \MADGRAPH, \MGvATNLO, {\SHERPA 2}, and \BLACKHAT{}+\SHERPA (corrected for hadronization and multiple-parton interactions). Black circular markers with the gray hatched band represent the unfolded data measurements and their total uncertainties. Overlaid are the predictions together with their uncertainties. The lower plots show the ratio of each prediction to the unfolded data.}
    \label{xsec_dPhiJets}
\end{figure*}

\begin{figure}[!htbp]
\centering
    	 \includegraphics[width=0.48\textwidth]{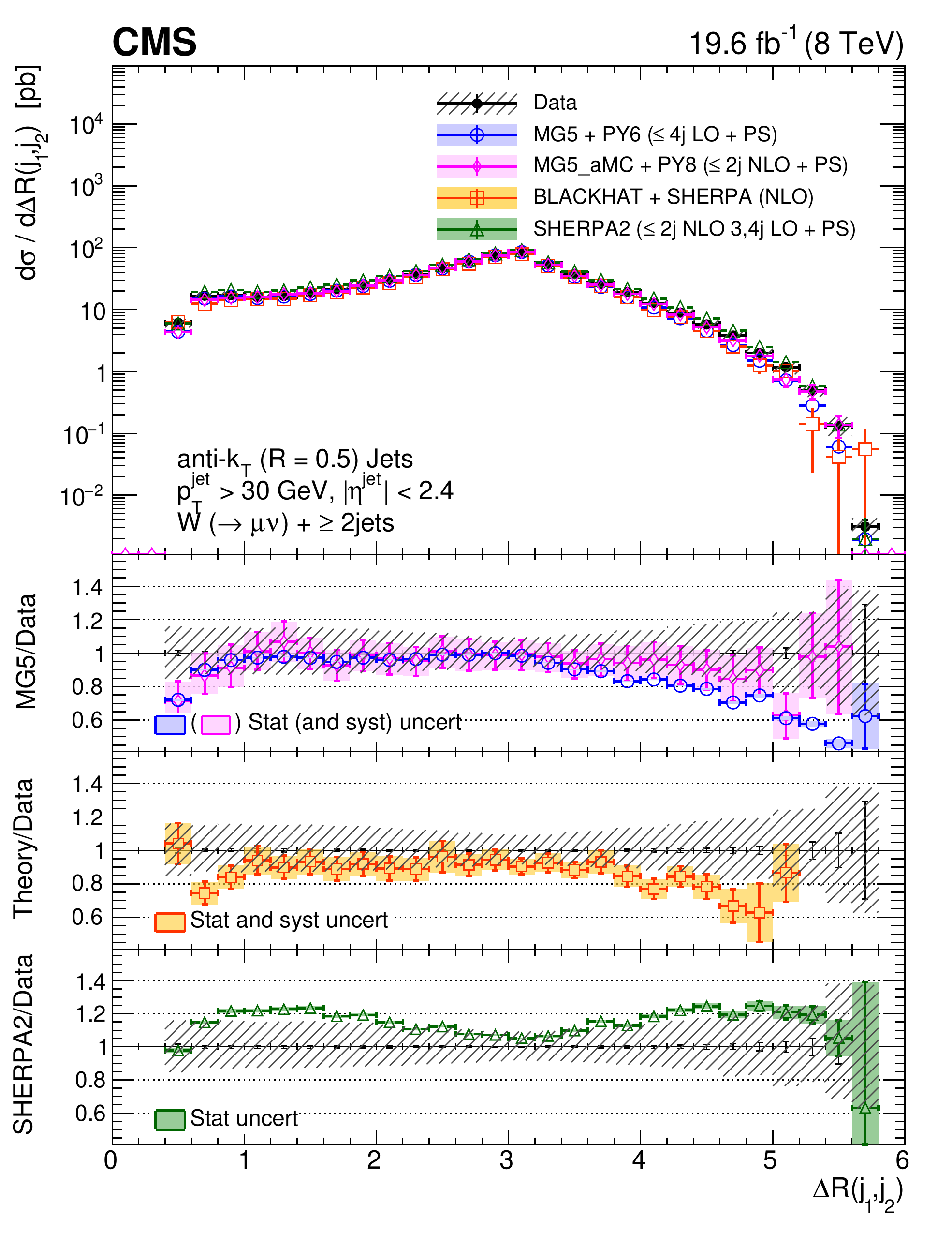}
    	\caption{Cross section differential in $\Delta R(j_{1},j_{2})$ for an inclusive jet multiplicity of 2, compared to the predictions of \MADGRAPH, \MGvATNLO, {\SHERPA 2}, and \BLACKHAT{}+\SHERPA (corrected for hadronization and multiple-parton interactions). Black circular markers with the gray hatched band represent the unfolded data measurements and their total uncertainties. Overlaid are the predictions together with their uncertainties. The lower plots show the ratio of each prediction to the unfolded data.}
    \label{xsec_dRJets}

\end{figure}

\begin{figure*}[!htbp]
\centering
         \includegraphics[width=0.48\textwidth]{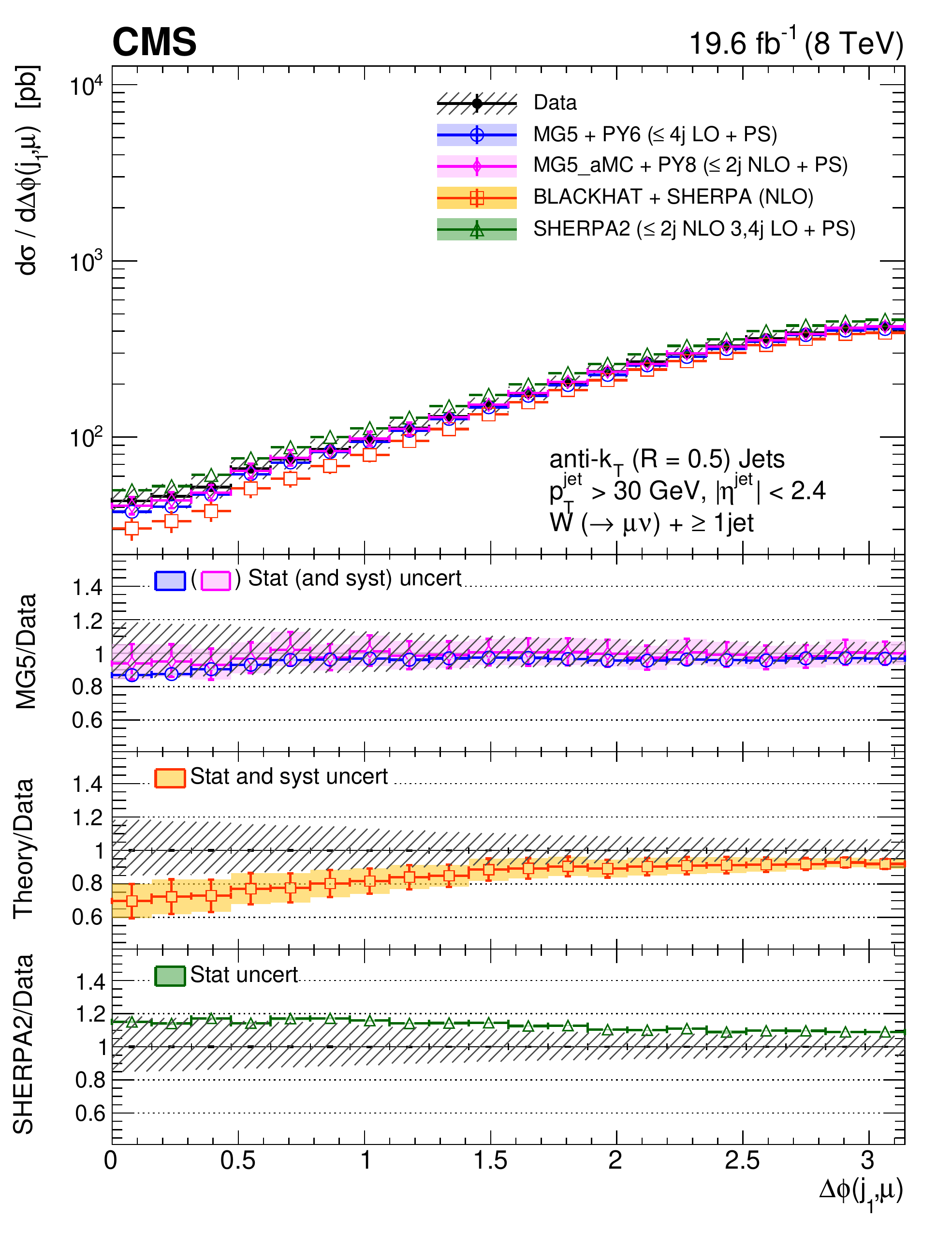}
         \includegraphics[width=0.48\textwidth]{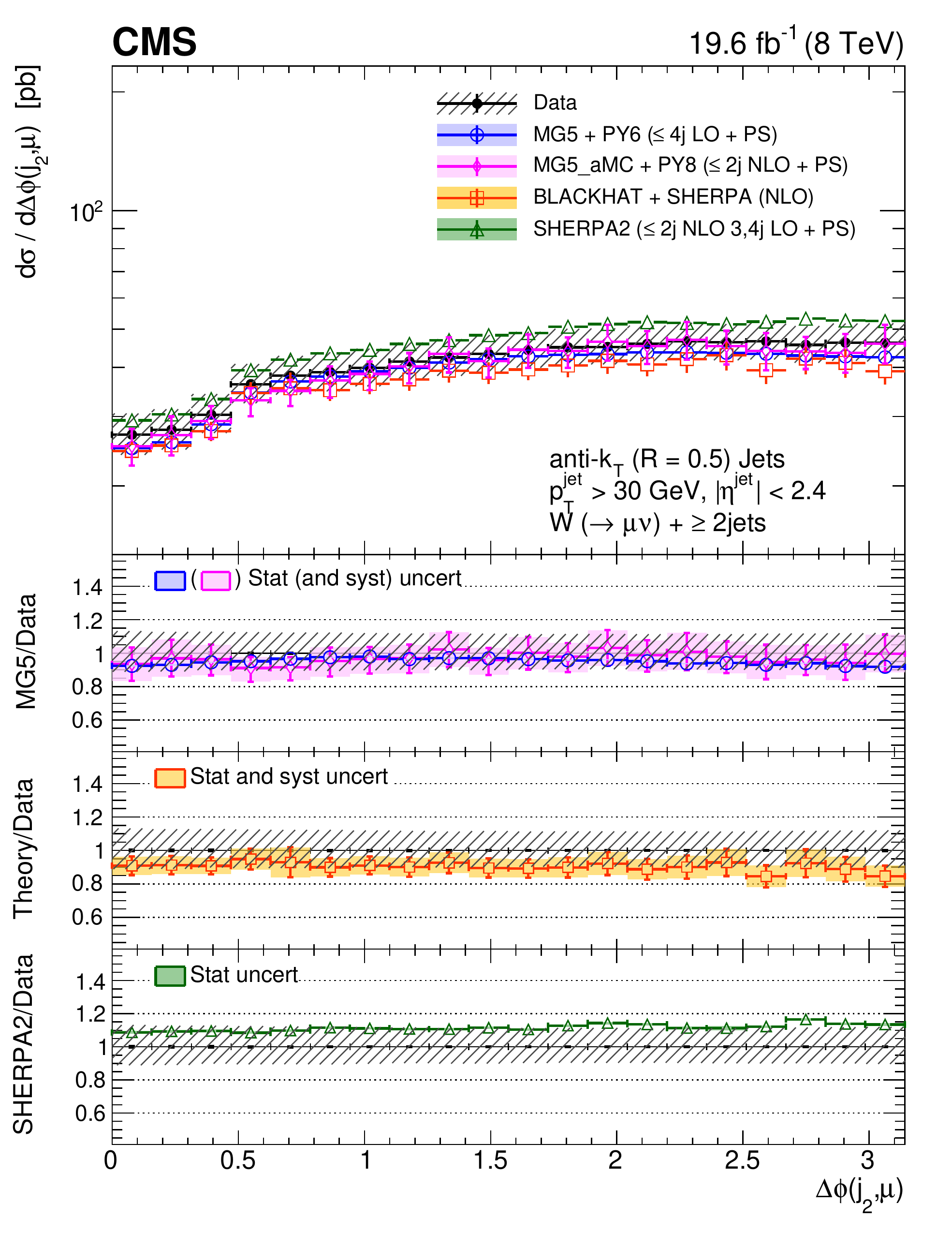}
         \includegraphics[width=0.48\textwidth]{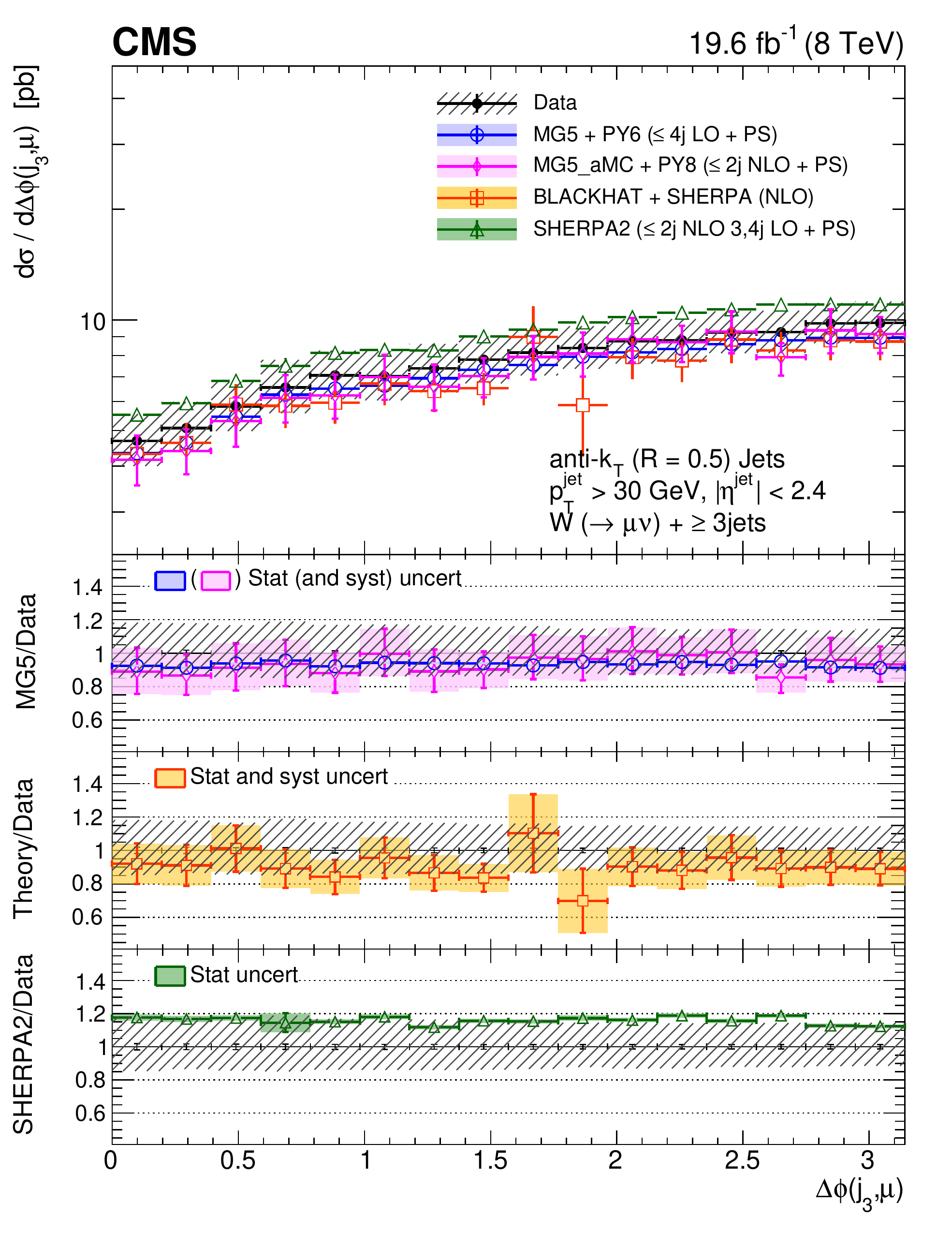}
         \includegraphics[width=0.48\textwidth]{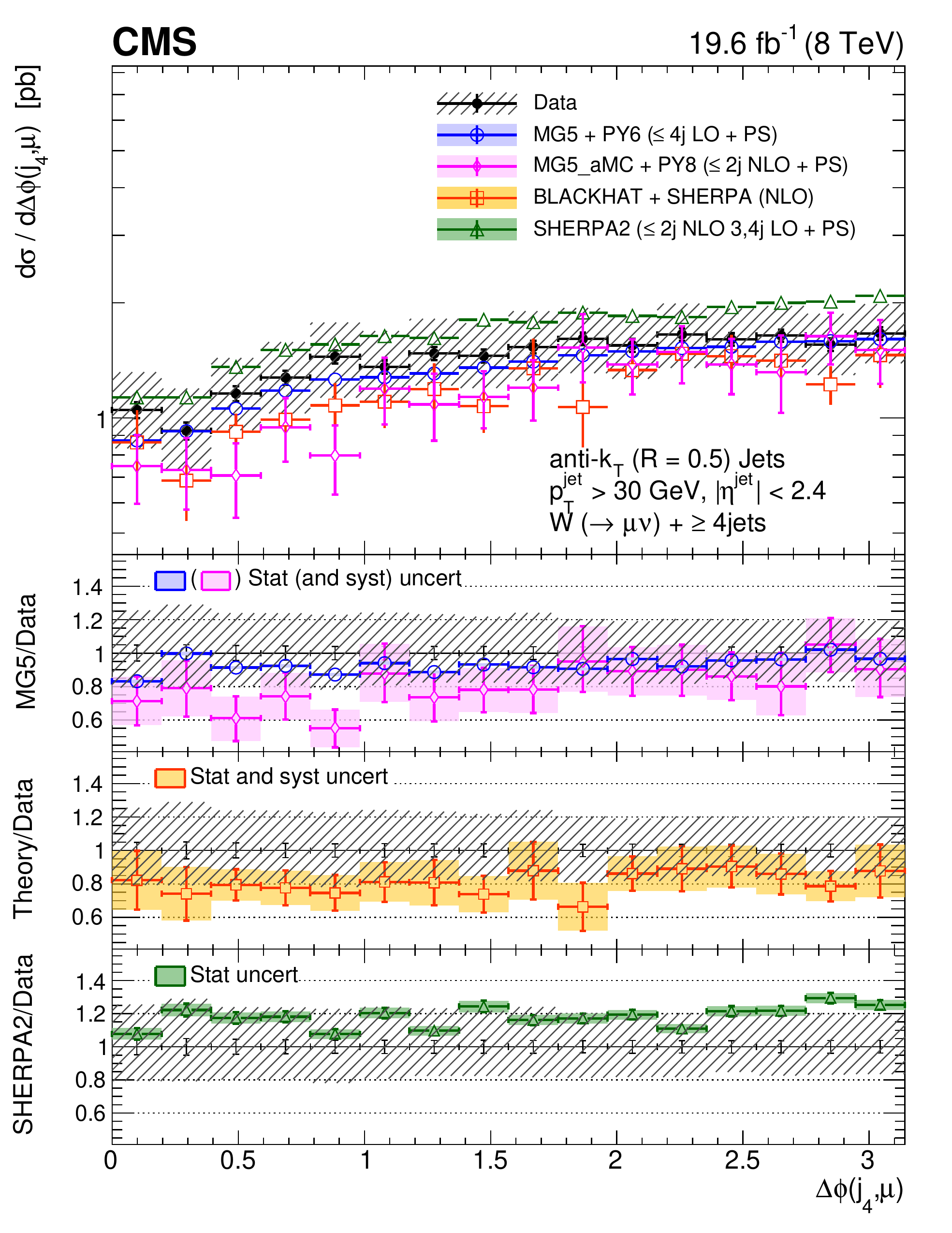}
         \caption{Cross sections differential in $\Delta\phi(j_{n},\mu)$ for inclusive jet multiplicities $n = 1$--4, compared to the predictions of \MADGRAPH, \MGvATNLO, {\SHERPA 2}, and \BLACKHAT{}+\SHERPA (corrected for hadronization and multiple-parton interactions). Black circular markers with the gray hatched band represent the unfolded data measurements and their total uncertainties. Overlaid are the predictions together with their uncertainties. The lower plots show the ratio of each prediction to the unfolded data.}
    \label{xsec_dPhiJetsMu}
\end{figure*}

\begin{figure*}[!htbp]
\centering
    	 \includegraphics[width=0.48\textwidth]{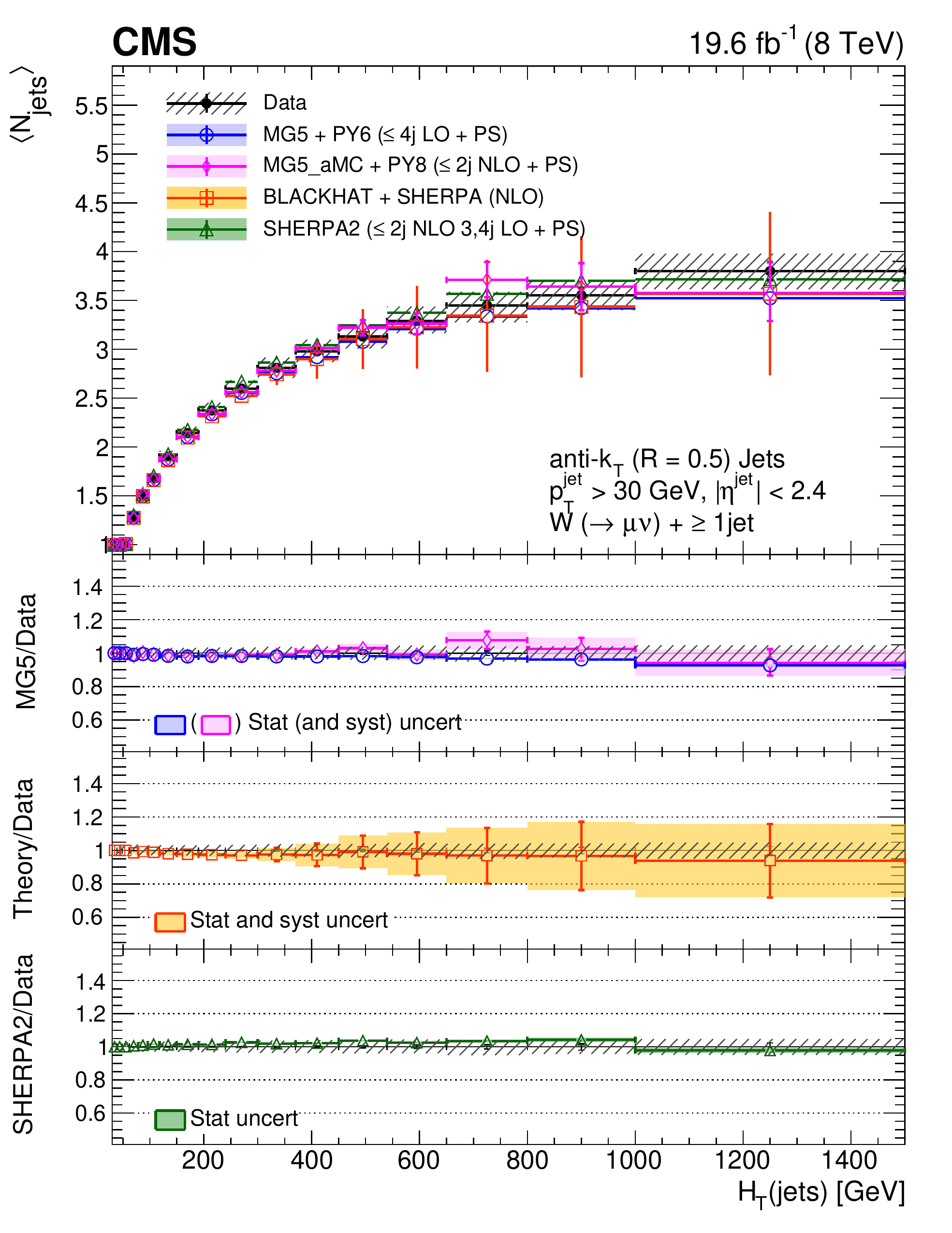}
    	 \includegraphics[width=0.48\textwidth]{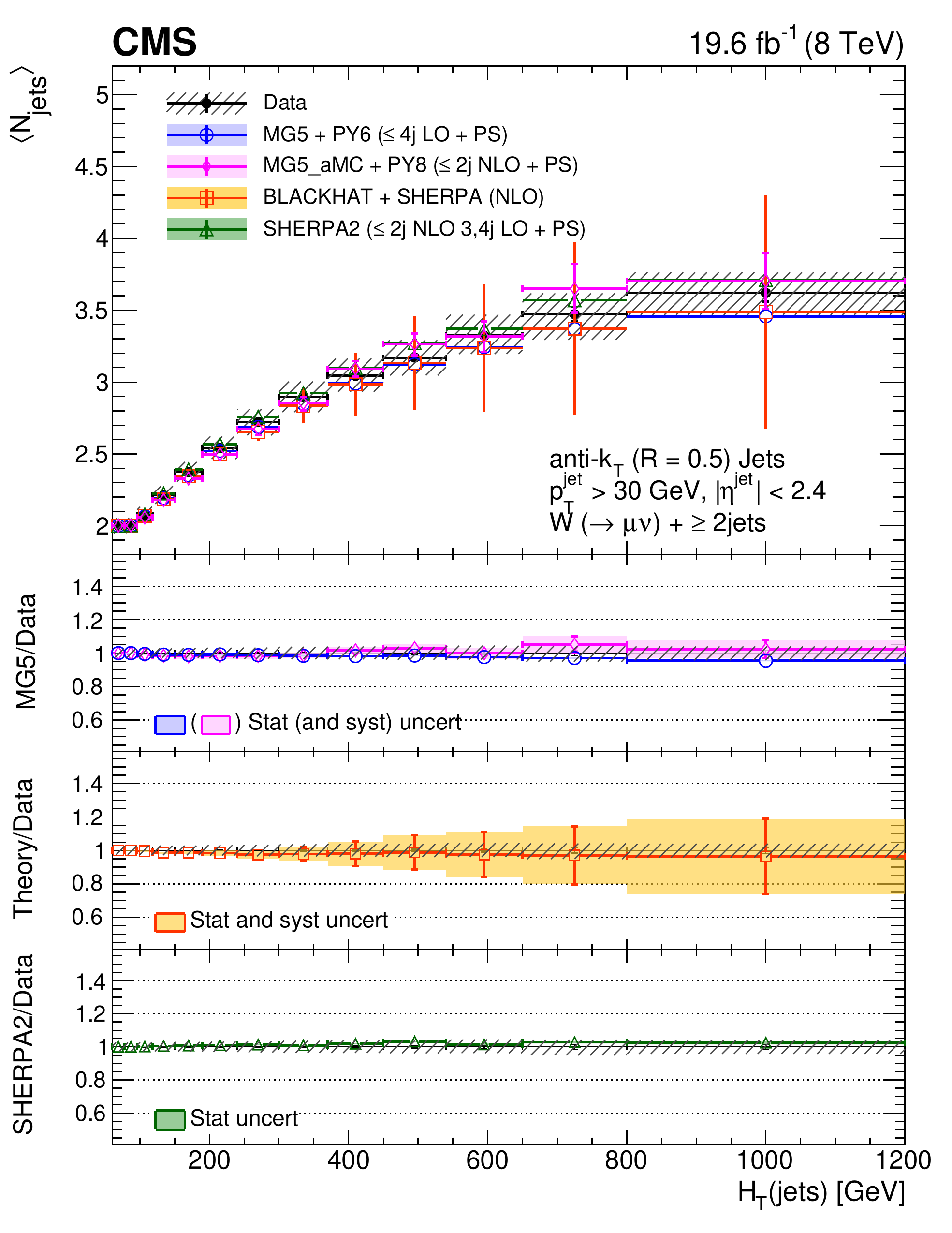}
    	 \includegraphics[width=0.48\textwidth]{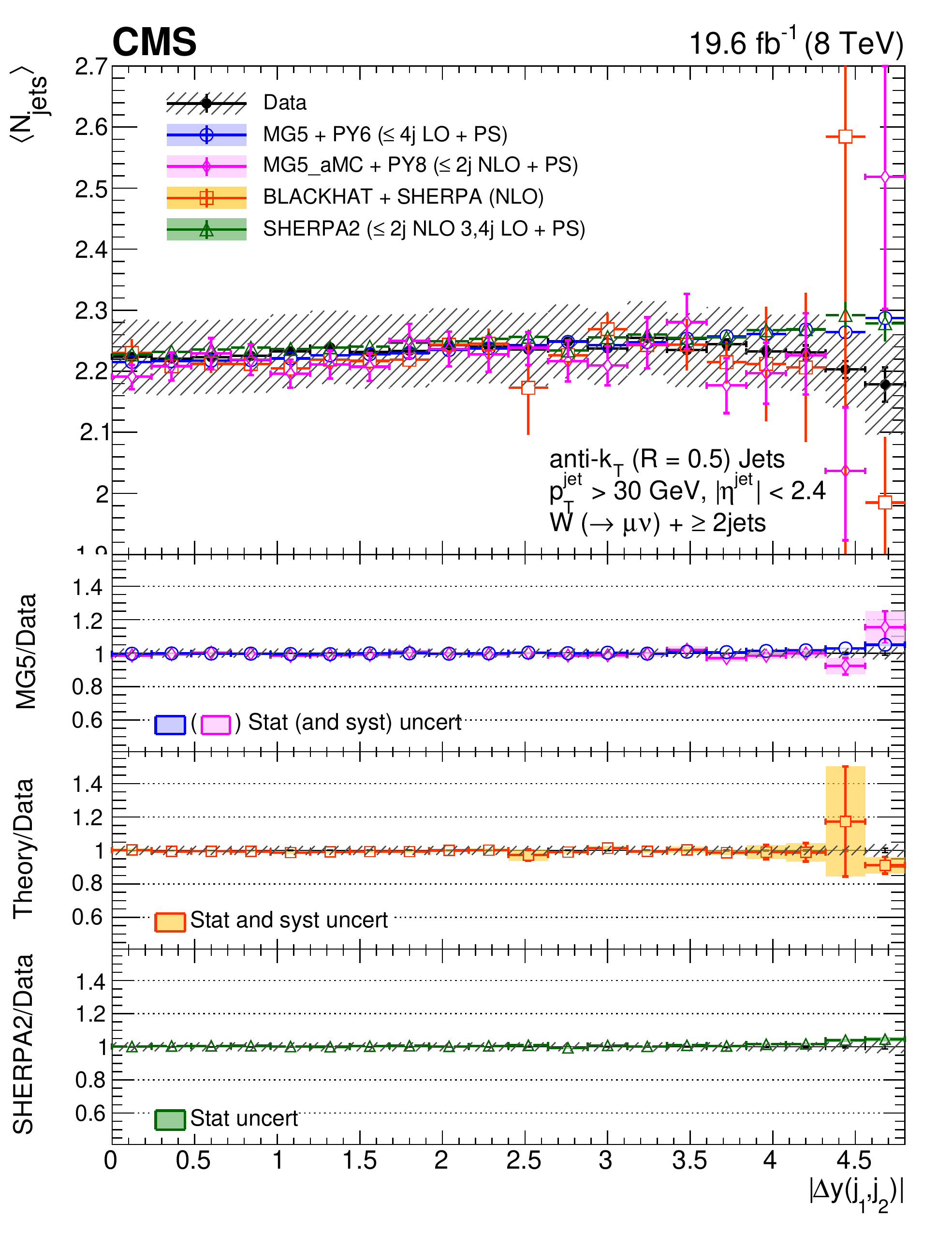}
    	 \includegraphics[width=0.48\textwidth]{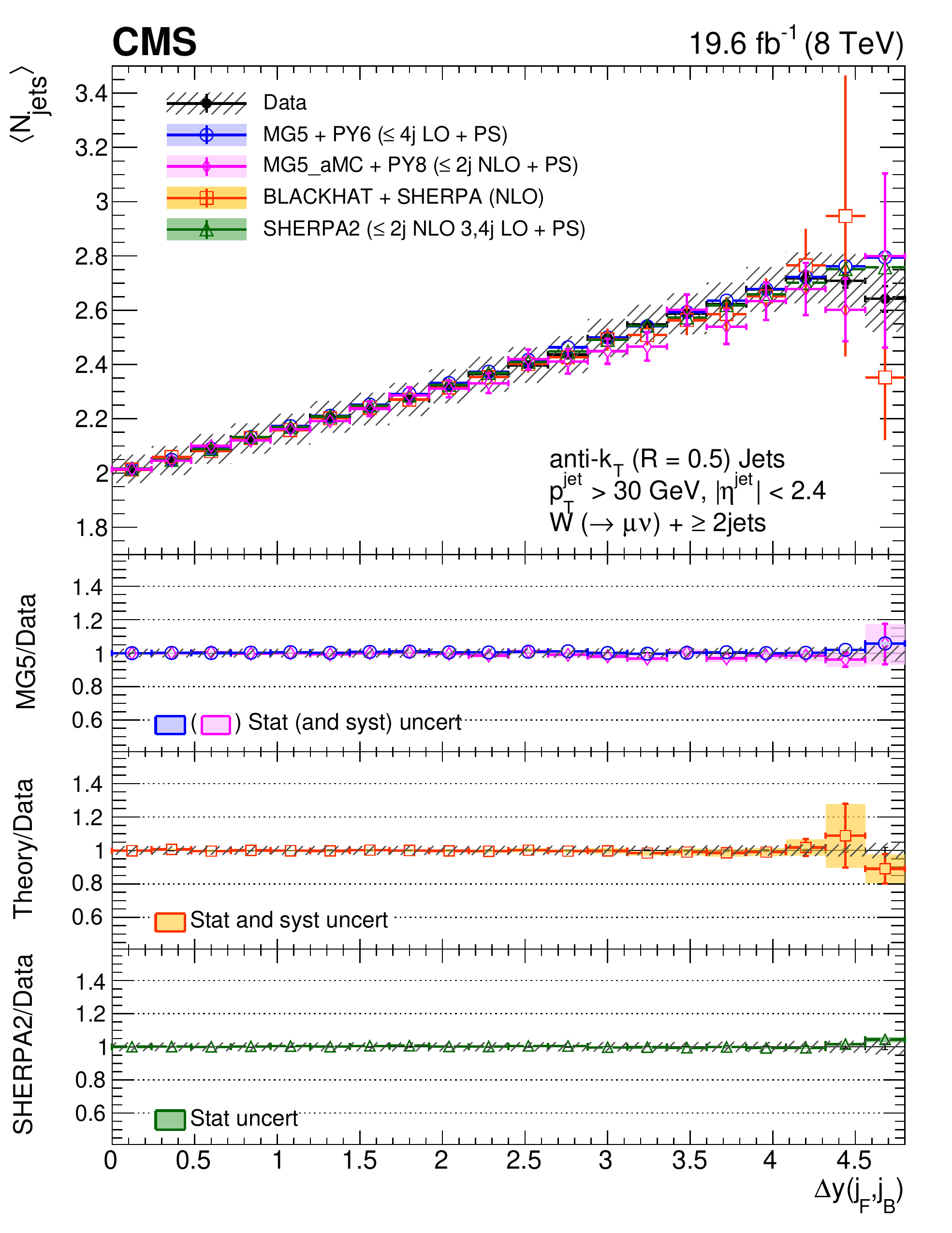}
    	\caption{Average number of jets $\langle N_\text{jets}\rangle$ as a function of $\HT$ for inclusive jet multiplicities 1--2 (top row) and as a function of $\Delta y(j_1,j_2)$ and $\Delta y(j_F,j_B)$ for an inclusive jet multiplicity of 2 (bottom row), compared to the predictions of \MADGRAPH, \MGvATNLO, {\SHERPA 2}, and \BLACKHAT{}+\SHERPA (corrected for hadronization and multiple-parton interactions). Black circular markers with the gray hatched band represent the unfolded data measurements and their total uncertainties. Overlaid are the predictions together with their uncertainties. The lower plots show the ratio of each prediction to the unfolded data.}
    \label{xsec_MeanNJets}
\end{figure*}

The measured $\PW(\to \mu\nu)$+jets fiducial cross sections are shown in Figs.~\ref{xsec_Njets}--\ref{xsec_MeanNJets} and compared
to the predictions of the LO MC generator \MADGRAPH{}5+\PYTHIA{}6 (described in Section~\ref{samples}), to those of \MGvATNLO and {\SHERPA 2} NLO MC generators,
and to the fixed-order theoretical predictions provided by \BLACKHAT{}+\SHERPA~\cite{BLACKHAT} and by a $\PW +$1 jet NNLO calculation~\cite{Boughezal:2015dva,Boughezal:2016dtm}. The 8\TeV data sample allows us to determine the cross sections for jet multiplicities up to 7 and to study the fiducial cross sections as functions of most kinematic observables for up to four jets.

{\tolerance=1200
An NLO prediction is provided by \MGvATNLO version 2.2.1~\cite{madgraphamcnlo}, a MC generator with up to three final-state partons,
with ME computation for up to two jets at NLO accuracy, which uses the NNPDF3.0 PDF set~\cite{nnpdf30}.
The generator is interfaced with {\PYTHIA 8}~\cite{PYTHIA8} for parton showering and hadronization, and the corresponding sample is denoted \MGvATNLO{}\-+\PYTHIA{}8 (denoted as MG5$\_$aMC+PY8 in the figure legends).
The merging of parton shower and ME is done with the {FxFx} merging scheme~\cite{fxfx} and the merging scale is set at 30\GeV.
The NNPDF2.3 PDF set~\cite{nnpdf23} and the {CUETP8M1} tune~\cite{cuetp8m1tune} are used in \PYTHIA{}8.
Using the weighting methods available in the generator~\cite{Frederix:2011ss}, PDF and scale uncertainties are assigned to the \MGvATNLO{}\-+\PYTHIA{}8 predictions
by considering the NNPDF3.0 PDF uncertainties, and by independently varying the factorization and renormalization scales by a factor of 0.5 or 2, excluding the combinations where one scale is varied by a factor of 0.5 and the other one by a factor of 2.
\par}

Another NLO prediction is provided by {\SHERPA } version 2.1.1, a multileg NLO MC generator with parton showering interfaced with \BLACKHAT~\cite{Berger:2008ag,Berger:2010gf} for the one-loop corrections. This sample of events is produced with the CT10 PDF set. The corresponding sample is denoted {\SHERPA~2}.
The {\SHERPA 2}~matrix element calculations include the production of up to four parton jets, with NLO accuracy for up to two jets and LO accuracy for three and four jets. The merging of parton showers and MEs is done with the MEPS@NLO method~\cite{Hoeche:2012yf,Hoeche:2011fd} and the merging scale set at $20\GeV$. The predictions from \MADGRAPH{}5+\PYTHIA{}6 and {\SHERPA 2} are shown with statistical uncertainties only.

The \BLACKHAT{}+\SHERPA calculation yields fixed-order NLO predictions for 8\TeV $\PW+n$ jets at the level of ME partons, where $n=1$--4. The choice of renormalization and factorization scales for \BLACKHAT{}+\SHERPA is
$\hat{H}_\mathrm{T}'/2$, where $\hat{H}_\mathrm{T}' = \sum_{m}^{} \pt^m + E_{\mathrm{T}}^\PW$, the sum running over final-state partons, and $E_{\mathrm{T}}^\PW$ being the transverse energy of the \PW{}~boson. A nonperturbative correction is applied to the \BLACKHAT{}+\SHERPA distributions to account for the effects of multiple-parton interactions and hadronization. This correction is determined with \MADGRAPH 5.1.1 interfaced with \PYTHIA 6.426 with and without hadronization and multiple-parton interactions. The nonperturbative correction factor is mostly in the range 0.90--1.20.
A PDF uncertainty is assigned to the predictions of \BLACKHAT{}+\SHERPA by considering the error sets of CT10 PDFs. A factorization and renormalization scale uncertainty is also assigned to \BLACKHAT{}+\SHERPA predictions, as determined by varying the scales simultaneously by a factor of 0.5 or 2.

An NNLO calculation of \PW+jet production in perturbative QCD ($N_\text{jetti}$ NNLO) is also used for comparisons with certain measured distributions (leading jet \pt, $\HT$, and $\abs{\eta}$, Figs.~\ref{xsec_JetPt1to4}, \ref{xsec_JetHT1to4}, and \ref{xsec_Jeteta1to4}) for $N_\text{jets} \geq 1$. The CT14 NNLO PDF set is used in the calculation. A nonperturbative correction is applied to this prediction, as in the case of \BLACKHAT{}+\SHERPA, as well as an additional correction factor of about 1.01 due to the effect of final-state radiation from the muon. A factorization and renormalization scale uncertainty is assigned to this prediction, as determined by varying the central scale $\sqrt{\smash[b]{m_{\ell\nu}^2+\bigl(\sum_\text{jet}^{} \pt^\text{jet}\bigr)}^2}$ by a factor of 0.5 or 2.

The measured exclusive and inclusive jet multiplicity distributions, shown in Fig.~\ref{xsec_Njets}, are in agreement with the predictions of the \MGvATNLO{}\-+\PYTHIA{}8 generators and with the calculation of \BLACKHAT{}+\SHERPA. For multiplicities above 5, {\SHERPA~2} starts to deviate upward from the measurement.

The cross sections differential in jet \pt for inclusive jet multiplicities from 1 to 4 are shown in Fig.~\ref{xsec_JetPt1to4}. The jet \pt and $\HT$ distributions are sensitive to the effects of higher-order processes. The current results extend to 1.0 and 1.5\TeV in the leading-jet \pt and $\HT$ distributions, respectively, for at least one jet. The predictions from \BLACKHAT{}+\SHERPA (jets 1 through 4) are in agreement with the measured distributions within the systematic uncertainties. The predictions from \MADGRAPH{}5+\PYTHIA{}6 show reasonable agreement with data, with the largest discrepancy being an overestimate of up to 20\% for the leading and second-leading jet \pt distributions in the intermediate-\pt region. In comparison to the corresponding measurements of the leading and second-leading jet \pt spectra made by CMS with 7\TeV data~\cite{Wjets7Tev}, we observe a smaller slope in the ratio of the \MADGRAPH{}5+\PYTHIA{}6 prediction to the measurement.
The predictions from \MGvATNLO{}\-+\PYTHIA{}8 are in agreement with data within uncertainties. The NNLO prediction for at least one jet agrees with the unfolded jet \pt cross section within the systematic uncertainties. At low \pt values (below 50 \GeV), the predictions for the first-, second-, and third-leading jet \pt from {\SHERPA 2} overestimate the data.

The $\HT$ distributions for inclusive jet multiplicities of 1 to 4 are shown in Fig.~\ref{xsec_JetHT1to4}. The $\HT$ distributions are best modeled by the NNLO prediction for an inclusive jet multiplicity of 1, and by \MADGRAPH{}5+\PYTHIA{}6 and \MGvATNLO{}\-+\PYTHIA{}8 for inclusive jet multiplicities of 1 and 2. For higher jet multiplicities, the \MADGRAPH{}5+\PYTHIA{}6 and \MGvATNLO{}\-+\PYTHIA{}8 predictions underestimate the data at low values of $\HT$ (below 200\GeV). The {\SHERPA 2} predictions for $\HT$ consistently overestimate the data for all inclusive jet multiplicities and display a harder $\HT$ spectrum.
The \BLACKHAT{}+\SHERPA prediction underestimates the data $\HT$ distribution
for $N_\text{jets} \geq 1$, as expected because the NLO prediction for $\HT$ for $N_\text{jets} \geq 1$ is a fixed-order prediction with up to two real partons, and contributions from higher jet multiplicities are missing~\cite{Maitre:2013wha}.

The dijet \pt and invariant mass spectra for inclusive jet multiplicities of 2, 3, and 4 are shown in Figs.~\ref{xsec_diJetPt}~and~\ref{xsec_diJetMass}. Dijet quantities are based on the two leading jets in the event, and they constitute an important test of the modeling of \pt correlations among jets, whose correct accounting is crucial for searches for physics beyond the SM in dijet final states. All of the predictions agree reasonably well with data, but {\SHERPA 2} consistently overestimates the data for high values of dijet \pt and invariant mass, particularly in the dijet \pt spectrum for $N_\text{jets} \geq 2$. The \MGvATNLO{}\-+\PYTHIA{}8 prediction also underestimates the data for values of the invariant mass below 200\GeV in the inclusive four-jet distribution.

The dependence of the cross section on several angular variables and angular correlations between jets is also studied. The pseudorapidity distributions for the four leading jets in each event are shown in Fig.~\ref{xsec_Jeteta1to4}. The cross sections are best predicted by \MADGRAPH{}5+\PYTHIA{}6 and \MGvATNLO+\PYTHIA{}8. All predictions agree with the data, with some variations in the overall normalization and a slight underestimation for large values of $\abs{\eta}$.

The distributions of the rapidity difference and the azimuthal angles between \pt-ordered and rapidity-ordered jets are shown in Figs.~\ref{xsec_dRapidityJets}--\ref{xsec_dPhiJets}.
The measurement of the rapidity difference between \pt-ordered jets is shown for different jet pairings: the two leading jets $\Delta y(j_1,j_2)$ and the first- (second-) and third-leading jets $\Delta y(j_1,j_3)$ ($\Delta y(j_2,j_3)$). The measurement of the rapidity difference between rapidity-ordered jets makes use of the most forward and most backward jets, $\Delta y(j_{\rm F},j_{\rm B})$. The quantities $\Delta y(j_1,j_2)$ and $\Delta y(j_{\rm F},j_{\rm B})$ are studied for inclusive jet multiplicities of 2 to 4, while $\Delta y(j_1,j_3)$ and $\Delta y(j_2,j_3)$ are studied for $N_\text{jets} \geq 3$. A study of the rapidity difference between the two leading jets is helpful in testing the wide-angle soft parton radiation and the implementation of parton showering. The measurement of the rapidity differences between the forward/backward jets is also instrumental in understanding QCD radiation and wide-angle particle emission. The distribution of the azimuthal angle difference is sensitive to higher-order processes and is shown for \pt-ordered and rapidity-ordered jets for an inclusive multiplicity of 2.  Overall, the predicted distributions of the rapidity difference between \pt-ordered jets are in agreement with the measurements, with \MADGRAPH{}5+\PYTHIA{}6 and \BLACKHAT{}+\SHERPA underestimating the data for $\abs{\Delta y}$ values above 2. A similar observation can be made for the rapidity difference between the most forward and most backward jets. This behavior is also reflected in the $\Delta R(j_{1},j_{2})$ measurement, shown in Fig.~\ref{xsec_dRJets}. All predictions for the azimuthal angle difference between jets are in agreement with data, with some variations in the overall normalization.

The distributions of the azimuthal angle between the muon and the leading jet, for inclusive jet multiplicities 1 to 4, are shown in Fig.~\ref{xsec_dPhiJetsMu}. Overall, the predictions are in agreement with the measurements, except for \BLACKHAT{}+\SHERPA, which disagrees with the data at low values of the $\Delta\phi$ for an inclusive jet multiplicity of 1.

Finally, the average number of jets, $\langle N_\text{jets}\rangle$, is shown as a function of $\HT$, $\Delta y(j_1,j_2)$, and $\Delta y(j_{\rm F},j_{\rm B})$ in the inclusive two-jet events in Fig.~\ref{xsec_MeanNJets}. In the high-$\HT$ region, which is particularly sensitive to higher-order processes, the average number of jets plateaus around a value of 3.5. Although \MADGRAPH{}5+\PYTHIA{}6 tends to underestimate $\langle N_\text{jets}\rangle$  and {\SHERPA 2} tends to overestimate it, the deviations are not significant and both generators appear to adequately reproduce the data. Good agreement is observed between the data and all predictions for the dependence of $\langle N_\text{jets}\rangle$ on the \pt-ordered and rapidity-ordered rapidity differences.
These measurements provide an important test of the implementation and modeling of wide-angle gluon emission in the MC generators and NLO calculations. Overall, the accuracy of the predictions for $\langle N_\text{jets}\rangle$ is much better than was found at the Tevatron~\cite{PhysRevD.88.092001}.

\clearpage

\section{Summary}
\label{summary}

Differential cross sections for a $\PW$ boson produced in association with jets in pp collisions at a center-of-mass energy of 8\TeV were measured.
The data correspond to an integrated luminosity of 19.6\fbinv and were collected with the CMS detector at the LHC.

Cross sections measured using the muonic decay mode of the $\PW$ boson were presented
as functions of the jet multiplicity, the transverse momenta and
pseudorapidities of the four leading jets, $\HT$ for jet multiplicities up to four, and the dijet \pt and invariant mass. Cross sections were also presented as functions of several angular correlation variables: rapidity difference, azimuthal angle difference, and $\Delta R$ between \pt-ordered and rapidity-ordered jets, and azimuthal angle difference between the muon and the leading jet. The dependence of the average number of jets on $\HT$ and on rapidity differences between jets was examined.

The results were corrected for detector effects by means of regularized unfolding and
compared with particle-level simulated predictions using \MADGRAPH{}5+\PYTHIA{}6; {\SHERPA 2} and \MGvATNLO{}\-+\PYTHIA{}8 (multileg NLO); \BLACKHAT{}+\SHERPA (NLO); and $N_\text{jetti}$ NNLO. We expect that predictions made at higher order from NLO and NNLO generally give a better description of our data.

The NNLO predictions for $\PW$+1 jet production were compared with the measured cross sections differential in leading jet \pt, $\HT$, and leading jet $\abs{\eta}$ and agree with data within uncertainties.

The predictions generally describe the jet multiplicity within the uncertainties, with increasing deviations observed in {\SHERPA 2} for jet multiplicities greater than 4. The cross sections differential in the \pt of the three leading jets are overestimated by \MADGRAPH{}5+\PYTHIA{}6 in a region of intermediate \pt, and by {\SHERPA 2} at low \pt. The cross sections as functions of jet \pt predicted by \BLACKHAT{}+\SHERPA and by \MGvATNLO{}\-+\PYTHIA{}8 agree with the measurements within uncertainties.

The cross section as a function of \HT is underestimated by \BLACKHAT{}+\SHERPA for $N_\text{jets} \geq 1$ because the contribution from $\PW+ {\geq}3$~jets is missing from an NLO prediction of $\PW+ {\geq}1$~jet. The corresponding predictions from {\SHERPA 2} overestimate the cross section, particularly at high~$\HT$.

The predictions for the jet $\abs{\eta}$ distribution deviate from the measurements for large values of $\abs{\eta}$, as do the predictions for the angular correlation distributions in rapidity for large rapidity differences. Improvement in describing the data at high rapidity difference and at low azimuthal angle difference between muon and jet is observed when considering \MGvATNLO{}\-+\PYTHIA{}8 versus tree-level \MADGRAPH{}5+\PYTHIA{}6. The distribution of the azimuthal angle between the muon and the leading jet is not well modeled by \BLACKHAT{}+\SHERPA for $N_\text{jets} \geq 1$. The predictions for the correlation distributions in azimuthal angle between jets agree with the measurements, as well as the dependence of the average number of jets on angular correlation variables and on $\HT$.

\begin{acknowledgments}
\hyphenation{Bundes-ministerium Forschungs-gemeinschaft Forschungs-zentren Rachada-pisek}
We extend our thanks to Daniel Ma{\^i}tre for the \BLACKHAT{}+\SHERPA predictions, and to Radja Boughezal for the NNLO predictions.

We congratulate our colleagues in the CERN accelerator departments for the excellent performance of the LHC and thank the technical and administrative staffs at CERN and at other CMS institutes for their contributions to the success of the CMS effort. In addition, we gratefully acknowledge the computing centers and personnel of the Worldwide LHC Computing Grid for delivering so effectively the computing infrastructure essential to our analyses. Finally, we acknowledge the enduring support for the construction and operation of the LHC and the CMS detector provided by the following funding agencies: the Austrian Federal Ministry of Science, Research and Economy and the Austrian Science Fund; the Belgian Fonds de la Recherche Scientifique, and Fonds voor Wetenschappelijk Onderzoek; the Brazilian Funding Agencies (CNPq, CAPES, FAPERJ, and FAPESP); the Bulgarian Ministry of Education and Science; CERN; the Chinese Academy of Sciences, Ministry of Science and Technology, and National Natural Science Foundation of China; the Colombian Funding Agency (COLCIENCIAS); the Croatian Ministry of Science, Education and Sport, and the Croatian Science Foundation; the Research Promotion Foundation, Cyprus; the Secretariat for Higher Education, Science, Technology and Innovation, Ecuador; the Ministry of Education and Research, Estonian Research Council via IUT23-4 and IUT23-6 and European Regional Development Fund, Estonia; the Academy of Finland, Finnish Ministry of Education and Culture, and Helsinki Institute of Physics; the Institut National de Physique Nucl\'eaire et de Physique des Particules~/~CNRS, and Commissariat \`a l'\'Energie Atomique et aux \'Energies Alternatives~/~CEA, France; the Bundesministerium f\"ur Bildung und Forschung, Deutsche Forschungsgemeinschaft, and Helmholtz-Gemeinschaft Deutscher Forschungszentren, Germany; the General Secretariat for Research and Technology, Greece; the National Scientific Research Foundation, and National Innovation Office, Hungary; the Department of Atomic Energy and the Department of Science and Technology, India; the Institute for Studies in Theoretical Physics and Mathematics, Iran; the Science Foundation, Ireland; the Istituto Nazionale di Fisica Nucleare, Italy; the Ministry of Science, ICT and Future Planning, and National Research Foundation (NRF), Republic of Korea; the Lithuanian Academy of Sciences; the Ministry of Education, and University of Malaya (Malaysia); the Mexican Funding Agencies (BUAP, CINVESTAV, CONACYT, LNS, SEP, and UASLP-FAI); the Ministry of Business, Innovation and Employment, New Zealand; the Pakistan Atomic Energy Commission; the Ministry of Science and Higher Education and the National Science Centre, Poland; the Funda\c{c}\~ao para a Ci\^encia e a Tecnologia, Portugal; JINR, Dubna; the Ministry of Education and Science of the Russian Federation, the Federal Agency of Atomic Energy of the Russian Federation, Russian Academy of Sciences, and the Russian Foundation for Basic Research; the Ministry of Education, Science and Technological Development of Serbia; the Secretar\'{\i}a de Estado de Investigaci\'on, Desarrollo e Innovaci\'on and Programa Consolider-Ingenio 2010, Spain; the Swiss Funding Agencies (ETH Board, ETH Zurich, PSI, SNF, UniZH, Canton Zurich, and SER); the Ministry of Science and Technology, Taipei; the Thailand Center of Excellence in Physics, the Institute for the Promotion of Teaching Science and Technology of Thailand, Special Task Force for Activating Research and the National Science and Technology Development Agency of Thailand; the Scientific and Technical Research Council of Turkey, and Turkish Atomic Energy Authority; the National Academy of Sciences of Ukraine, and State Fund for Fundamental Researches, Ukraine; the Science and Technology Facilities Council, UK; the US Department of Energy, and the US National Science Foundation.

Individuals have received support from the Marie-Curie program and the European Research Council and EPLANET (European Union); the Leventis Foundation; the A. P. Sloan Foundation; the Alexander von Humboldt Foundation; the Belgian Federal Science Policy Office; the Fonds pour la Formation \`a la Recherche dans l'Industrie et dans l'Agriculture (FRIA-Belgium); the Agentschap voor Innovatie door Wetenschap en Technologie (IWT-Belgium); the Ministry of Education, Youth and Sports (MEYS) of the Czech Republic; the Council of Science and Industrial Research, India; the HOMING PLUS program of the Foundation for Polish Science, cofinanced from European Union, Regional Development Fund, the Mobility Plus program of the Ministry of Science and Higher Education, the National Science Center (Poland), contracts Harmonia 2014/14/M/ST2/00428, Opus 2013/11/B/ST2/04202, 2014/13/B/ST2/02543 and 2014/15/B/ST2/03998, Sonata-bis 2012/07/E/ST2/01406; the Thalis and Aristeia programs cofinanced by EU-ESF and the Greek NSRF; the National Priorities Research Program by Qatar National Research Fund; the Programa Clar\'in-COFUND del Principado de Asturias; the Rachadapisek Sompot Fund for Postdoctoral Fellowship, Chulalongkorn University and the Chulalongkorn Academic into Its 2nd Century Project Advancement Project (Thailand); and the Welch Foundation, contract C-1845. \end{acknowledgments}
\bibliography{auto_generated}

\cleardoublepage \appendix\section{The CMS Collaboration \label{app:collab}}\begin{sloppypar}\hyphenpenalty=5000\widowpenalty=500\clubpenalty=5000\input{SMP-14-023-authorlist.tex}\end{sloppypar}
\end{document}

%% file: SMP-14-023-authorlist.tex
\textbf{Yerevan Physics Institute,  Yerevan,  Armenia}\\*[0pt]
V.~Khachatryan, A.M.~Sirunyan, A.~Tumasyan
\vskip\cmsinstskip
\textbf{Institut f\"{u}r Hochenergiephysik,  Wien,  Austria}\\*[0pt]
W.~Adam, E.~Asilar, T.~Bergauer, J.~Brandstetter, E.~Brondolin, M.~Dragicevic, J.~Er\"{o}, M.~Flechl, M.~Friedl, R.~Fr\"{u}hwirth\cmsAuthorMark{1}, V.M.~Ghete, C.~Hartl, N.~H\"{o}rmann, J.~Hrubec, M.~Jeitler\cmsAuthorMark{1}, A.~K\"{o}nig, I.~Kr\"{a}tschmer, D.~Liko, T.~Matsushita, I.~Mikulec, D.~Rabady, N.~Rad, B.~Rahbaran, H.~Rohringer, J.~Schieck\cmsAuthorMark{1}, J.~Strauss, W.~Treberer-Treberspurg, W.~Waltenberger, C.-E.~Wulz\cmsAuthorMark{1}
% ...
\vskip\cmsinstskip
% ...
1:~~Also at Vienna University of Technology, Vienna, Austria\\
% ...
